\documentclass[twoside,11pt]{article}

\usepackage{blindtext}

\usepackage{jmlr2e}
\hypersetup{hidelinks}

\usepackage{algorithm}      
\usepackage[noend]{algpseudocode} 
\usepackage{amsmath}
\usepackage{mathrsfs}
\usepackage{multirow}
\usepackage{ulem}
\usepackage{color}

\def\V{\mathbf{V}}
\def\W{\mathbf{W}}
\def\X{\mathbf{X}}

\def\Z{\mathbf{Z}}
\def\T{\mathbf{T}}

\def\Ssc{\mathcal{S}}
\def\Tsc{\mathcal{T}}

\def\e{\mathbf{e}}

\def\z{\mathbf{z}}
\def\Vsc{\mathcal{V}} 

\usepackage{mathrsfs}
\def\Dscr{\mathscr{D}}

\def\bSigma{\boldsymbol{\Sigma}}
\def\bPhi{\boldsymbol{\Phi}}
\def\bbeta{\boldsymbol{\beta}}
\def\bxi{\boldsymbol{\xi}}
\def\balpha{\boldsymbol{\alpha}}
\def\bgamma{\boldsymbol{\gamma}}
\def\btheta{\boldsymbol{\theta}}

\def\bzero{\mathbf{0}}

\def\Eschat{\widehat{\mathbb{E}}}

\def\Msc{\mathcal{M}}

\def\bepsilon{\mathbf{\epsilon}}

\def\D{\mathbf{D}}
\def\argmindum{\mathop{\mbox{argmin}}}
\def\argmin#1{\argmindum_{#1}}

\def\subTsc{_{\scriptscriptstyle \Tsc}}

\def\bbetatilde{\widetilde{\bbeta}}
\def\Omegahat{\widehat{\boldsymbol{\Omega}}}
\def\bOmega{\boldsymbol{\Omega}}

\def \R {\mathbb{R}}
\def \P {\mathbb{P}}
\def \E {\mathbb{E}}
\def \I {\mathbb{I}}
\def\trans{^{\scriptscriptstyle \sf T}}

\def\Trsc{{\rm Thre}}

\def\Var{\mathrm{Var}}
\def\subTsc{_{\scriptscriptstyle \Tsc}}
\def\subSsc{_{\scriptscriptstyle \Ssc}}
\def\Cupsub{{\scriptscriptstyle \cup}}
\def\Tscsub{{\scriptscriptstyle \Tsc}}
\def\Sscsub{{\scriptscriptstyle \Ssc}}

\def\supone{^{[1]}}
\def\supzero{^{[0]}}

\newtheorem{assume}{Assumption}

\usepackage{lastpage}
\jmlrheading{27}{2026}{1-\pageref{LastPage}}{7/25; Revised 6/26}{6/26}{25-1766}{Doudou Zhou, Mengyan Li, Yun Wang, Tianxi Cai and Molei Liu}

\ShortHeadings{Domain Adaptation Targeting Heterogeneous and Imbalanced Subgroups}{Zhou, Li, Wang, Cai and Liu}
\firstpageno{1}

\begin{document}

\title{Domain Adaptation Targeting Heterogeneous and Imbalanced Subgroups}

\author{\name Doudou Zhou \email ddzhou@nus.edu.sg \\
       \addr Department of Statistics and Data Science\\ National University of Singapore, 
       Singapore
       \AND 
       \name Mengyan Li \email mengyanli@bentley.edu \\
       \addr Department of Mathematical Sciences\\
       Bentley University, Waltham, MA, USA 
       \AND
       \name Yun Wang\thanks{Zhou, Li, and Wang are equal contributors.} \email yunwang@stu.pku.edu.cn  \\
       \addr  Department of Biostatistics\\
        Peking University Health Science Center, Beijing, China
       \AND
       \name Tianxi Cai \email tcai@hsph.harvard.edu \\
       \addr Departments of Biostatistics and Biomedical Informatics\\
       Harvard T.H. Chan School of Public Health and Harvard Medical School, Boston, MA, USA 
       \AND
       \name Molei Liu\thanks{Corresponding author. }\email moleiliu@bjmu.edu.cn \\
       \addr Department of Biostatistics, Peking University Health Science Center;\\
       Beijing International Center for Mathematical Research Peking University, Beijing, China}

\editor{Brian Kulis}

\maketitle

\begin{abstract}
Domain adaptation enables generalizable and efficient data-driven research. However, existing work has largely focused on domain adaptation for some intrinsically homogeneous target cohort, overlooking inherent heterogeneity within the target, which can exacerbate biases and unfairness in the presence of subgroups with imbalanced sample sizes. We develop a novel domain adaptation framework that addresses more complicated target data that consists of heterogeneous and data-sparse subgroups and lacks gold-standard label observations. Our method simultaneously handles high-dimensionality, covariate shift, and outcome model heterogeneity by combining a model-assisted debiasing step used for covariate shift correction with an adaptive knowledge-guided sparsification procedure used to mitigate the issue of sample disparity. We also introduce a new model selection strategy to avoid negative knowledge transfer in the absence of labels in the target data. Our method is theoretically justified for being robust to nuisance model misspecification and adaptive to heterogeneity between the subgroups. Numerical experiments and two real-world applications, including genetic risk modeling of type II diabetes and prediction of mutation-induced protein stability changes, demonstrate the practical advantages of our method.
\end{abstract}

\begin{keywords}
Covariate shift; sampling disparity; model heterogeneity; double robustness; model calibration; knowledge transfer.
\end{keywords}

\section{Introduction}

\subsection{Background}\label{sec:bg}

Domain adaptation has been increasingly used when accurate labels are scarce or unavailable in a target population, but related labeled data are available from a source population. This setting arises broadly in scientific prediction problems where gold-standard target outcomes are expensive, delayed, or infeasible to collect. Existing work often focuses on adapting to an intrinsically homogeneous target cohort \citep[e.g.,][]{liu2023augmented,li2022translasso}. In many real-world application scenarios, however, the target population contains heterogeneous subgroups with imbalanced sample sizes, creating a harder problem: one must correct source--target distributional shift while protecting underrepresented subgroups from bias and negative transfer. Here, negative transfer means that borrowing information from a larger or related subgroup worsens estimation or prediction relative to the underrepresented-subgroup-only procedure.

Biobank studies linked to electronic health records (EHR) serve as a representative motivating example. Genetic disease risk modeling is critical for decision-making and knowledge discovery, especially in precision medicine \citep{moons2012risk}. In these studies, the scientific target is often a risk model based on prespecified genetic variants or baseline clinical factors, while gold-standard disease labels are available only for a limited subset of samples because manual chart review is costly. Meanwhile, the target cohort may contain highly imbalanced ancestry or demographic subgroups, so a model transferred from the labeled source cohort can perform unevenly across these subgroups.


The prediction of protein stability changes \(\Delta\Delta G\) provides another example. In our real-world study, S461 \citep{hern23} is used as the labeled source collection and S8754 \citep{xu24} as the target collection whose labels are withheld during training. The two collections differ in data source, curation workflow, and sequence filtering, making source--target distributional shift plausible. We define majority and minority subgroups by assay environment: near-neutral-pH assays (pH 6--7) are substantially more represented than acidic-pH assays (pH 1--5). This distinction is important because the measured \(\Delta\Delta G\) can depend on experimental conditions. Hence, a model dominated by near-neutral assays may not transfer reliably to the acidic-pH subgroup. 


The same structure appears not just in biomedical applications. In remote sensing, labels may be sufficient for well-surveyed regions or common land-cover types, but scarce for rare habitats or imagery from new sensors. In industrial reliability assessment, failure labels are often available under standard laboratory protocols, but limited under extreme operating conditions or new production lines. These examples share the same structure: labeled outcomes are available in a related source population, target labels are unavailable during training, and the target population contains subgroups with both sample-size imbalance and different outcome models. Therefore, model heterogeneity and sampling disparity are not simply descriptive features of the data. If they are ignored, a transferred model or decision can be dominated by the majority subgroup, leading to biased and poor minority-subgroup prediction. In addition, naive borrowing from the majority subgroup can harm the minority subgroup when the subgroup-specific outcome models differ. These challenges motivate domain adaptation methods that explicitly account for target-subgroup heterogeneity, sample disparity, and negative transfer protection.


\subsection{Related literature}\label{sec:related}

Recent work relevant to our setting has studied two distinct issues: source--target covariate shift and outcome-model heterogeneity across related domains or subgroups. Covariate shift is a central challenge in domain adaptation and is often addressed by importance weighting, including kernel mean matching \citep{huang2007correcting} and robust weighting schemes under strong shift \citep{reddi2015doubly}. In the broader machine learning literature, feature alignment methods provide another important class of unsupervised domain adaptation approaches. For example, CORAL \citep{sun2017correlation} reduces the domain discrepancy by aligning second-order feature statistics without using target labels. Pseudo-labeling methods instead exploit unlabeled target samples by generating surrogate labels for target domain learning. Recent theory has studied this idea for kernel ridge regression \citep{wang2023pseudo,kim2025transfer}. A related statistical line combines importance weighting with outcome imputation in
doubly robust estimating equations \citep{chakrabortty2019high, liu2023augmented, qiu2023efficient, tian2024semi,
zhou2022doubly}. These methods improve robustness to nuisance model misspecification, but are mainly designed for homogeneous target cohorts and low-dimensional target models, whereas our setting involves high-dimensional sparse target models and imbalanced heterogeneous target subgroups.

Beyond covariate shift, the heterogeneity of the outcome model poses challenges to domain adaptation. Knowledge-guided transfer methods address this issue by assuming structural similarity between outcome models across related domains or tasks. Examples include source-guided shrinkage \citep{bastani2021predicting}, Trans-Lasso and its extensions \citep{li2022translasso, tian2022transfer, li2023targeting}, and semi-supervised transfer learning with covariate and model shift \citep{cai2022semi}. In a related high-dimensional setting, TransFusion \citep{he2024transfusion} addresses covariate and model shifts through a fused-regularization strategy. These methods are effective for model shift problems, but generally require labeled target data or treat the target population as homogeneous. Thus, they do not directly address our label-free target setting with subgroup imbalance.

Several recent works study domain adaptation problems with multiple sources. Specifically, \citet{zhan2024transfer} and \citet{guha2025enhancing} combine heterogeneous sources using individual probabilistic weights. \citet{li2025multi} formulates domain adaptation with complex structures as a problem of tensor completion. These works are conceptually related because the four source/target--subgroup pairs in our setting can be viewed as distinct distributions. Our problem setup is more structured and asymmetric than in the existing work. Specifically, our target is the high-dimensional sparse model for the underrepresented target subgroup; the corresponding labeled source data are used for domain adaptation without labels in the target; and the majority subgroup is used only as auxiliary knowledge whose transferability must be assessed. Existing multi-source domain adaptation methods do not directly address subgroup imbalance, target labels unavailable for training, high-dimensionality, and potential negative transfer simultaneously.

Our work is also related to source-free domain adaptation, where one adapts a source model to an unlabeled target domain without direct access to the source data \citep{wang2024unveiling}. This literature is motivated in part by storage, privacy, and deployment constraints. Our setting differs in both data access and inferential target: we use labeled source data together with unlabeled target data, and our goal is to estimate a subgroup-specific target parameter rather than only to adapt a black-box source predictor. Nevertheless, source-free and multi-source domain adaptation raise related questions about when and how information should be transferred, and we discuss possible extensions of our setting in Section~\ref{sec:discussion}.

Finally, our theoretical framework is related to model-assisted semiparametric estimation with complex nuisance models, particularly in the context of conditional models and heterogeneous treatment effects (HTE). Recent works have extended the double machine learning (DML) framework \citep{chernozhukov2018double} to this setting, including \cite{fan2022estimation} and \cite{semenova2021debiased} for low-dimensional effect modifiers with
high-dimensional adjustment covariates, \cite{kennedy2020towards} for smooth or sparse HTE structures, and \cite{kato2024triple, baybutt2023doubly} for calibrated or debiased inference \citep[e.g.,][]{tan2020model} under model misspecification. However, these methods are not applicable to our setting involving a high-dimensional sparse target model under covariate shift, where we seek both robustness to nuisance model misspecification and tolerance to slow nuisance convergence. Simultaneously achieving these is technically challenging due to the difficulty in controlling regularization bias and establishing asymptotic linearity.

\subsection{Our contribution}

To mitigate the methodological gap in domain adaptation for targets with heterogeneous and imbalanced subgroups, we study a practically important and technically challenging problem that involves three interrelated difficulties: (1) covariate shift between the source and target domains, (2) sampling disparity and model heterogeneity across subgroups within the target domain, and (3) high-dimensionality of the outcome model. These challenges are further compounded by the lack of target labels available for training. To address them, we propose a new framework for domain Adaptation to targets with heterogeneous and IMbalanced Subgroups (AIMS). AIMS simultaneously delivers doubly robust covariate shift correction, knowledge transfer between target subgroups, and negative transfer protection within a single unified approach for high-dimensional target domains without outcome labels available for training.

\paragraph{Methodological contributions.} AIMS consists of three components, each addressing a specific challenge that prior work cannot handle jointly:

\begin{itemize}
\item[(i)] \textit{Doubly robust covariate shift correction with bifold debiasing (Algorithm \ref{alg:math}).} 
AIMS combines importance weighting and imputation in a doubly robust estimating equation, then uses bifold bias correction, consisting of nodewise Lasso debiasing and moment calibration, to remove regularization and nuisance estimation bias. 
A subsequent thresholding step recovers $\ell_2$-consistency. Existing doubly robust covariate shift methods \citep{liu2023augmented, zhou2022doubly} are restricted to low-dimensional outcome models. 

\item[(ii)] \textit{Knowledge transfer from majority to minority through thresholded contrast estimation (Algorithm \ref{alg:2}).} 
AIMS transfers knowledge from the majority to the minority subgroup without using target labels for training. It uses the majority model as an ``offset'' and estimates only the subgroup contrast. When this contrast is sparse, the minority estimator gains accuracy by borrowing information from the majority subgroup. This differs from existing knowledge transfer methods in the model shift setting \citep{cai2022semi, he2024transfusion}, which require labeled target samples.

\item[(iii)] \textit{Label-free negative transfer protection via model ensemble (Algorithm \ref{alg:3}).} 
AIMS protects against negative transfer when the subgroup contrast is dense and the majority cannot help in learning the minority model. It constructs a dense and debiased reference vector and uses cross-fitting and exponential weighting to aggregate the minority-only and knowledge transfer estimators. Existing negative transfer protection methods \citep{tian2022transfer, cai2022semi} compare candidate estimators using labeled target data, which is unavailable in our setting.

\end{itemize}

\paragraph{Theoretical contributions.}
We establish three main results that demonstrate the theoretical advantages of AIMS over the existing work. 
First, our calibrated estimator is model doubly robust. The closest comparator, \cite{zhou2022doubly}, establishes a similar property but only under a low-dimensional outcome model, and their analysis does not extend to our error rates when the dimension diverges.
Second, for the estimation of the minority model, AIMS matches, under a mild nuisance error condition, the rate of supervised knowledge transfer methods that use target labels for training \citep{tian2022transfer, he2024transfusion, cai2022semi}.
Third, our adaptive estimator achieves the faster one among the convergence rates of the minority-only and knowledge transfer estimators, up to a negative transfer protection cost. This is comparable to existing work that requires labeled target samples for model selection \citep[e.g.][]{cai2022semi}.

\paragraph{Practical relevance.} By jointly addressing covariate shift, model shift, and sample imbalance when target outcome labels are unavailable for training, AIMS provides a novel and practical solution for fair and equitable domain adaptation. This is particularly relevant in biomedical research, where minority subgroups often lack labeled outcomes, yet demand accurate and fair risk prediction.
Similarly, in protein engineering, available stability measurements may be collected under heterogeneous and imbalanced experimental conditions. In this scenario, variants relevant to the setting of the target design can be underrepresented in training data and may lack reliable target-setting labels, despite the need for accurate predictions of stability in the downstream design.

\def\bdelta{\boldsymbol{\delta}}
\def\bDelta{\boldsymbol{\Delta}}
\def\bzeta{\boldsymbol{\zeta}}
\def\Fsc{\mathcal{F}}
\def\Gsc{\mathcal{G}}
\def\Lsc{\mathcal{L}}
\def\Hsc{\mathcal{H}}

\section{Formulation of the Problem}\label{sec:setup}

Let $Y \in \R$ be the outcome of interest, $R \in \{0,1\}$ be the subgroup indicator ($R=1$ for the majority, $R=0$ for the minority), and $\Z = (\X\trans,\W\trans)\trans \in \R^{q+p}$ be the covariate vector, where $\X = (X_1,\dots,X_q)\trans$ are risk factors ($X_1 = 1$ {is the intercept term}) for predicting $Y$, and $\W = (W_1,\dots,W_p)\trans$ are auxiliary covariates that may be informative about $Y$ and the source--target shift but are not included in the final target risk model. Both $\X$ and $\W$ can be high-dimensional. Throughout the paper, the terms majority and minority refer to subgroup sample sizes rather than to the class distribution of $Y$. The subgroup indicator $R$ is assumed to be observed and may be determined by an observed attribute, such as ancestry, sex, age group, or experimental condition. In our formulation, $R$ indexes subgroup-specific distributions and target parameters rather than being treated as an ordinary component of $\X$ in a pooled model.

Our primary goal is to construct a prediction model of $Y\sim\X$ for the minority ($R=0$) in the target population $\Tsc$:
\begin{equation}
\E\subTsc[Y \mid \X,  R = 0]  = g(\X\trans\bar\bbeta\supzero),
\label{equ:target:model}
\end{equation}
where $\E\subTsc[\cdot]$ denotes the expectation operator on $\Tsc$, $g(\cdot)$ is a known link function, and $\bar\bbeta\supzero$ is a high-dimensional and sparse coefficient vector.  We do not require the {\textit{working}} model (\ref{equ:target:model}) to hold and define the population parameter $\bar\bbeta\supzero$ as the solution to 
\begin{equation}
\E\subTsc[\X\{Y-g(\X\trans \bbeta)\} \mid R = 0] = \bzero.
\label{equ:target:ee}
\end{equation}
This solution corresponds to the ordinary least squares regression when $g(x)=x$ and logistic regression when $g(x) = 1/(1+e^{-x})$. The generalized linear form is used as a working target model rather than a strict assumption on the true conditional mean. By (\ref{equ:target:ee}), $\bar\bbeta\supzero$ represents the projection of the conditional mean onto the risk factors $\X$. This choice is motivated by biomedical risk modeling, where GLMs provide interpretable adjusted associations and deployable risk scores. In high-dimensional and imbalanced settings, sparsity further yields a tractable structure.

We assume $Y$ is only observed in the source cohort $\Ssc$, not in the target $\Tsc$, and observe a dataset $\Dscr = \{\D_i =(S_i Y_i, \Z_i\trans , S_i,R_i)\trans: i = 1, 2, \dots,n \}$. Here, $R_i\in\{0,1\}$ indicates the group and $S_i$ is the source/target indicator ($S_i = 1$ for $\Ssc$, $S_i = 0$ for $\Tsc$). The sample size of each stratum is  $n_{\scriptscriptstyle \Ssc,r}:= \sum_{i=1}^{n} \I(S_i =1, R_i = r)$ and $n_{\scriptscriptstyle \Tsc,r}:= \sum_{i=1}^{n} \I(S_i =0, R_i = r)$ for $r \in \{0,1\}$, where $\I(\cdot)$ is the indicator function. In addition, we use $n\subSsc:=n_{\scriptscriptstyle \Ssc,0}+n_{\scriptscriptstyle \Ssc,1}$ and $n\subTsc:=n_{\scriptscriptstyle \Tsc,0}+n_{\scriptscriptstyle \Tsc,1}$ to represent the sample sizes of $\Ssc$ and $\Tsc$ so we have $n=n\subSsc + n\subTsc$. We consider an imbalanced sampling scenario on both $\Ssc$ and $\Tsc$ where $n_{\scriptscriptstyle \Ssc,1}\gg n_{\scriptscriptstyle \Ssc,0}$ and $n_{\scriptscriptstyle \Tsc,1}\gg n_{\scriptscriptstyle \Tsc,0}$. A common example is that $R=1$ represents the European majority ancestry group and $R=0$ represents the underrepresented African group. Other examples include better-represented versus underrepresented protein families, mutation classes, or assay platforms in protein mutation studies.

We assume that for $(s,r)\in\{0,1\}^2$, the samples of $(Y,\Z)$ are generated by
\begin{equation}
Y,\Z\mid S=s, R=r\sim  p_{\Z|S=s,R=r}(\z)\cdot p_{Y|\Z,R=r}(y),
\label{equ:cs:assume}
\end{equation}
where $p_{\Z|S=s,R=r}(\cdot)$ and $p_{Y|\Z,R=r}(\cdot)$ represent the distribution function of $\Z$ given $S=s,R=r$, and that of $Y$ conditional on $\Z$ and $R=r$, respectively. This defines a domain adaptation problem where, given $R=r$, the distribution of covariates $\Z=(\X\trans,\W\trans)\trans$ is assumed to be different between $\Ssc$ (source) and $\Tsc$ (target), while the conditional distribution of $Y\mid\Z,R=r$ is assumed to be the same. Thus, within each subgroup, the source and target samples may have different covariate distributions, but share the same conditional outcome mechanism given the full covariate vector $\Z$. Conditional on $R=r$, this is a typical covariate shift scenario frequently encountered in practice. For example, EHR cohorts can vary in their coding systems and versions \citep{guo2021systematic}, and, notably, MGB underwent a change in its coding system around 2015, resulting in distributional shifts of EHR features. Covariate shifts can also be caused by the variation of demographic characteristics between different institutions or time intervals and the evolving practice patterns of healthcare professionals \citep{braithwaite2018changing}. In protein applications, covariate shift may arise from differences between assay platforms, experimental protocols, protein families, or mutation distributions.

We introduce the density ratio $h_r^{\star}(\z):= p_{\Z|S=0, R=r}(\z)/p_{\Z|S=1, R=r}(\z)$ to characterize the covariate shift between the source and target within subgroup $r \in \{0,1\}$. Under such an assumption in (\ref{equ:cs:assume}), the conditional mean function $m_r^{\star}(\z):= \E[Y \mid \Z=\z, R=r]$ remains to be the same between $\Ssc$ and $\Tsc$. Nevertheless, we have $\E\subTsc[Y \mid \X,  R = 0]\neq\E\subSsc[Y \mid \X,  R = 0]$ due to the joint distributional shift of $\X$ and the auxiliary $\W$. Consequently, directly regressing $Y$ against $\X$ on $\Ssc$ will typically produce a biased estimator for the target $\bar\bbeta\supzero$, no matter the model (\ref{equ:target:model}) is correct or not. Our first technical challenge is to correct this bias when using labeled observations from $\Ssc$ to estimate the model parameters defined on $\Tsc$.  Standard domain adaptation via importance weighting or imputation is challenged by high-dimensionality and subgroup disparities.

\begin{remark}
\label{rem:xw}
In EHR studies such as the application in Section \ref{sec:real:1}, $\X$ includes genetic variants and demographic variables to model disease risk  $Y$, while $\W$ contains EHR proxies such as diagnostic codes and laboratory tests. The focus is on predicting $Y$ using $\X$ rather than all covariates in $\Z$ because the auxiliary $\W$ is not available at baseline, whereas biomedical discoveries and deployment decisions are often meant to be based on baseline risk factors. More generally, the distinction between $\X$ and $\W$ is determined by the scientific or deployment target, not by their predictive strength. Auxiliary variables $\W$ may be strongly associated with $Y$ but excluded from the target model because they are derived from post-baseline information, unavailable at deployment, or difficult to interpret. Such $\W$ can appear in diverse domains. For example, in longitudinal clinical studies, $\W$ can include an early response or endpoint, such as short-term biomarker changes or tumor response that are predictive of the long-term endpoint $Y$, but not available at the decision time \citep{prentice1989surrogate}. In the industrial reliability examination, $\W$ may include accelerated stress-test readouts or early degradation signals. 

Usually, the distributional shift between the source and target populations, such as the temporal shift in the clinical profile of the biobank participants, cannot be fully captured by the difference in $\X$. Thus, it is important to include auxiliary variables $\W$ to adjust for their covariate shift and transfer the knowledge of $Y\mid \X,\W$ from source to target, since $Y \mid \X, R=0$ can be inherently different between the source and target populations. For this purpose, $\W$ is used directly in nuisance models for the density ratio and conditional mean, while the final target model remains defined on $\X$. Importantly, our framework also allows for the case without auxiliary variables: if no such $\W$ is available or needed, we set $\W=\emptyset$ and apply the same framework with $\Z=\X$. 
\end{remark}

The preceding discussion addresses source--target covariate shift within a fixed subgroup. We next describe the second component of the problem: borrowing information from the larger subgroup to improve estimation for the smaller subgroup while avoiding negative transfer. Analogously to (\ref{equ:target:ee}), we define the model coefficient in the majority subgroup of $\Tsc$, $\bar\bbeta\supone$ as the solution to the equations:
$$
\E\subTsc[\X\{Y-g(\X\trans \bbeta)\} \mid R = 1] = \bzero.
$$
Due to the distributional heterogeneity between the two subgroups, $\bar\bbeta\supone$ and $\bar\bbeta\supzero$ tend to be different. The larger $n_{\scriptscriptstyle \Ssc,1}$ and $n_{\scriptscriptstyle \Tsc,1}$ can lead to a more precise estimate of $\bar\bbeta\supone$ compared to $\bar\bbeta\supzero$ learned with small minority samples. It is important to mitigate this disparity and a tentative strategy is to leverage the estimate of $\bar\bbeta\supone$ as external knowledge to guide the learning of $\bar\bbeta\supzero$. This is motivated by the observation that genome-wide associations or models of a wide range of diseases and traits tend to have similar patterns between different ancestry groups, as revealed in recent studies \citep[e.g.,][]{lam2019comparative,verma2023diversity}.  More generally, this motivates modeling the difference between subgroup-specific target coefficients as sparse: the two subgroups may share many active effects, while only a smaller set of effects differs meaningfully. For these purposes, we introduce the assumption
\begin{equation}
(\bar\bbeta\supzero,\bar\bbeta\supone)\in \left\{\bar\bbeta\supzero,\bar\bbeta\supone:\|\bar\bbeta\supzero\|_0\leq s_{\beta}\supzero,~\|\bar\bbeta\supzero-\bar\bbeta\supone\|_v\leq R_{\delta,v}\right\}
\label{equ:asu:sparse},
\end{equation}
where $\|\bbeta\|_0$ represents the number of non-zero entries in the vector $\bbeta$ and $\|\bbeta\|_v$ is the $\ell_v$-norm of $\bbeta$ for some fixed $v\in[0,1]$. Here, $\|\cdot\|_v$ corresponds to the exact ($v=0$) or approximate ($v\in(0,1]$) sparsity norm, and $s_{\beta}\supzero$ and $R_{\delta,v}$ represent the sparsity levels of the coefficients $\bar\bbeta\supzero$ and the difference between the two subgroups $\bar\bbeta\supzero-\bar\bbeta\supone$.

\begin{figure}[htb!]
\centering
\includegraphics[width=0.55\linewidth]{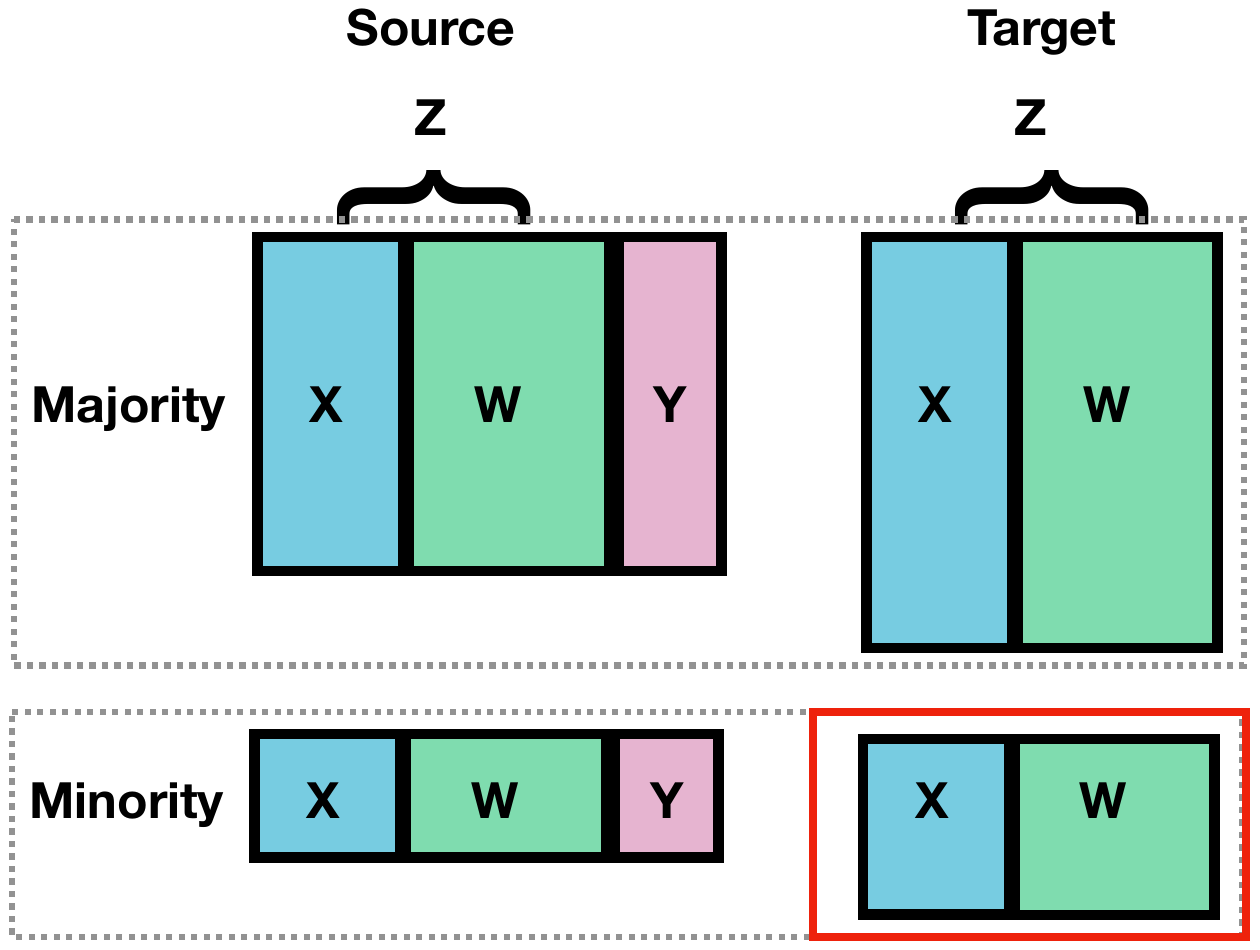} 
\caption{\label{fig:setup} Structure of the observed data in our problem setup.}
\end{figure}

Since outcome models across sub-populations are presumed to be similar for most risk factors, we expect $R_{\delta,v}\ll s_{\beta}\supzero$ when taking $v=0$ as a special case. This promises an efficiency gain by using $\bar\bbeta\supone$ to assist in learning $\bar\bbeta\supzero$ in the underrepresented group, as their difference is sparser and easier to estimate compared to $\bar\bbeta\supzero$ itself. On the other hand, given that $s_{\beta}\supzero$ and $R_{\delta,v}$ are unknown in practice, our objective is to maintain adaptivity and robustness to excessive model heterogeneity, i.e., the case where $R_{\delta,v}$ is large compared to $s_{\beta}\supzero$. In this scenario, the knowledge of $\bar\bbeta\supone$ can be non-informative or misleading to the target parameters $\bar\bbeta\supzero$, and it is desirable to avoid the potential negative transfer caused by this. To facilitate understanding, the problem setup described above is illustrated in Figure~\ref{fig:setup}.

\section{Method}
\label{sec:method}

We first give a brief overview of our method. AIMS consists of three main components. 
First, for each subgroup \(r\in\{0,1\}\), we construct a doubly robust estimating equation  to correct the source--target covariate shift using the density ratio and 
conditional mean nuisance models. In this step, we apply a bifold bias correction to 
reduce both the regularization bias in the target coefficient estimator and 
the first-order bias from regularized nuisance estimation. The resulting 
dense debiased vector is then thresholded to recover an \(\ell_2\)-consistent 
sparse estimator. 
Second, for the underrepresented subgroup, we use the 
majority-subgroup estimator as auxiliary knowledge and estimate the sparse difference between the two subgroup-specific models. Third, a final 
surrogate-loss-based aggregation step protects against negative transfer 
when the majority model is insufficiently transferable.


\subsection{Model-assisted domain adaptation for high-dimensional target models}\label{sec:method:cs}

Define the empirical mean operators on the source samples, target samples, and their union for group $r \in \{0,1\}$ as: 
$$
\begin{aligned}
&    \Eschat\subSsc^{ [r]} f(\D) = n_{\scriptscriptstyle \Ssc,r}^{-1} \sum_{i=1}^n  \I(S_i = 1, R_i = r) f(\D_i), \\
&    \Eschat\subTsc^{ [r]} f(\D) = n_{\scriptscriptstyle \Tsc,r}^{-1} \sum_{i=1}^n  \I(S_i = 0, R_i = r) f(\D_i), \text{ and } \\
&    \Eschat^{[r]}_{\Sscsub\Cupsub\Tscsub} f(\D) = (n_{\scriptscriptstyle \Ssc,r} + n_{\scriptscriptstyle \Tsc,r})^{-1} \sum_{i=1}^n  \I(R_i = r) f(\D_i).
\end{aligned}
$$

We specify the nuisance models for density ratio and conditional mean imputation  as: 
$h_r(\z) = \exp\big( \phi(\z)\trans\balpha^{[r]}\big)$ and $m_r(\z)=b\big(\phi(\z)\trans\bgamma^{[r]}\big)$,
where $\phi: \R^{p+q} \rightarrow \R^{d}$ is a flexible basis function of $\Z$. For example, one could simply take $\phi(\z) = \z$ or set $\phi(\z)$ as a concatenation of non-linear bases of the components of $\z$. More generally, \(\phi(\cdot)\) may include nonlinear transformations, interactions, splines, or fixed representations extracted from external models. The vectors $\balpha^{[r]}$ and $\bgamma^{[r]}$ are the coefficients for the nuisance models that need to be estimated. The function $b(\cdot)$ is a prespecified, monotonically increasing link function, which may or may not be the same as the link function $g(\cdot)$. Denote by $\bPhi = \phi(\Z)$. 

The nuisance models above are working models used for covariate shift correction and outcome imputation. They are not intended to impose rigid linearity in the raw covariates, since the feature map \(\phi(\cdot)\) can enlarge the nuisance feature space. At the same time, the sparse high-dimensional form keeps nuisance estimation feasible when the labeled source sample is limited, especially in the minority subgroup. As shown below, the doubly robust construction does not require both nuisance models to be correctly specified simultaneously: the target estimating equation remains valid when either the density ratio model or the conditional mean model is correctly specified within the working model class.

When $m_r(\z) = m_r^{\star}(\z)$, the target parameter $\bar\bbeta^{[r]}$ is the solution to the following imputation-based (IM) estimating equation: $
\E\subTsc[\X\{m_r(\Z)-g(\X\trans \bbeta)\} \mid R = r] = \bzero$. If $h_r(\z)=h_r^{\star}(\z)$,  $\bar\bbeta^{[r]}$ solves the importance weighted (IW) estimating equation: $\E\subSsc[h_r(\Z) \X\{Y-g(\X\trans \bbeta)\} \mid R = r] = \bzero$.

Empirically, $\bar\bbeta^{[r]}$ can be estimated using sample versions of these equations. However, these simple strategies are prone to potential model misspecification. Specifically, the IW method leads to inconsistency if the density ratio model $h_r$ is misspecified, and the IM method fails when  $m_r$ is misspecified. To overcome this challenge, we combine the two nuisance models to construct a doubly robust domain adaptation estimating equation for $\bar\bbeta^{[r]}$ as
\begin{equation}
 \E\subSsc[h_r(\Z)\X\{m_r(\Z)-Y\} \mid R = r] - \E\subTsc[\X\{m_r(\Z)-g(\X\trans \bbeta)\} \mid R = r]  = \bzero.
\label{eq:dr}   
\end{equation}
Proposition \ref{prop:1} shows that (\ref{eq:dr}) provides a consistent solution for $\bar\bbeta^{[r]}$ when either the density ratio or imputation model is correctly specified. 
\begin{proposition}
Let $\bbeta^{[r]}$ be the solution of the population estimating equation \eqref{eq:dr}. We have $\bbeta^{[r]} = \bar \bbeta^{[r]}$ when either $h_r^{\star}(\z) = \exp\big( \phi(\z)\trans\balpha^{[r]}\big)$ or $m_r^{\star}(\z)=b\big(\phi(\z)\trans\bgamma^{[r]}\big)$.
\label{prop:1}    
\end{proposition}

Motivated by this, we propose an empirical doubly robust loss function. Specifically, given the two nuisance parameters $\balpha^{[r]}$ and $\bgamma^{[r]}$ whose estimation will be discussed later, we define the doubly robust covariate shift-corrected sparse regression for $\bar\bbeta^{[r]}$ as:  
\begin{equation}
\widehat \bbeta^{[r]}   = \arg \min_{\bbeta \in \R^q } \Lsc_r(\bbeta; \balpha^{[r]},\bgamma^{[r]} )+\lambda^{[r]}\|\bbeta\|_1,
   \label{equ:dr}
\end{equation}
where the loss function is defined as
\[
\Lsc_r(\bbeta; \balpha^{[r]}, \bgamma^{[r]} ) =  
\left[\Eschat_{\scriptscriptstyle \Ssc}^{[r]} \X\trans \exp(\bPhi\trans\balpha^{[r]})\{  b(\bPhi\trans\bgamma^{[r]}) - Y \} -  \Eschat_{\scriptscriptstyle \Tsc}^{[r]}\X\trans b(\bPhi\trans\bgamma^{[r]}) \right]\bbeta + \Eschat_{\scriptscriptstyle \Tsc}^{[r]} G(\X\trans \bbeta), 
\]
where $G(a)=\int_{0}^ag(u) {\rm d} u$, and $\lambda^{[r]}$ is a penalization parameter. 
We will present the optimal rate for all tuning parameters, including $\lambda^{[r]}$, in Section \ref{sec:theory}, with empirical tuning strategies in Supplementary S.2.3. A summary of the theoretical orders and practical tuning rules for all tuning parameters is also provided in Supplementary S.2.3. By construction, $\E \partial_{\bbeta} \Lsc_r(\bbeta; \balpha^{[r]}, \bgamma^{[r]})$ matches the left-hand side of \eqref{eq:dr}. Thus, by Proposition \ref{prop:1},  $\Lsc_r(\bbeta; \balpha^{[r]}, \bgamma^{[r]})$ is doubly robust to the misspecification of the two nuisance models in the sense that either a correct density ratio or a correct conditional mean imputation model can lead to the target solution $\bar \bbeta^{[r]}$. 

Motivated by the above-introduced construction, a natural strategy for the estimation of $\bar \bbeta^{[r]}$ is to first estimate $\balpha^{[r]}$ and $\bgamma^{[r]}$  through regularized regression
\begin{equation}
\begin{aligned}
\widetilde\balpha^{[r]} = &  \underset{\balpha\in\mathbb{R}^{d}}{\arg \min } \Eschat_{\Sscsub\Cupsub\Tscsub}^{[r]}\{\rho_{\scriptscriptstyle \Tsc,r} S \exp (\bPhi \trans \balpha) - \rho_{\scriptscriptstyle \Ssc,r} (1-S)\bPhi \trans \balpha  \} + \lambda_{\alpha}^{[r]}  \|\balpha\|_1;\\
\widetilde \bgamma^{[r]} = & \underset{\bgamma \in\mathbb{R}^{d}}{\arg \min } \Eschat_{\Sscsub}^{[r]} \{ -Y  \bPhi\trans \bgamma + B(\bPhi\trans \bgamma)\}    + \lambda_{\gamma}^{[r]}  \|\bgamma\|_1,
\end{aligned}    
\label{eq:initial alpha}
\end{equation}
where
$\rho_{\scriptscriptstyle \Tsc,r} =(n_{\scriptscriptstyle \Ssc,r}+n_{\scriptscriptstyle \Tsc,r})/n_{\scriptscriptstyle \Tsc,r}$, $\rho_{\scriptscriptstyle \Ssc,r}=(n_{\scriptscriptstyle \Ssc,r} + n_{\scriptscriptstyle \Tsc,r})/n_{\scriptscriptstyle \Ssc,r}$, $B(a)=\int_{0}^a b(u) {\rm d} u$, and $\lambda_{\alpha}^{[r]}$, $\lambda_{\gamma}^{[r]}$ are two tuning parameters, then obtain the  preliminary estimator $\widetilde \bbeta^{[r]}$ as 
\begin{equation}
\widetilde \bbeta^{[r]} =   \underset{\bbeta \in \mathbb{R}^{q}}{\arg \min }  \Lsc_r(\bbeta, \widetilde \balpha^{[r]}, \widetilde \bgamma^{[r]} )+\lambda^{[r]}\|\bbeta\|_1 \,.
\label{eq:init beta}
\end{equation}
According to our above discussion, $\widetilde \bbeta^{[r]}$ tends to be consistent when either the density ratio or the conditional mean model is correctly specified. However, it is unlikely to achieve a desirable convergence rate due to the excessive biases in $\widetilde \balpha^{[r]}$ and $\widetilde \bgamma^{[r]}$ introduced by regularization; see Remark \ref{rem:5} for more discussion. To address this issue, we propose a bifold bias correction approach. Denote by $\bar\bSigma^{[r]}_{\bbeta} = \E\subTsc[\dot{g}(\X\trans \bbeta) \X \X\trans \mid R = r]$ and $\widehat{\bSigma}^{[r]}_{\bbeta} = \Eschat_{\scriptscriptstyle \Tsc}^{[r]}[ \dot{g}(\X\trans \bbeta) \X \X\trans]$ where $\dot g(\cdot)$ is the derivative of $g(\cdot)$. As the first fold of bias correction, we address the bias caused by the regularization term in (\ref{eq:init beta}) through a one-step debiased construction for each $\beta_j^{[r]} = \e_j \trans\bbeta^{[r]}$ ($j \in\{1, \ldots,q\}$):
\begin{equation}
 \widehat{\beta}_{{\rm Deb},j}^{[r]}(\balpha^{[r]},\bgamma^{[r]}) =\widetilde{\beta}_j^{[r]}  +\Omegahat^{[r]}_{j}\partial_{\bbeta} \Lsc_r(\widetilde\bbeta^{[r]}; \balpha^{[r]}, \bgamma^{[r]}),   
\label{bical}
\end{equation}
where $\e_j$ is the $j$-th unit vector in $\mathbb{R}^q$, $\widetilde{\beta}_j^{[r]}$ is the $j$-th element of $\widetilde \bbeta^{[r]}$, 
$\Omegahat^{[r]}_{j}$ is a regularized estimation of $\bar \bOmega^{[r]}_j$, the $j$-th row of $ \bar \bOmega^{[r]}=\big[\bar\bSigma^{[r]}_{\bar\bbeta^{[r]}}\big]^{-1}$, and $\partial_{\bbeta}$ is the partial derivative operator with respect to $\bbeta$. To construct $\Omegahat^{[r]}_{j}$, we use the node-wise lasso method proposed by \cite{van2014asymptotically}. See Supplementary S.2.1 for the detailed expression of $\Omegahat^{[r]}_{j}$.

This step only reduces the biases arising from the regularization on $\bbeta$ in (\ref{eq:init beta}). In the second fold of bias correction, we further mitigate the biases due to the excessive regularization errors in $\widetilde\balpha^{[r]}$ and $\widetilde\bgamma^{[r]}$. Our key idea is to further calibrate the two nuisance models and make them satisfy the moment conditions: 
\begin{equation}
\E\big[\partial_{\balpha}\partial_{\bbeta} \Lsc_r(\bar\bbeta^{[r]}; \balpha^{[r]}, \bgamma^{[r]})\big]=\bzero;\quad \E\big[\partial_{\bgamma}\partial_{\bbeta} \Lsc_r(\bar\bbeta^{[r]}; \balpha^{[r]}, \bgamma^{[r]})\big]=\bzero.
\label{equ:ideal:mom}
\end{equation}
Under these two conditions, the first-order errors in $\balpha^{[r]}$ and $\bgamma^{[r]}$ can be removed through concentration in a similar spirit with the idea of Neyman orthogonality \citep{chernozhukov2018double}. This motivates us to obtain the calibrated nuisance estimators as $\widehat\balpha^{[r]}_j = \widetilde \balpha^{[r]}+ \widehat\bxi^{[r]}_j$ and $\widehat\bgamma^{[r]}_j = \widetilde \bgamma^{[r]} + \widehat\bzeta^{[r]}_j$ with
\begin{equation}
\begin{split}
\widehat\bxi^{[r]}_j &=\argmin{\bxi\in\mathbb{R}^{d}}\Eschat_{\Sscsub\Cupsub\Tscsub}^{[r]}\widehat w_{j}^{[r]}\dot{b}( \bPhi\trans\widetilde \bgamma^{[r]})\Fsc^{[r]}(\bxi;\widetilde \balpha^{[r]}) + \lambda_{\alpha_j}^{[r]} \|\bxi\|_1 \, ;\\
\widehat\bzeta^{[r]}_j&=\argmin{\bzeta\in\mathbb{R}^{d}}\Eschat_{\Sscsub}^{[r]} \widehat w_{j}^{[r]} \exp(\bPhi \trans\widetilde\balpha^{[r]})\Gsc(\bzeta;\widetilde \bgamma^{[r]})+ \lambda^{[r]}_{\gamma_j} \|\bzeta\|_1 \,,
\end{split}
\label{equ:cal:beta:alpha}
\end{equation}
where $\widehat w_{j}^{[r]}=\Omegahat^{[r]}_j \X$ and $\Fsc^{[r]}(\bxi;\balpha)= \rho_{\scriptscriptstyle \Tsc,r}  S \exp \{\bPhi \trans (\balpha+\bxi)\} - \rho_{\scriptscriptstyle \Ssc,r}  (1-S)\bPhi \trans (\balpha+\bxi)$, $\Gsc(\bzeta;\bgamma)=-Y \bPhi \trans (\bgamma+\bzeta) + B\{\bPhi\trans (\bgamma+\bzeta)\}$. The Lasso problems in \eqref{equ:cal:beta:alpha} are designed such that their Karush--Kuhn--Tucker (gradient) conditions empirically match our desirable moment conditions in (\ref{equ:ideal:mom}). We then plug the calibrated $\widehat\balpha^{[r]}_j$ and $\widehat\bgamma^{[r]}_j$ into (\ref{bical}) for a bifold bias-corrected estimator of each $\beta_j^{[r]}$, denoted as $\widehat{\beta}_{{\rm Deb},j}^{[r]} = \widehat{\beta}_{{\rm Deb},j}^{[r]}(\widehat\balpha^{[r]}_j,\widehat\bgamma^{[r]}_j)$. We also denote by $\widehat{\bbeta}^{[r]}_{\rm Deb} = \big( \widehat{\beta}_{{\rm Deb},1}^{[r]},\ldots,\widehat{\beta}_{{\rm Deb},q}^{[r]}\big)\trans$. The above-introduced covariate shift correction approach is summarized in Algorithm \ref{alg}.

\begin{algorithm}[htb!]
    \caption{\label{alg} Covariate shift correction with bifold debiasing}  
    \textbf{Input:} $\Dscr^{[r]} = \{\D_i =(S_i Y_i, \X_i\trans, \W_i\trans, S_i, R_i)\trans:  R_i = r,~i \in [n] \}$;

    \quad \textbf{1:} Obtain the preliminary estimators $\widetilde \balpha^{[r]}$ and $\widetilde\bgamma^{[r]}$ by \eqref{eq:initial alpha} and $\widetilde \bbeta^{[r]}$ by \eqref{eq:init beta}. 
    This step fits the initial density ratio model, imputation model, and target model coefficients.
    
     \quad \textbf{2:} For $j=1, \dots, q$: derive the form of the first fold bias correction for $\beta_j^{[r]}$ by (\ref{bical}).
     This step removes the leading regularization bias in the sparse target estimator.

    \quad \textbf{3:} For $j=1, \dots, q$: conduct the second fold calibration and obtain $\widehat\bgamma^{[r]}_j$ and $\widehat\balpha^{[r]}_j$ by \eqref{equ:cal:beta:alpha}.
    This step calibrates the nuisance estimators to reduce first-order nuisance-estimation bias.

    \quad \textbf{4:} For $j=1, \dots, q$: plug $\widehat\bgamma^{[r]}_j$ and $\widehat\balpha^{[r]}_j$ into (\ref{bical}) to obtain $\widehat{\beta}_{{\rm Deb},j}^{[r]}=\widehat{\beta}_{{\rm Deb},j}^{[r]}(\widehat\balpha^{[r]}_j,\widehat\bgamma^{[r]}_j)$.

    \textbf{Output:} The debiased coefficient vector $\widehat{\bbeta}^{[r]}_{\rm Deb} = ( \widehat{\beta}_{{\rm Deb},1}^{[r]},\ldots,\widehat{\beta}_{{\rm Deb},q}^{[r]})\trans$. 
    \label{alg:math}
\end{algorithm}

\begin{remark}
Note that the weight $\widehat w_{j}^{[r]}$ used in (\ref{equ:cal:beta:alpha}) may {take negative values}, which could make the loss function in  (\ref{equ:cal:beta:alpha}) irregular and ill-posed. To handle this, one can divide the samples into two sets with positive and negative only $\widehat w_{ij}$'s respectively and solve \eqref{equ:cal:beta:alpha} on them separately. Details of this stratification strategy are presented in Supplementary S.2.2. 
\label{rmk:split}
\end{remark}

Although the bias-corrected $\widehat{\bbeta}^{[r]}_{\rm Deb}$ is element-wise consistent to $\bar\bbeta^{[r]}$ with a desirable convergence rate, its overall $\ell_2$-error $\big\|\widehat{\bbeta}^{[r]}_{\rm Deb}-\bar\bbeta^{[r]}\big\|_2$ does not converge to zero since $\widehat{\bbeta}^{[r]}_{\rm Deb}$ is a dense estimator in contrast to the sparse $\bar\bbeta^{[r]}$, with its $\ell_2$-error growing fast with the large dimension $q$. For the purpose of risk prediction, we further construct an $\ell_2$-consistent estimator for $\bar\bbeta^{[r]}$ through a thresholding and sparsifying approach:
\begin{equation}
\widehat\bbeta_{\rm Thr}^{[r]} = \big\{\Trsc\big( \widehat{\beta}_{{\rm Deb},j}^{[r]},\tau^{[r]}\big)\big\}_{j=1,\ldots,q},
\label{equ:thre}
\end{equation}
where $\Trsc(z,c)= z \I(|z|\geq c)$, and $\tau^{[r]}$ is a tuning parameter. In Theorem \ref{thm:beta_thresh}, we prove the $\ell_2$-consistency of $\widehat\bbeta_{\rm Thr}^{[r]}$ and its  convergence rate. 
Related to our above discussion, we also comment in Remarks \ref{rem:thm1} and \ref{rem:5} that $\widehat\bbeta_{\rm Thr}^{[r]}$ is substantially more robust to regularization errors in the nuisance estimators compared with the preliminary estimator $\widetilde{\bbeta}^{[r]}$. For practical implementation, \(\tau^{[r]}\) is chosen on the theoretical scale derived in Theorem~\ref{thm:beta_thresh}; practical tuning details and sensitivity analyses are provided in Supplementary S.2.3.

\subsection{Adaptive knowledge transfer between the subgroups}

The construction of $\widehat\bbeta_{\rm Thr}^{[0]}$ enables us to borrow information from the source minority population, which may still suffer from high variability due to small sample size. Thus, we propose a knowledge-guided transfer learning procedure using the majority group estimator $\widehat\bbeta_{\rm Thr}^{[1]}$ to aid the estimation of $\bar \bbeta^{[0]}$. Similar to recent work like \cite{li2022translasso}, our approach relies on the presumption that $\bar\bdelta=\bar \bbeta^{[1]}-\bar \bbeta^{[0]}$ is sparser and easier to estimate well than $\bar \bbeta^{[0]}$, e.g., $R_{\delta,v}\ll s_{\beta}\supzero$ when $v=0$ in (\ref{equ:asu:sparse}). The idea is to take $\widehat\bbeta_{\rm Thr}^{[1]}$ as a baseline (offset) and estimate the difference term $\bar\bbeta^{[0]}-\widehat\bbeta_{\rm Thr}^{[1]}$ instead of $\bar\bbeta^{[0]}$ through thresholding and sparsifying.  This strategy is beneficial when the majority and minority target models share most relevant predictors, but it may be harmful when their difference is not sufficiently sparse, i.e., being denser than the target coefficients themselves. The details are described in Algorithm \ref{alg:2}. 

\begin{algorithm}[htb!]
\caption{Majority-knowledge-guided thresholding estimation}\label{alg:2}
\raggedright
\textbf{Input:} The debiased coefficient vector $\widehat{\bbeta}^{[0]}_{\rm Deb}$ obtained by implementing Algorithm \ref{alg} on the minority group, and the sparsified estimator $\widehat\bbeta_{\rm Thr}^{[1]}$ obtained on the majority group using (\ref{equ:thre}).   

\textbf{Output:} Knowledge transfer estimator $\widehat\bbeta^{[0]}_{\rm KTr}=\widehat\bbeta_{\rm Thr}^{[1]}+\widehat\bdelta$ where $\widehat\bdelta=(\widehat\delta_1,\ldots,\widehat\delta_q)\trans$, $\widehat\delta_j=\Trsc\big(\widehat{\beta}^{[0]}_{{\rm Deb},j}-\widehat\beta_{{\rm Thr},j}^{[1]},\tau_{\scriptscriptstyle{\rm KTr }}\big)$ and $\tau_{\scriptscriptstyle{\rm KTr }}$ is a tuning parameter.
\end{algorithm} 

As shown in Section \ref{sec:theory}, when $\bar\bdelta$ is sparser than $\bar \bbeta^{[0]}$, Algorithm \ref{alg:2} effectively leverages knowledge from the majority group to improve the estimation of $\bar \bbeta^{[0]}$. However, if $\bar\bdelta$ is denser than $\bar \bbeta^{[0]}$, i.e., $\bar \bbeta^{[1]}$ differs significantly from $\bar \bbeta^{[0]}$, $\widehat\bbeta^{[0]}_{\rm KTr}$ given by Algorithm \ref{alg:2} could be less efficient than the minority-only estimator $\widehat\bbeta_{\rm Thr}^{[0]}$, suffering from negative knowledge transfer. In this sense, negative transfer refers to the case where majority-guided borrowing worsens estimation relative to the minority-only estimator. 

To prevent this, we propose a novel negative transfer protection approach that ensures the final estimator for $\bar\bbeta^{[0]}$ to be no worse than $\widehat\bbeta_{\rm Thr}^{[0]}$ and is adaptive to the unknown transferability level between the two sub-populations (i.e., sparsity level of $\bar\bdelta$). For similar purposes, recent work like \cite{tian2022transfer} uses hold-out samples from the minority group to select between or ensemble the minority-only and knowledge transfer estimators. However, such a strategy is not directly applicable to our case due to the absence of labeled data on the target site $\Tsc$. To handle this challenge, we introduce Algorithm \ref{alg:3} that relies on a surrogate loss 
$
\widehat{\mathcal{Q}}(\bbeta^{[0]};\widehat{\bbeta}^{[0]}_{\rm Deb}):=\big\|\bbeta^{[0]}-\widehat{\bbeta}^{[0]}_{\rm Deb}\big\|_2^2$ to evaluate the accuracy of $\bbeta^{[0]}$. In this formulation, the dense vector $\widehat{\bbeta}^{[0]}_{\rm Deb}$ can be essentially viewed as a Gaussian vector with mean $\bar\bbeta^{[0]}$ (neglecting higher-order error terms) and, thus, serving as an appropriate criterion for model selection. The data splitting in Algorithm~\ref{alg:3} ensures that the dense debiased vector used for evaluation is constructed using samples independent from the candidate estimators being compared. We demonstrate this point more rigorously in Lemma \ref{lemma:approx}.

\begin{algorithm}[htb!]
\caption{\label{alg:3} Protect against negative knowledge transfer}

\textbf{Input:} $\Dscr^{[r]} = \{\D_i =(S_i Y_i, \X_i\trans, \W_i\trans, S_i, R_i)\trans:  R_i = r,~i \in [n] \}, r=0,1$;  

\quad \textbf{1:} Randomly split the minority data $\Dscr^{[0]}$ (including both the source and target samples) into two disjoint sets $\Dscr^{[0]}_1$ and $\Dscr^{[0]}_2$ of equal size.

\quad \textbf{2:} Implement Algorithm \ref{alg} on $\Dscr^{[0]}_k$ to derive $\widehat{\bbeta}^{[0]}_{{\rm Deb},k}$ for $k=1,2$ and on $\Dscr^{[1]}$ for $\widehat{\bbeta}^{[1]}_{{\rm Deb}}$.

\quad \textbf{3:} Implement thresholding on the dense vectors to obtain $\widehat\bbeta_{{\rm Thr},k}^{[0]}$ for $k=1,2$ and $\widehat\bbeta_{\rm Thr}^{[1]}$.

\quad \textbf{4:} Implement Algorithm \ref{alg:2} with $\widehat\bbeta_{\rm Thr}^{[1]}$ and $\widehat{\bbeta}^{[0]}_{{\rm Deb},k}$ to obtain $\widehat{\bbeta}^{[0]}_{{\rm KTr},k}$ for $k=1,2$.

\quad \textbf{5:}  For $k=1,2$, aggregate the estimators as
\begin{equation}
\begin{aligned}
& \widehat{\bbeta}_{{\rm AIMS},k}^{[0]} =  w_k \widehat\bbeta_{{\rm Thr},k}^{[0]}+ (1-w_k)\widehat{\bbeta}^{[0]}_{{\rm KTr},k}, \\
& \text{ where } w_k=\frac{e^{-a\widehat{\mathcal{Q}}(\widehat\bbeta_{{\rm Thr},k}^{[0]};\widehat{\bbeta}^{[0]}_{{\rm Deb},3-k})}}{e^{-a\widehat{\mathcal{Q}}(\widehat\bbeta_{{\rm Thr},k}^{[0]};\widehat{\bbeta}^{[0]}_{{\rm Deb},3-k})}+e^{-a\widehat{\mathcal{Q}}(\widehat\bbeta_{{\rm KTr},k}^{[0]};\widehat{\bbeta}^{[0]}_{{\rm Deb},3-k})}}.    
\end{aligned}
\label{equ:ensemb}
\end{equation}

\textbf{Output:} The final AIMS estimator 
$
\widehat{\bbeta}_{\rm AIMS}^{[0]} =\big(\widehat{\bbeta}_{{\rm AIMS},1}^{[0]} + \widehat{\bbeta}_{{\rm AIMS},2}^{[0]}\big)/2.
$
\end{algorithm}

In Algorithm \ref{alg:3}, we split the minority data and obtain two independent bias-corrected dense vectors,  $\widehat{\bbeta}^{[0]}_{\rm Deb,1}$ and $\widehat{\bbeta}^{[0]}_{\rm Deb,2}$. One of them, say $\widehat{\bbeta}^{[0]}_{\rm Deb,1}$, is used to derive the sparse minority-only estimator $\widehat\bbeta_{\rm Thr,1}^{[0]}$ and the knowledge transfer estimator $\widehat\bbeta_{\rm KTr,1}^{[0]}$. Then, the other one $\widehat{\bbeta}^{[0]}_{\rm Deb,2}$ is used to ensemble these two estimators through the surrogate loss function $\widehat{\mathcal{Q}}(\cdot;\widehat{\bbeta}^{[0]}_{\rm Deb,2})$. We use cross-fitting to leverage the two folds more effectively. For ensembling, we use the exponential weighting strategy \citep[e.g.,][]{dalalyan2007aggregation} with some prespecified temperature parameter $a>0$ in (\ref{equ:ensemb}). When $a=\infty$, $\widehat{\bbeta}_{{\rm AIMS},k}^{[0]}$ will be the one chosen from $\widehat\bbeta_{{\rm Thr},k}^{[0]}$ and $\widehat{\bbeta}^{[0]}_{{\rm KTr},k}$ that has smaller loss $\widehat{\mathcal{Q}}$ contrasted with the debiased vector from the other fold of data. For finite \(a>0\), the exponential weights provide a smooth aggregation between the minority-only and knowledge transfer estimators. In practice, \(a\) is selected from a fixed grid, and sensitivity analyses are reported in Supplementary S.2.3. 
As will be shown in Theorem \ref{thm:makeup}, the resulting estimator, denoted as AIMS, achieves the better convergence rate between the minority-only and knowledge transfer estimators up to a mild detection error incurred by the aggregation step.

\section{Theoretical Justification}
\label{sec:theory}

To introduce the key sparsity assumptions and the convergence results, we define the population-level model parameters as follows. Let $\bar\balpha^{[r]} =  \underset{\balpha\in\mathbb{R}^{d}}{\arg \min } \E_{\Sscsub\Cupsub\Tscsub}^{[r]}\{\rho_{\scriptscriptstyle \Tsc,r} S \exp (\bPhi \trans \balpha) - \rho_{\scriptscriptstyle \Ssc,r} (1-S)\bPhi \trans \balpha  \}$ and $\bar \bgamma^{[r]} =  \underset{\bgamma \in\mathbb{R}^{d}}{\arg \min } \E_{\Sscsub}^{[r]} \{ -Y  \bPhi\trans \bgamma + B(\bPhi\trans \bgamma)\}$ 
be the population-level preliminary nuisance model coefficients. 
For $j = 1,\ldots,q$, let $\bar\balpha_{j}^{[r]}  = \bar\balpha^{[r]} + \bar\bxi^{[r]}_j $ and $\bar\bgamma_{j}^{[r]}  = \bar\bgamma^{[r]} + \bar \bzeta^{[r]}_j$ be the population-level calibrated coefficients corresponding to equation (\ref{equ:cal:beta:alpha}), where 
\begin{equation}
\label{equ:mom:cons}
\begin{split}
\bar\bxi^{[r]}_j &=\argmin{\bxi\in\mathbb{R}^{d}}\E_{\Sscsub\Cupsub\Tscsub}^{[r]}\bar w_{j}^{[r]}\dot{b}( \bPhi\trans\bar\bgamma^{[r]})\Fsc^{[r]}(\bxi;\bar \balpha^{[r]}) \,,\\
\bar\bzeta^{[r]}_j & =\argmin{\bzeta\in\mathbb{R}^{d}}\E_{\Sscsub}^{[r]} \bar w_{j}^{[r]} \exp(\bPhi \trans\bar\balpha^{[r]})\Gsc(\bzeta;\bar \bgamma^{[r]}) \,,
\end{split}
\end{equation}
and $\bar w_j^{[r]} = \bar \bOmega_j ^{[r]}\X$.
If the density ratio model is correct, we can show that $\bar\bxi^{[r]}_j=\bzero$ and $\bar\balpha^{[r]} = \bar\balpha_{j}^{[r]}$ for all $j$.
Similarly, if the imputation model is correct, we can show that
$\bar\bzeta^{[r]}_j=\bzero$ and $\bar\bgamma^{[r]} = \bar\bgamma_{j}^{[r]}$ for all $j$. Define $s_{\beta}^{[r]} = \big\|\bar\bbeta^{[r]}\big\|_0$ and 
$s_{\rm nui}^{[r]}=\max\Big\{ \big\|\bar\balpha^{[r]}\big\|_0,\big\|\bar\bgamma^{[r]}\big\|_0, 
{\rm max}_{j\in\{1,\ldots,q\}}\big\|\bar\bOmega^{[r]}_j\big\|_0\Big\}$ 
to represent the (exact) sparsity levels of the target and nuisance model coefficients, respectively. Since the nuisance parameters $\bar\bxi^{[r]}_j$ and $\bar\bzeta^{[r]}_j$ are defined through the reweighted regression in (\ref{equ:mom:cons}), we introduce $R_{{\rm nui},v}^{[r]}=\max\big\{ \|\bar\bxi^{[r]}_j \|_v ,\|\bar\bzeta^{[r]}_j\|_v \big\}$ for an arbitrary $v\in[0,1]$ to accommodate both the exact (i.e., $v=0$) and approximate sparsity (i.e., $v\in(0,1]$) regimes. In addition, we define $R_{\delta, v} =\|\bar\bbeta^{[1]}-\bar\bbeta^{[0]}\|_v =\|\bar\bdelta\|_v$ for any $v\in[0,1]$ to characterize the model heterogeneity between the majority and minority sub-populations on the target. Again, our analyses cover both the exact and approximate sparse $\bar\bdelta$ scenarios. 

We use $a_n = o(b_n)$ if $\lim_{n \rightarrow \infty} a_n/b_n = 0$, $a_n = O(b_n)$ if $\lim\sup_{n \rightarrow \infty} |a_n/b_n| \leq C$ for some constant $C$, and $a_n \asymp b_n$ if  $a_n = O(b_n)$ and $b_n = O(a_n)$. We denote convergence in probability by  $\overset{\P}{\to}$  and  convergence in distribution by $\stackrel{\mathcal{D}}{\rightarrow}$. For a sequence of random variables $Z_n$, we use $Z_n = o_p(a_n)$ if $|Z_n|/a_n \overset{\P}{\to} 0$ and $Z_n = O_p(a_n)$ if $\lim_{C \rightarrow \infty} \lim\sup_{n\rightarrow \infty} \P( |Z_n/a_n| > C) = 0$. Let $n_r = n_{\subSsc, r} \wedge n_{\subTsc, r} = \min\{n_{\subSsc, r} , n_{\subTsc, r} \}$, and assume $n_{\subSsc, 0} =o(n_{\subSsc, 1})$ and $n_{\subTsc, 0}=o(n_{\subTsc, 1})$, i.e., the minority sample sizes are much smaller than the corresponding majority sample sizes.

We begin by justifying the robustness and effectiveness of the model-assisted covariate shift correction step in Algorithm \ref{alg} and its subsequent thresholding estimator $\widehat\bbeta_{\rm Thr}^{[r]}$.
First, we introduce Assumption \ref{asu:model}, which states that at least one nuisance model is correctly specified. This assumption is referred to as the model-doubly-robust assumption in the semiparametric literature \citep{smucler2019unifying}.
\begin{assume}
For $r\in\{0,1\}$, either $h_r^{\star}(\z):= p_{\Z|S=0, R=r}(\z)/p_{\Z|S=1, R=r}(\z)= \exp \{\phi(\z)\trans \bar\balpha^{[r]}\}$ or  $m_r^{\star}(\z):= \E[Y \mid \Z=\z, R=r]=b\{\phi(\z)\trans \bar\bgamma^{[r]}\}$ holds.
\label{asu:model}
\end{assume}

All technical assumptions are presented in Supplementary S.1.1 and are standard in the literature on high-dimensional inference \citep[e.g.,][]{van2014asymptotically} and doubly robust inference \citep[e.g.,][]{tan2020model}. In summary, Assumption S.1 rules out heavy tails and singularity in the predictors $\X$ and nuisance covariates $\bPhi$.  In this assumption, we introduce $K>0$ such that $\|\X_{i}\|_{\infty}\leq K$ for all $i\in\{1,\ldots,n\}$ with probability approaching $1$, where $\|(x_1,\ldots,x_q)\|_{\infty}:=\max_{j\in\{1,\ldots,q\}}|x_j|$. For bounded designs, we have $K=O(1)$ and for Gaussian or sub-Gaussian design, $K=O[\{\log (q)\}^{1/2}]$. 
Assumption S.2 imposes sparsity constraints on $s_{\beta}^{[r]}$, $s_{\rm nui}^{[r]}$, $R_{{\rm nui},v}^{[r]}$ and $R_{\delta, v}$, ensuring they are not excessively large relative to sample sizes.
These constraints are crucial for the consistency of the nuisance and target model estimators, as in the sparse GLM literature  \citep{wainwright2019high}. Assumption S.3 imposes regularity conditions on link functions $g(\cdot)$ and $b(\cdot)$, which are satisfied in broad GLM settings. Assumption S.4 specifies the optimal tuning parameter rates. 

Under Assumption \ref{asu:model} and Assumptions S.1 to S.4, the convergence results for $\Omegahat^{[r]}_j$, $\widetilde{\bbeta}^{[r]}$, and $\{\widehat\balpha^{[r]}_j,\widehat\bgamma^{[r]}_j\}$ are established in Lemmas S.1, S.2, and S.3; see Supplementary Section S.1.1. We then use these results to establish the properties of each bias-corrected estimator $\widehat{\beta}_{{\rm Deb},j}^{[r]}$ in Lemma \ref{thm:beta}.


\begin{lemma}
\label{thm:beta}
Under Assumption \ref{asu:model} and Assumptions S.1 -- S.4, we have 
\begin{align*}
\label{eq:asyexp_beta}
    \widehat{\beta}_{{\rm Deb},j}^{[r]}  - \bar\beta^{[r]}_{j}= \bar\bOmega^{[r]}_j\partial_{\bbeta}\Lsc_r(\bar\bbeta^{[r]}; \bar\balpha^{[r]}_j, \bar\bgamma^{[r]}_j)  +  O_p( \Delta_r),
\end{align*}
where $n_{r}^{1/2}\bar\bOmega^{[r]}_j\partial_{\bbeta}\Lsc_r(\bar\bbeta^{[r]}; \bar\balpha^{[r]}_j, \bar\bgamma^{[r]}_j)\stackrel{\mathcal{D}}{\rightarrow} N(0,\sigma_j^2)$ for some $\sigma_j^2=O(1)$ and
\begin{equation}
\Delta_r= \frac{{K^5}(s_{\rm nui}^{[r]} + s_\beta^{[r]}) \log (q)}{n_r}+  KR_{{\rm nui},v}^{[r]}\left\{\frac{\log (q)}{n_r} \right\}^{1- {v}/{2}}.
\label{equ:def:delta}
\end{equation}
\end{lemma}

In Lemma \ref{thm:beta}, we derive the asymptotic expansion $\widehat{\beta}_{{\rm Deb},j}^{[r]}  = \bar\beta_j^{[r]} + \epsilon_j + {\rm bias}_j$, where $\epsilon_j$ is a mean-zero stochastic term with variance of order $1/n_r$, and ${\rm bias}_j$ stands for a bias term free of any first-order errors from the nuisance estimators $\widehat\balpha^{[r]}_j$ and $\widehat\bgamma^{[r]}_j$. Thus, $\widehat{\beta}_{{\rm Deb},j}^{[r]}$ is insensitive to the nuisance errors. This point will be further discussed in Remark \ref{rem:thm1}.

\begin{remark}
Assumption S.2 alone is not sufficient to guarantee the $n_r^{-1/2}$ convergence rate and asymptotic normality of $\widehat{\beta}_{{\rm Deb}, j}^{[r]}$ in Lemma \ref{thm:beta}.  This rate is also not required for the subsequent convergence
theorems. Nonetheless, under moderately stronger sparsity conditions $s_{\rm nui}^{[r]}+s_\beta^{[r]} = o\big[n_r^{1/2}/\{K^5 \log (q)\}\big]$ and $R_{\rm nui,\nu}^{[r]} = o\big(n_r^{(1-v)/2}/[K\{\log(q)\}^{1-v/2}]\big)$, $n_{r}^{1/2}(\widehat{\beta}_{{\rm Deb},j}^{[r]}  - \bar\beta^{[r]}_{j})$ will converge to a zero-mean normal distribution, enabling the interval estimation for $\bar\beta^{[r]}_{j}$.
\label{rem:lemma:1}
\end{remark}

Although Lemma \ref{thm:beta} establishes element-wise convergence for $\widehat{\bbeta}_{{\rm Deb}}^{[r]}=(\widehat{\beta}_{{\rm Deb},1}^{[r]},\ldots,\widehat{\beta}_{{\rm Deb},q}^{[r]})\trans$, it can still fail to achieve $\ell_2$-consistency, i.e.,  $\big\|\widehat{\bbeta}_{{\rm Deb}}^{[r]}-\bar\bbeta^{[r]}\big\|_2=o_p(1)$. As pointed out at the end of Section \ref{sec:method:cs}, this is because $\widehat{\bbeta}_{{\rm Deb}}^{[r]}$ is dense, so its estimation errors can accumulate as $q$ increases. This motivates us to employ the thresholding procedure to obtain $\widehat \bbeta_{\rm Thr}^{[r]}$. The $\ell_1$- and $\ell_2$-convergence rates of $\widehat \bbeta_{\rm Thr}^{[r]}$ are given in Theorem \ref{thm:beta_thresh}.

\begin{theorem}
 \label{thm:beta_thresh}
 Under Assumption \ref{asu:model} and Assumptions S.1 -- S.4 with 
 $\tau^{[r]} 
\asymp \Big\{  \sqrt{\log(q)}\big(
 n_r^{-1/2}+ \Delta_r
\big) \Big\}$, we have $ \|\widehat \bbeta_{\rm Thr}^{[r]} - \bar\bbeta^{[r]}\|_1 =O_p \big(s_\beta^{[r]}\tau^{[r]} \big)$ and $\|\widehat \bbeta_{\rm Thr}^{[r]} - \bar\bbeta^{[r]}\|_2 =O_p \big(\sqrt{s_\beta^{[r]}}\tau^{[r]} \big)$. 
\end{theorem}

\begin{remark}
The error term $\Delta_r$ defined in (\ref{equ:def:delta}) encodes the impact of the nuisance estimators on the convergence rate of $\widehat{\beta}_{{\rm Deb},j}^{[r]}$ and $\widehat\bbeta_{\rm Thr}^{[r]}$. 
If $\Delta_r=O(n_r^{-1/2})$, then the rate in Theorem \ref{thm:beta_thresh} matches the oracle rate for sparse estimation of $\bar\bbeta^{[r]}$.
Particularly, when $v=0$ and $K= O(1)$,
\[
\Delta_r
=
O\left\{
\frac{(s_{\rm nui}^{[r]}+s_\beta^{[r]}+R_{{\rm nui},0}^{[r]})\log(q)}{n_r}
\right\},
\]
which has an $n_r^{-1}$ factor rather than an $n_r^{-1/2}$ factor. Hence, $\Delta_r=o(n_r^{-1/2})$ provided that the parameters are sufficiently sparse, i.e., 
$(s_{\rm nui}^{[r]}+s_\beta^{[r]}+R_{{\rm nui},0}^{[r]})\log(q)=o(n_r^{1/2})$.

\label{rem:thm1}
\end{remark}

\begin{remark}
Recall that our preliminary estimator $\widetilde \bbeta^{[r]}$ is also based on a doubly robust formulation but does not include the calibration steps 2 -- 4 in Algorithm \ref{alg}. Unlike $\widehat \bbeta_{\rm Thr}^{[r]}$, it does not remove the first order nuisance estimation error $\{s_{\rm nui}^{[r]}\log(q)/n_r\}^{1/2}$; see Lemma S.2. This implies that our proposed calibration steps can effectively reduce the sensitivity to errors in the nuisance estimators, which is in a similar spirit to the rate-double-robustness property studied in \cite{kennedy2020towards}. Nevertheless, \cite{kennedy2020towards} does not establish any model-double-robustness analogous to our Assumption \ref{asu:model}.

\citet{zhou2022doubly} also established model-double-robustness while correcting the bias from the two nuisance functions. However, their analysis is restricted to a low-dimensional outcome model (i.e., $q$ is fixed) and does not extend to the error rates in Lemma~\ref{thm:beta} and Theorem~\ref{thm:beta_thresh} when $q$ diverges.  
For fixed $q$, the rate in Lemma~\ref{thm:beta} is comparable to that in Theorem~1 of \citet{zhou2022doubly}; in this low-dimensional setting, the lasso penalty used to estimate $\bbeta^{[r]}$ and the subsequent lasso-based debiasing step are no longer necessary.

 \label{rem:5}   
\end{remark}

In the following results, we establish the convergence rates of the knowledge transfer estimator $\widehat\bbeta^{[0]}_{\rm KTr}$ and the final estimator $\widehat{\bbeta}_{\rm AIMS}^{[0]}$, showing their potential improvement over the minority-only estimator $\widehat \bbeta_{\rm Thr}^{[0]}$.

\begin{lemma}\label{lemma:KTr}
Under Assumption \ref{asu:model} and Assumptions S.1 -- S.4, suppose that
$\tau^{[r]}$ is of the order given in Theorem \ref{thm:beta_thresh} for $r\in\{0,1\}$ and
$\tau_{\scriptscriptstyle{\rm KTr }} \asymp\{\sqrt{\log(q)}(n_0^{-1/2} + \Delta)\}$, 
where $\Delta = \max_{r\in\{0, 1\}} \Delta_r$. Then
we have 
$
\|\widehat\bdelta -  \bar\bdelta\|_1 =O_p \Big[\{\log(q)/n_0\}^{-v/2} R_{\delta, v}  \tau_{\scriptscriptstyle{\rm KTr }} \Big]$ and $\|\widehat\bdelta -  \bar\bdelta\|_2 =O_p \Big[\{\log(q)/n_0\}^{-v/4} R_{\delta, v}^{1/2} \tau_{\scriptscriptstyle{\rm KTr }} \Big]$. 
\end{lemma}

Lemma \ref{lemma:KTr} is an important intermediate result for the estimation error of $\bar\bdelta=\bar\bbeta^{[1]}-\bar\bbeta^{[0]}$. Based on this lemma, we derive the $\ell_2$-convergence rate of the knowledge transfer estimator in Algorithm \ref{alg:2} in Theorem \ref{thm:KTr}.
\begin{theorem}\label{thm:KTr}
Under Assumption \ref{asu:model} and Assumptions S.1 -- S.4, 
with 
$\tau^{[r]}$ for $r\in\{0,1\}$ and $\tau_{\scriptscriptstyle{\rm KTr }}$ of the orders given in Lemma \ref{lemma:KTr},
we have
\begin{equation}
 \big\|\widehat \bbeta_{\rm KTr}^{[0]} - \bar\bbeta^{[0]}\big\|_2 = O_p  
\left[\big\{s_{\beta}^{[1]} \log(q)\big\}^{1/2}\big(
 n_1^{-1/2}+ \Delta_1\big) + 
 \frac{\{\log(q)\}^{1/2-v/4}}{n_0^{-v/4}} R_{\delta, v}^{1/2} \big(n_0^{-1/2} + \Delta\big)\right].
 \label{equ:thm:2}
\end{equation}
\end{theorem}

The first error term in (\ref{equ:thm:2}) arises from the majority estimator $\widehat\bbeta_{\rm Thr}^{[1]}$ and the second term corresponds to the error of $\widehat\bdelta$ given in Lemma \ref{lemma:KTr}.
We discuss the potential efficiency gain from the knowledge transfer step in Algorithm \ref{alg:2} in Remark \ref{rem:eff:gain}. 
\begin{remark}
For simplicity, 
we consider a scenario where 
$\bdelta$ is exactly sparse ($v =0$), $s_{\beta}^{[1]}\asymp s_{\beta}^{[0]}$,
$s_{\rm nui}^{[1]}\asymp s_{\rm nui}^{[0]}$, and $R_{{\rm nui},v}^{[1]}\asymp R_{{\rm nui},v}^{[0]}$, which implies that $\Delta_1=o(\Delta_0)$. 
When $R_{\delta, 0}=o(s_{\beta}^{[0]})$, i.e., the difference $\bar\bdelta$ is sparser than the target model coefficients, Theorem
\ref{thm:beta_thresh} and Lemma \ref{lemma:KTr} imply that the ratio of the
$\ell_2$-convergence rate of $\widehat\bdelta$ to that of
$\widehat \bbeta_{\rm Thr}^{[0]}$ is
$(R_{\delta,0}/s_\beta^{[0]})^{1/2}=o(1)$, indicating efficiency improvement of the knowledge transfer estimator $\widehat \bbeta_{\rm KTr}^{[0]}$ over the minority-data-only estimator $\widehat \bbeta_{\rm Thr}^{[0]}$.

Our knowledge transfer between the majority and minority subgroups does not rely on target labels. Despite this, under the additional assumption $\Delta_r = O(n_r^{-1/2})$ for $r\in \{0, 1\}$, the error rate of $\widehat \bbeta_{\rm KTr}^{[0]}$ is comparable to the standard transfer
learning rate obtained when $n_0$ labeled target observations are available. 
This comparison applies directly to supervised
transfer learning methods \citep{li2022translasso,tian2022transfer,he2021transfer}, and is also aligned with the intermediate result in Theorem 1 of \citet{cai2022semi} after accounting for the additional density ratio estimation error in their analysis.

\label{rem:eff:gain}
\end{remark}

Nevertheless, in a slightly simplified setting where
$\Delta_r=O(n_r^{-1/2})$ for $r\in\{0,1\}$, Theorem \ref{thm:KTr}
shows that the rate upper bound for $\widehat\bbeta_{\rm KTr}^{[0]}$ can be larger
than that for $\widehat\bbeta_{\rm Thr}^{[0]}$ whenever
$s_\beta^{[0]}/s_\beta^{[1]}=o(n_0/n_1)$ or
$s_\beta^{[0]}
=
o\left(R_{\delta,\nu}\{\log(q)/n_0\}^{-\nu/2}\right)$.
This phenomenon is referred to as negative knowledge transfer. 
To avoid this issue, we propose the model selection Algorithm \ref{alg:3} to obtain the final estimator $\widehat\bbeta^{[0]}_{{\rm AIMS}}$ and justify its effectiveness below.
We begin with  Lemma \ref{lemma:approx}, which shows that the $\widehat{\bbeta}^{[0]}_{\rm Deb}$-based loss function $\widehat{\mathcal{Q}}\big(\bbeta;\widehat{\bbeta}^{[0]}_{\rm Deb}\big)$ offers a good approximation of the ideal model selection criterion $\|\bbeta-\bar\bbeta^{[0]}\big\|_2^2$.

\begin{lemma}
\label{lemma:approx}
Under Assumption \ref{asu:model} and Assumptions S.1 -- S.4, we have $\widehat{\mathcal{Q}}\big(\bbeta;\widehat{\bbeta}^{[0]}_{\rm Deb}\big)=\|\bbeta-\bar\bbeta^{[0]}\big\|_2^2+2\big(\bbeta-\bar\bbeta^{[0]}\big)\trans\big(\mathcal{E} + {\rm Rem}\big) + C_2$,
where
\[
\mathcal{E}= \big\{\bar \bOmega_{1}^{[0]} 
\partial_{\bbeta}\Lsc_r(\bar\bbeta_1^{[0]}; \bar\balpha_1^{[0]}, \bar\bgamma_1^{[0]}), \dots, \bar \bOmega_{q}^{[0]} 
\partial_{\bbeta}\Lsc_r(\bar\bbeta_q^{[0]}; \bar\balpha_q^{[0]}, \bar\bgamma_q^{[0]})\big\}\trans
\]
is a mean-zero empirical process satisfying $\|\mathcal{E}\|_{\infty}=O_p[\{\log(q)/n_0\}^{1/2}]$, ${\rm Rem}$ denote the remainder approximation-error term satisfying 
\[
\|{\rm Rem}\|_{\infty}=O_p[\{\log(q)\}^{1/2} \Delta_0],
\]
and $C_2$ is a constant independent of $\bbeta$.
\end{lemma}

As discussed in Remark \ref{rem:thm1}, the term ${\rm Rem}$, whose
$\ell_\infty$ norm is controlled by $\{\log(q)\}^{1/2}\Delta_0$,
captures the second order influence of nuisance estimation errors. Meanwhile,
the term $2(\bbeta-\bar\bbeta^{[0]})^\top\mathcal E$ can be controlled by
the concentration of $\mathcal E$. Combining this lemma with Theorems
\ref{thm:beta_thresh} and \ref{thm:KTr}, we derive the convergence rate of
the final estimator $\widehat\bbeta_{\rm AIMS}^{[0]}$ in Theorem
\ref{thm:makeup}.

\begin{theorem}
Under Assumption \ref{asu:model} and Assumptions S.1 -- S.4 with 
$\tau^{[r]}$ for $r\in\{0,1\}$ and $\tau_{\scriptscriptstyle{\rm KTr }}$ of the orders given in Lemma \ref{lemma:KTr}, we have 
\begin{align*}
\big\|\widehat\bbeta^{[0]}_{{\rm AIMS}} - \bar\bbeta^{[0]}\big\|_2 = O_p \Big\{
\Big(\big\|\widehat\bbeta^{[0]}_{{\rm Thr} }- \bar\bbeta^{[0]} \big\|_2 + {\rm DetectErr}\Big)
\wedge  \big\|\widehat\bbeta_{\rm KTr}^{[0]} - \bar\bbeta^{[0]}\big\|_2\Big\}, 
\end{align*}
with the detection error term
$
{\rm DetectErr} = \Big\{(n_0^{-1/2}  + \Delta_0 ) ( \big\|\widehat\bbeta^{[0]}_{{\rm KTr}}- \bar\bbeta^{[0]} \big\|_2 + \big\|\widehat\bbeta^{[0]}_{{\rm Thr}}  -  \bar\bbeta^{[0]} \big\|_2) \Big\}^{1/2}. 
$
When {$\Delta_r=o(n_r^{-1/2})$} for $r\in\{0,1\}$, we have
$$
{\rm DetectErr} = O_p\left(\Big[n_0^{-1/2} \max_{r\in\{0,1\}} \Big\{ \frac{s_\beta^{[r]}\log(q)}{n_r}\Big\}^{1/2} +  n_0^{-1/2}  \Big\{\frac{\log(q)}{n_0} \Big\}^{1/2 - v/4}R_{\delta, v}^{1/2} \Big]^{1/2}\right).
$$
When further assuming $R_{\delta,v} = o\big[\{n_0/\log(q)\}^{1/2 - v/2}\big]$, 
$
{\rm DetectErr} = o_p [n_0^{-3/8}{\{\log(q)\}^{1/8}}].
$
\label{thm:makeup}    
\end{theorem}

In Theorem \ref{thm:makeup}, we show that the $\ell_2$ error of
$\widehat\bbeta_{\rm AIMS}^{[0]}$ is controlled by the better of two quantities:
the error rate of the knowledge transfer estimator $\widehat\bbeta_{\rm KTr}^{[0]}$,
and the error rate of the minority-only estimator $\widehat\bbeta_{\rm Thr}^{[0]}$
plus the detection error incurred by model selection in Algorithm \ref{alg:3}. We further
show that this detection price is
$o_p[n_0^{-3/8}\{\log(q)\}^{1/8}]$ under additional sparsity conditions and the
negligibility of the higher order nuisance error $\Delta_r$. In particular,
when $\nu=0$, the condition on $R_{\delta,\nu}$ requires $\bar\bdelta$ to be
ultra sparse. This price is relatively mild, since $n_0^{-3/8}$ is slower than,
but still close to, the parametric rate $n_0^{-1/2}$.

Similar data-splitting-based procedures are employed in \cite{tian2022transfer}, \cite{cai2022semi}, and related work to protect against negative transfer.
These procedures use target labels, whereas our detection step is implemented without observed labels in the target sample.
In particular, the procedure of \citet{cai2022semi} incurs an additional selection cost of order $n_0^{-1/4}$ on the $\ell_2$ scale for adaptive selection between the knowledge transfer estimator and the target-only estimator.
Under standard sparsity conditions in high-dimensional regression \citep{negahban2012unified, wainwright2019high}, i.e., $R_{\delta, v} = O \big[\{n_0/\log(q)\}^{1-v/2}\big]$, our detection error is also of order at most  $n_0^{-1/4}$. 

\section{Simulation Studies}\label{sec:sim}

We conduct simulation studies under three related data-generating mechanisms. Across the three settings, the sparse linear components are fixed, and the majority-group coefficients are generated as local perturbations of the minority-group coefficients. This creates a regime in which majority-to-minority transfer is plausible, but not guaranteed to outperform the minority-only estimator. For $r\in\{0,1\}$, we consider the following three settings:
\[
\begin{array}{ll}
\text{Setting I:} &
\begin{aligned}[t]
\P(Y=1\mid \Z,R=r)&={\rm expit}\{\Z\trans\bgamma^{[r]}\},\\
\P(S=1\mid \Z,R=r)&={\rm expit}\{\Z\trans\balpha^{[r]}\};
\end{aligned}\\[0.6em]
\text{Setting II:} &
\begin{aligned}[t]
\P(Y=1\mid \Z,R=r)&={\rm expit}\{\Z\trans\bgamma^{[r]}+u_Y(\Z)\},\\
\P(S=1\mid \Z,R=r)&={\rm expit}\{\Z\trans\balpha^{[r]}\};
\end{aligned}\\[0.6em]
\text{Setting III:} &
\begin{aligned}[t]
\P(Y=1\mid \Z,R=r)&={\rm expit}\{\Z\trans\bgamma^{[r]}\},\\
\P(S=1\mid \Z,R=r)&={\rm expit}\{\Z\trans\balpha^{[r]}+u_S(\Z)\}.
\end{aligned}
\end{array}
\]
Setting I is the correctly specified reference case. Setting II tests robustness when the outcome working model is misspecified but the density ratio model remains correctly specified, whereas Setting III reverses this pattern by misspecifying the density ratio model while keeping the imputation model correctly specified. Exact coefficient vectors, perturbation functions, and full data-generating formulas are provided in Supplementary S.3.2.

We set $q=100$, $n_{\scriptscriptstyle \Ssc,1}=2000$, and $n_{\scriptscriptstyle \Tsc,1}=3000$. The primary target-unlabeled minority sample size is $n_{\scriptscriptstyle \Tsc,0}=2000$; the additional cases $n_{\scriptscriptstyle \Tsc,0}\in\{1000,3000\}$ are reported in Supplementary Table~\ref{tab:sim:k10:nt0:sensitivity}. The main results are organized around two one-dimensional grids. Along the sample-size axis, $p=400$ and $n_{\scriptscriptstyle \Ssc,0}\in\{300,400,500,600\}$. Along the dimensionality axis, $n_{\scriptscriptstyle \Ssc,0}=400$ and $p\in\{50,100,200,400,800\}$. Unless otherwise stated, the AIMS implementation follows Algorithm~\ref{alg:3} with $C_\tau=2$ in $\tau^{[r]}=C_\tau\{\log(q)/n_{\scriptscriptstyle \Ssc,r}\}^{1/2}$, bootstrap quantile $q_\tau=0.8$ with $B=500$ Gaussian multiplier bootstrap draws for the calibrated nuisance and transfer thresholds, and aggregation temperature $a=5$. Performance is measured by the squared $\ell_2$ error $\|\widehat\bbeta-\bar\bbeta^{[0]}\|_2^2$.

These choices define a moderately imbalanced high-dimensional reference design. The majority strata are larger than the minority source stratum, reflecting the sampling regime for which transfer from the majority group can be useful but must be protected against bias. The two main grids then vary the minority labeled sample size and the target-model dimension separately, while keeping the nuisance dimension and the majority sample sizes fixed so that changes in performance can be attributed to the displayed axis. The algorithmic constants $C_\tau$, $q_\tau$, and $a$ are used as default tuning values for the main comparison; Supplementary Tables~\ref{tab:sim:k10:qboot:sensitivity} and \ref{tab:sim:k10:ctau:sensitivity} report sensitivity analyses for $q_\tau$ and $C_\tau$, Supplementary Table~\ref{tab:sim:k10:aims:ablation} reports ablations of the minority-side, majority-guided, and aggregation-temperature variants of AIMS, and Supplementary Table~\ref{tab:sim:k10:majority:reference} reports the majority-side estimator performance.

We compare AIMS with three groups of benchmarks. The first group consists of single-nuisance covariate shift estimators derived from the estimating equations in Section~\ref{sec:method:cs}. Importance weighting strategy (IW; Supplementary S.3.1, Algorithm S.1) estimates the minority source-to-target density ratio and then fits the working target model using an $\ell_1$-penalized weighted logistic loss on labeled minority-source observations, following the classical covariate shift correction idea \citep{huang2007correcting}. Imputation strategy (IM, also referred to as pseudo-labeling; Supplementary S.3.1, Algorithm S.2) estimates the minority conditional mean, evaluates it on minority-target covariates as pseudo-outcomes, and then fits the same working target model on the target covariate distribution. IW and IM use only one of the two nuisance models in the doubly robust estimating equation \citep{chakrabortty2019high,liu2023augmented,tian2024semi}. IW$_{\rm aLasso}$ and IM$_{\rm aLasso}$ replace the ordinary $\ell_1$ penalty in the final target-model fit by an adaptive weighted $\ell_1$ penalty, while IM$_{\rm RF}$ and IM$_{\rm XGB}$ replace the parametric conditional mean fit in IM by random forests and XGBoost.

The second group contains transfer and distribution-alignment baselines. Since TransGLM \citep{tian2022transfer} requires labeled observations from the recipient population, it is not directly applicable to our unlabeled-target setting. For comparison, we use the source-labeled adaptation in Supplementary S.3.1, Algorithm S.3: the labeled minority-source sample is treated as the recipient sample, and the labeled majority-source sample is treated as the auxiliary source sample. TransGLM$_{\rm iw}$ further weights these two labeled samples by the estimated source-to-target density ratios to account for covariate shift.  We adapt TransFusion \citep{he2024transfusion} in the same transfer format, using labeled minority-source data as the pseudo-target task and labeled majority-source data as the auxiliary task. CORAL \citep{sun2017correlation} is used as a minority-only distribution-alignment baseline: it aligns the second-order covariance structure of the labeled minority-source covariates to that of the unlabeled minority-target covariates before fitting the sparse logistic target model. The third group contains diagnostic AIMS variants: AIMS$_{\rm min\textit{-}o}$ is the minority-only thresholded estimator from the covariate shift correction step in Algorithm~\ref{alg:math}, AIMS$_{{\rm maj}\textit{-}{\rm g}}$ is the majority-guided transfer estimator from Algorithm~\ref{alg:2}, and the aggregation-temperature variants correspond to the final negative-transfer-protected aggregation in Algorithm~\ref{alg:3}. For readability, the main figures focus on AIMS, IM$_{\rm aLasso}$, TransGLM$_{\rm iw}$, and TransFusion; complete numerical comparisons are reported in Supplementary Tables~\ref{tab:sim:k10:sample:full} and \ref{tab:sim:k10:dimension:full}.

\begin{figure}[htb!]
\centering
\includegraphics[width=0.96\textwidth]{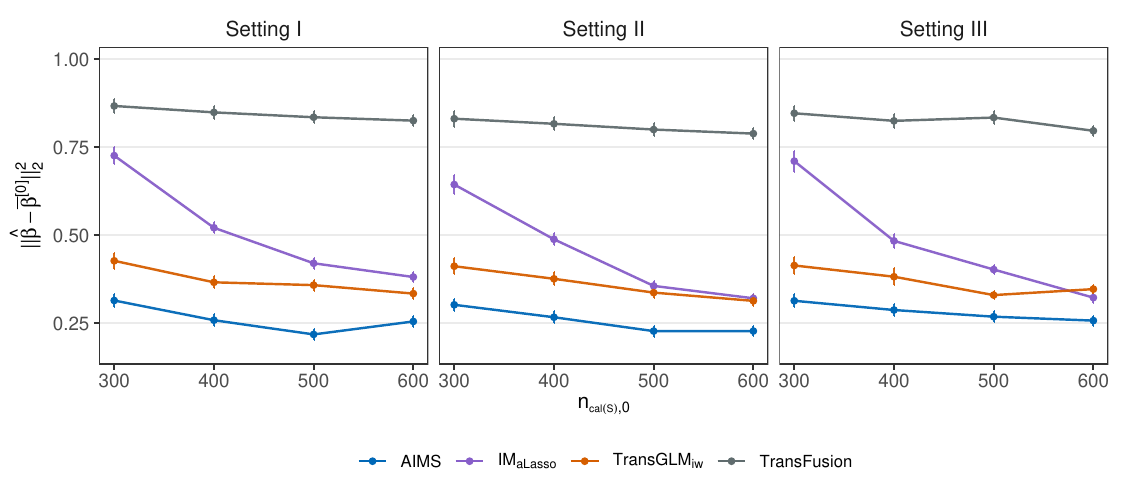}
\caption{Simulation performance along the minority source-sample-size axis, varying $n_{\scriptscriptstyle \Ssc,0}$ with $p=400$ and fixed $n_{\scriptscriptstyle \Tsc,0}=2000$. Points show empirical mean squared $\ell_2$ errors and error bars show one standard error across Monte Carlo repetitions. Across the displayed methods and grid points, the empirical standard deviations are all below $0.30$ (standard errors below $0.03$).}
\label{fig:sim:k10:sample}
\end{figure}

\begin{figure}[htb!]
\centering
\includegraphics[width=0.96\textwidth]{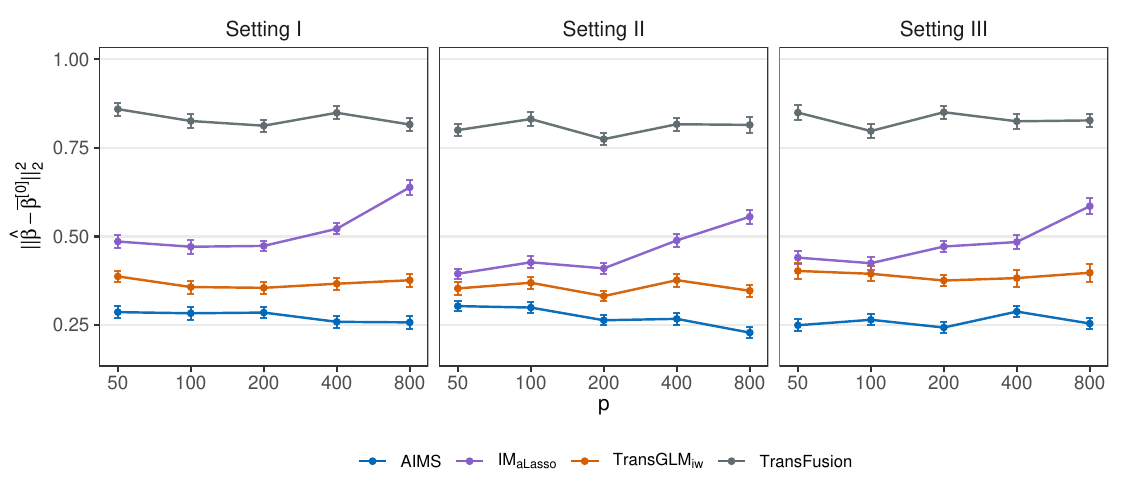}
\caption{Simulation performance along the dimensionality axis, varying $p$ with fixed $n_{\scriptscriptstyle \Ssc,0}=400$ and $n_{\scriptscriptstyle \Tsc,0}=2000$. The horizontal axis is displayed on a log scale with tick labels at the original $p$ values. Points show empirical mean squared $\ell_2$ errors and error bars show one standard error across Monte Carlo repetitions. Across the displayed methods and grid points, the empirical standard deviations are all below $0.30$ (standard errors below $0.03$).}
\label{fig:sim:k10:dimension}
\end{figure}

Figures~\ref{fig:sim:k10:sample} and \ref{fig:sim:k10:dimension} show that AIMS is consistently the leading or near-leading method among the plotted competitors across the three nuisance-model regimes. The gap relative to IM$_{\rm aLasso}$ illustrates the gain from using both the imputation and weighting information rather than relying on the imputation component alone. The improvement over TransFusion is most visible in settings where direct transfer is sensitive to the source--target covariate shift. TransGLM$_{\rm iw}$ is the closest external competitor in several cases, but AIMS remains more stable across the sample-size and dimensionality axes. The overall pattern supports the role of the calibrated nuisance correction and the adaptive transfer-protection step in high-dimensional covariate shift settings. To check that the comparison is not specific to binary outcomes, we also repeat both grids under a continuous Gaussian outcome model; the results are reported in Supplementary Tables~\ref{tab:sim:continuous:sample} and \ref{tab:sim:continuous:dimension}.

\section{Real Data Analysis}\label{sec:real}

\subsection{Genetic risk model for type II diabetes}\label{sec:real:1}
Type II diabetes (T2D), a common chronic disease caused by insulin insufficiency, places a substantial economic and social burden on society. In the United States alone, the total estimated cost of diagnosed diabetes reached \$412.9 billion in 2022 \citep{parker2024economic}. Accurately predicting the risk of developing type II diabetes is crucial for early intervention, prevention, and reducing the long-term health and economic impacts associated with the disease. Extensive genetic studies have suggested that genetic factors play a significant role in the risk of T2D and some non-White racial groups including African Americans and Hispanic and Latin Americans have a higher genetic predisposition to T2D \citep[e.g.,][]{mercader2017genetic,mahajan2018fine,armstrong2024variant}.

In our study, we aim to use EHR-linked biobank data from Mass General Brigham (MGB) biobank \citep{castro2022mass} {to derive} genetic risk prediction models optimized for underrepresented populations by transferring knowledge from the majority population. On one hand, both the demographic variables and single nucleotide polymorphisms (SNPs) $\X$ and the EHR features $\W$ are readily available for all MGB biobank patients. On the other hand, the gold-standard label for the T2D status, $Y\in\{0,1\}$, has only been collected for a subset of $n_{\scriptscriptstyle \Ssc,1} = 375$ for the majority (White race)  and  $n_{\scriptscriptstyle \Ssc,0} = 77$ for the minority (non-White race) patients, whose medical records were manually reviewed in a biomedical study in 2014. In this study, to obtain gold-standard labels for multiple phenotypes more efficiently, patients with International Classification of Diseases (ICD) codes for several phenotypes, such as rheumatoid arthritis and coronary artery disease, were sampled and reviewed simultaneously, leading to sampling that was not completely at random. {The target data consist of}  $n_{\scriptscriptstyle \Tsc,1} = 5000$  and  $n_{\scriptscriptstyle \Tsc,0} = 1000$ patients drawn from the MGB biobank participants with their EHR features updated in 2021. Importantly, the EHR system at MGB, as well as its ICD version, changed around 2015. Thus, the shift in EHR and genomic features between the source and target samples can be attributed to both variation in the sampling method and the difference in the time window of data collection. 

Our goal is to construct a genetic risk model for the T2D status $Y$ using the high-dimensional demographic and genetic variants $\X\in \mathbb{R}^{272}$. To adjust for the distributional shift between the source and target samples, we include $66$ EHR features denoted as $\W$, consisting of T2D-relevant diagnostic and procedure codes provided by \cite{hong2021clinical} and a measure of total health utilization. The two nuisance models are fitted on $\Z=(\X\trans,\W\trans)\trans$, which is clearly high-dimensional given the labeled sample sizes $n_{\scriptscriptstyle \Ssc,0}$ and $n_{\scriptscriptstyle \Ssc,1}$. A validation dataset $\Vsc$ is created by randomly selecting and labeling $47$ individuals from the target minority population solely for evaluation; these validation labels are not used for model fitting, nuisance estimation, tuning, or transfer learning. To measure predictive performance, we calculate the following metrics on $\Vsc$: (1) Brier skill score (BSS), defined as $1-\Eschat_{\Vsc}\{Y-g(\X\trans\bbeta)\}^2/\widehat\Var_{\Vsc}(Y)$ for some estimator $\bbeta$, where $\Eschat_{\Vsc}$ and $\widehat\Var_{\Vsc}$ respectively denote the empirical mean and variance operators on the validation samples; (2) Goodness-of-fit (GOF), measured by negative deviance on the validation data: $1 - 2 \Eschat_{\Vsc}\{-Y\X\trans\bbeta + G(\X\trans\bbeta)\}$; and (3) area under the receiver operating characteristic curve (AUC) of the predictor $g(\X\trans\bbeta)$ on $\Vsc$. We report BSS and GOF together with AUC because these metrics capture complementary aspects of probabilistic risk prediction: AUC measures discrimination, whereas BSS and GOF assess calibration and deviance-based likelihood fit. In this binary-outcome analysis, all methods are evaluated through predicted probabilities \(g(\X^\top\bbeta)\in(0,1)\) on the same validation set, so larger differences in BSS and GOF reflect differences in probability calibration rather than ranking performance alone.

Following our simulation studies, we applied AIMS together with several representative baselines to construct the T2D risk model for the minority target subgroup, including IW, IM, IM$_{\rm XGB}$, CORAL, adapted TransFusion, and the phenotyping-based benchmark IM$_{\rm PheNorm}$. Here IM$_{\rm XGB}$ uses XGBoost-generated pseudo-outcomes in the imputation step, CORAL aligns labeled source and target covariates before sparse logistic regression, and TransFusion is adapted from \citet{he2024transfusion}. Additionally, we incorporated two unsupervised phenotyping algorithms, PheNorm \citep{yu2018enabling} and MAP \citep{liao2019high}, for comparison. We implemented these two algorithms with the EHR features $\W$ on the unlabeled target-minority covariates and used their imputed phenotypes as outcomes to regress against $\X$. This strategy is similar to the IM approach, and we denote the two benchmarks as IM$_{\rm PheNorm}$ and IM$_{\rm MAP}$. Full results for additional benchmark variants, including adaptive-Lasso versions, RF-based imputation, TransGLM variants, and IM$_{\rm MAP}$, are reported in Supplementary Table \ref{tab:eval:beta:full}.

We evaluated the performance of these methods on the validation set $\Vsc$, with results presented in Table \ref{tab:eval:beta}. AIMS achieved the best performance across all three evaluation metrics among the representative baselines. In particular, it substantially improved calibration-oriented performance relative to IM, increasing BSS from $0.23$ to $0.36$ and GOF from $-0.02$ to $0.16$, while also attaining the highest AUC of $0.89$. Among the representative baselines, IM$_{\rm PheNorm}$, IM, IM$_{\rm XGB}$, and TransFusion achieved relatively strong discrimination, but each yielded clearly worse BSS and GOF than AIMS. The full supplementary comparison in Table \ref{tab:eval:beta:full} leads to the same overall conclusion.

These results also provide diagnostic evidence about the role of the proposed components in the T2D application. The source and target samples differ in chart-review design, calendar time, and EHR coding system, making source--target shift plausible; the EHR-derived variables $\W$ therefore provide adjustment information not captured by the genetic covariates $\X$ alone. The largest gains of AIMS over IM occur in BSS and GOF rather than only in AUC, suggesting improved calibration of target-minority risk probabilities. In contrast, CORAL and adapted TransFusion have weaker BSS and GOF, indicating that feature alignment or source-labeled transfer alone is insufficient for this high-dimensional, subgroup-imbalanced setting with target labels unavailable for training. Because gold-standard target-majority labels are not available in the current T2D evaluation set, direct majority-side predictive evaluation is not feasible for this application; the full target-minority benchmark comparison is therefore reported here and in Supplementary Table~\ref{tab:eval:beta:full}.

\begin{table}[htb!]
\footnotesize
\centering

\begin{tabular}{lccccccc}
\hline \hline\\[-2.5ex] 
   & AIMS & IW & IM & IM$_{\rm XGB}$ & CORAL & TransFusion & IM$_{\rm PheNorm}$ \\ 
\hline\\[-2.5ex] 
BSS  & \textbf{0.36} & -0.11 & 0.23 & -1.07 & -0.11 & 0.01 & -0.20\\
GOF  & \textbf{0.16} & -0.85 & -0.02 & -1.46 & -0.88 & -0.25 & -0.45\\
AUC & \textbf{0.89} & 0.64 & 0.81 & 0.81 & 0.62 & 0.81 & 0.82\\
\hline\hline\\[-2ex] 
\end{tabular}

\caption{\label{tab:eval:beta}Predictive performance of representative T2D risk models evaluated on the validation data. Full benchmark results are reported in Supplementary Table \ref{tab:eval:beta:full}.}
\end{table}

\subsection{Prediction of mutation-induced Gibbs free energy changes}\label{sec:ddg}

Protein thermodynamic stability is commonly characterized by the Gibbs (un)folding free energy $\Delta G$, while mutation-induced stability changes are correspondingly quantified by $\Delta\Delta G$, the difference in $\Delta G$ between the mutant and wild-type proteins. Accordingly, $\Delta\Delta G$ is directly relevant to protein engineering and functional variant assessment. Our application study uses two curated protein-stability datasets. S8754 \citep{xu24}  is a large-scale $\Delta\Delta G$ resource containing 8,754 single-point mutations across 301 proteins, compiled from ProThermDB \citep{Nikam20} and ThermoMutDB \citep{Xavier20}  through extensive cleaning and manual verification. S461 \citep{hern23} is a cleaned benchmark subset retaining 461 single-point mutations with corrected annotations. Since S8754 and S461 were assembled under distinct curation pipelines and benchmark-oriented de-redundancy was used during dataset construction, nontrivial source--target distributional shift is to be expected in this setting.

Whereas standard supervised studies typically train on S8754 and evaluate on S461, the present analysis targets the labeled-scarce transfer learning regime studied by AIMS. We therefore treat S461 as the labeled source cohort and S8754 as the target cohort. The $\Delta\Delta G$ labels in S8754 are withheld throughout model fitting and used only for held-out evaluation. To construct the subgroup structure required by AIMS, we discretize assay conditions into coarse environment bins based on pH and temperature, as experimental assay conditions are known to significantly affect measured \(\Delta\Delta G\) values \citep{Vetriani1998, Tollinger03, Brien12, Leuenberger2017}. In our analysis, the majority subgroup consists of mutations measured at pH 6--7, and the minority subgroup consists of mutations measured at pH 1--5. Both groups are restricted to the common temperature range 20--30$^\circ$C. The resulting datasets contain 78 observations in the source minority subgroup, 237 in the source majority subgroup, 677 in the target minority subgroup, and 4,140 in the target majority subgroup.

For feature construction, both the wild-type and mutant sequences are encoded using the pretrained ESM-2-650M \citep{Lin23} model, and each mutation is represented by the difference between the final-layer CLS token embeddings of the mutant and wild-type sequences. These representations are pooled across source and target samples, standardized, and reduced by principal component analysis. The 300 principal components are retained as downstream covariates. Since no separate auxiliary covariates are available in this application, we set \(\W=\emptyset\) and \(\Z=\X\), so the same feature vector serves both as the target risk factors and as the nuisance covariates for covariate shift correction.


We then apply AIMS to the minority subgroup using labeled minority-source data and unlabeled minority-target covariates, while borrowing information from the majority subgroup. We report the final AIMS estimator with built-in negative transfer protection, and compare its performance with several baselines including importance weighting, imputation-only estimators based on linear, random forest, and XGBoost pseudo-outcomes, CORAL, adapted TransFusion, and TransGLM with and without importance weighting. All methods are evaluated on the target minority subgroup using held-out ground-truth labels, and performance is summarized by root mean square error (RMSE), mean absolute error (MAE), Pearson correlation, and Spearman correlation, with results presented in Table \ref{tab:eval:ddg}. When a method collapses to essentially constant predictions on the target minority subgroup, the correlation measures are undefined and are therefore reported as ``---''.

The results in Table~\ref{tab:eval:ddg} show that AIMS achieves the best performance on all four metrics. For point prediction accuracy, AIMS achieves the lowest RMSE ($2.67$) and MAE ($1.90$), with CORAL as the closest competitor (RMSE $2.69$, MAE $1.92$). The advantage is more pronounced for rank-order prediction: AIMS achieves a Spearman correlation of $0.27$, compared to $0.19$ for TransGLM, $0.16$ for TransFusion, and $0.14$ for CORAL. Methods including IM, IM$_{\rm RF}$, IM$_{\rm XGB}$, and TransGLM$_{\rm iw}$ collapse to essentially constant predictions and yield undefined correlations. This indicates that naive imputation-based and heavily regularized transfer estimators lose predictive signal entirely on the small and distributionally shifted minority target. AIMS retains meaningful predictive structure even in this challenging regime, supporting the generalizability of the framework beyond the T2D biobank application.

Also, for the prediction performance on the target majority subgroup, Supplementary Table~\ref{tab:eval:ddg2} reports three versions of AIMS: the preliminary estimator $\widetilde \bbeta^{[r]}$ (AIMS$_{\rm Init}$), the dense debiased estimator $\widehat{\bbeta}^{[r]}_{\rm Dense\_Deb}$ (AIMS$_{\rm Dense\_Deb}$), and the final thresholded estimator $\widehat\bbeta_{\rm Thr}^{[r]}$ (AIMS$_{\rm Thr}$). Within the target majority subgroup, AIMS$_{\rm Init}$ yields the smallest RMSE and MAE, whereas AIMS$_{\rm Thr}$ attains the strongest Pearson and Spearman correlations. AIMS$_{\rm Dense\_Deb}$ performs worst under the metrics RMSE and MAE, indicating that the debiased coefficient vector is better interpreted as an intermediate estimator than as a final prediction model. Relative to the minority-target results in Table~\ref{tab:eval:ddg}, prediction on the majority subgroup is empirically easier, with smaller errors and stronger correlations. This is consistent with the substantially larger labeled majority source sample.

\begin{table}[htb!]
\footnotesize
\centering

\begin{tabular}{lrrrrrrrrr}
\hline \hline\\[-2.5ex]
   & AIMS & IW & IM & IM$_{\rm RF}$ & IM$_{\rm XGB}$ & CORAL & TransFusion & TransGLM  & TransGLM$_{\rm iw}$ \\
\hline\\[-2.5ex] 
RMSE  & \textbf{2.67}  & 2.83 & 2.87 & 2.76 & 2.75 & 2.69 & 3.00 & 2.79 & 2.82\\
MAE  & \textbf{1.90}  & 2.06 & 2.09 & 2.00 & 2.00 & 1.92 & 2.19 & 2.00 & 2.05\\
Pearson & \textbf{0.14} & 0.09 & --- & --- & --- & 0.03 & 0.09 & 0.04 & ---\\
Spearman & \textbf{0.27} & 0.04  & --- & --- & --- & 0.14 & 0.16 & 0.19 & ---\\
\hline\hline\\[-2ex] 
\end{tabular}
\caption{\label{tab:eval:ddg} Predictive performance of the $\Delta\Delta G$ models evaluated on the target minority subgroup using held-out ground-truth labels.}

\end{table}

\section{Discussion}
\label{sec:discussion}

Unlike recent literature on doubly robust estimation with high-dimensional nuisance models and low-dimensional target parameters \citep[e.g.,][]{tan2020model,zhou2022doubly}, our target estimator is sparse-regularized and lacks asymptotic linearity, which makes calibrating the nuisance models more challenging. Our debiasing step (\ref{bical}) addresses this issue and enables proper calibration, while the downstream sparsifying step removes error accumulated in the dense debiased estimator. This approach can be generalized to other important settings, such as estimating the HTE with high-dimensional effect modifiers considered in \cite{kato2024triple} and others. Also, compared to existing literature in knowledge transfer \citep[e.g.,][]{li2022translasso}, we tackle a more challenging setup where {target labels are unavailable for training}, and the debiased coefficient vector obtained in the previous step is used as  a substitute for actual labeled samples in sparse regression and negative transfer protection. This makes our method and theory in this part more technically involved.

In biomedical and other fields like social science, both covariate shift and sampling disparity are prominent challenges that impede generalizable and responsible statistical learning. For example, leveraging a clinical trial to infer treatment effects on an observational cohort will also encounter the covariate shift between the two cohorts \citep[e.g.,][]{colnet2024causal}, and age and gender disparities have been frequently considered in recent  studies \citep[e.g.,][]{ting2017gender,ludmir2019factors}. Though these two challenges often co-occur in practice, existing analytical tools typically address only one at a time. Our work fills this methodological gap by simultaneously addressing both challenges in a coherent and complete framework, demonstrating potential for wide application. Finally, we note several challenges in applying AIMS, including high computational costs, potential privacy constraints, and extensions to multi-source settings. These warrant future research.


\acks{The authors thank the action editor and anonymous reviewers for their constructive comments and suggestions. Tianxi Cai was supported by NIH grants R01 LM013614 and R01 NS098023. Doudou Zhou was supported by the MOE AcRF Tier 1 Grant A-8003569-00-00 and the NUS Startup Grant A-0009985-00-00. The authors declare no competing interests.}

\clearpage
\appendix
\begin{center}
\textit{\large Technical supplementary material to}
\end{center}
\begin{center}
{\LARGE Domain Adaptation Targeting Heterogeneous and Imbalanced Subgroups}
\vskip10pt
\end{center}

\setcounter{section}{0}
\renewcommand{\thesection}{S.\arabic{section}}
\renewcommand{\theHsection}{supp.\arabic{section}}
\renewcommand{\theHsubsection}{supp.\arabic{section}.\arabic{subsection}}
\setcounter{equation}{0}
\counterwithout{equation}{section}
\renewcommand{\theequation}{S.\arabic{equation}}
\renewcommand{\theHequation}{supp.\arabic{equation}}
\setcounter{theorem}{0}
\counterwithout{theorem}{section}
\renewcommand{\thetheorem}{S.\arabic{theorem}}
\renewcommand{\theHtheorem}{supp.\arabic{theorem}}

\renewcommand{\theproposition}{S.\arabic{theorem}}
\renewcommand{\thelemma}{S.\arabic{theorem}}

\setcounter{algorithm}{0}
\renewcommand{\thealgorithm}{S.\arabic{algorithm}}
\renewcommand{\theHalgorithm}{supp.\arabic{algorithm}}

\setcounter{table}{0}
\renewcommand{\thetable}{S.\arabic{table}}
\renewcommand{\theHtable}{supp.\arabic{table}}

\newtheorem{assumption}{Assumption}
\renewcommand{\theassumption}{S.\arabic{assumption}}
\renewcommand{\theHassumption}{supp.assumption.\arabic{assumption}}

\section{Theoretical Justification}

\subsection{Technical Assumptions and Lemmas}\label{sec:app:asu}

We use $a_n = o(b_n)$ if $\lim_{n \rightarrow \infty} a_n/b_n = 0$, $a_n = O(b_n)$ if $\lim\sup_{n \rightarrow \infty} |a_n/b_n| \leq C$ for some constant $C$, and $a_n \asymp b_n$ if  $a_n = O(b_n)$ and $b_n = O(a_n)$. We denote convergence in probability by  $\overset{\P}{\to}$  and  convergence in distribution by $\stackrel{\mathcal{D}}{\rightarrow}$. For a sequence of random variables $Z_n$, we use $Z_n = o_p(a_n)$ if $|Z_n|/a_n \overset{\P}{\to} 0$ and $Z_n = O_p(a_n)$ if $\lim_{C \rightarrow \infty} \lim\sup_{n\rightarrow \infty} \P( |Z_n/a_n| > C) = 0$. Let $n_r = n_{\subSsc, r} \wedge n_{\subTsc, r} = \min\{n_{\subSsc, r}, n_{\subTsc, r} \}$, and assume $n_{\subSsc, 0} =o(n_{\subSsc, 1})$ and $n_{\subTsc, 0}=o(n_{\subTsc, 1})$, i.e., the minority sample sizes are much smaller than the majority sample sizes.
The sub-exponential norm of a random variable $X$ is defined as $\|X\|_{\psi_1} =\sup_{t\geq1} t^{-1}\{E(|X|^t)\}^{1/t}$.
Note that $\|X\|_{\psi_1} < C_1$ for some constant $C_1$, if $X$ is sub-exponential.
The sub-Gaussian norm of $X$ is defined as $\|X\|_{\psi_2} =
\sup_{t\geq1} t^{-1/2}\{E(|X|^t)\}^{1/t}$.
Note that $\|X\|_{\psi_2} < C_2$ for some constant $C_2$, if $X$ is sub-Gaussian.
Let 
$\mathcal I^{[r]}\subSsc = \{i:  R_i = r,~ S_i=1 \}$, $\mathcal I^{[r]}\subTsc = \{i:  R_i = r,~ S_i=0 \}$, $r\in \{0, 1\}$. 
Let $a\leq_p b$ be $a<b$ with probability going to one.

\begin{assumption} 
\label{asu:1}
We assume that 
\begin{enumerate}
    \item [(i)]
    $\X_i$ follows a sub-Gaussian distribution with $\max_{i,j}(|X_{ij}|) = O_p(K)$, and $\max_i(\|\bPhi_i\|_\infty)$ is finite. 
    \item [(ii)] For $r \in \{0,1\}$, all matrices listed below have eigenvalues that are bounded above and bounded away from zero:
    $\E\subTsc^{[r]} \{\X\X\trans\}$, 
    $\E\subSsc^{[r]} \{\X\X\trans\}$,
    $\bar\bSigma^{[r]}= \bar\bSigma^{[r]}_{\bar\bbeta^{[r]}}$ , $\E\subTsc^{[r]} \{\bPhi\bPhi\trans\}$,
    $\E\subSsc^{[r]} \{\exp(\bPhi\trans \bar\balpha^{[r]})\bPhi\bPhi\trans\}$, 
    $\E\subSsc^{[r]} \{\dot{b}(\bPhi\trans \bar\bgamma^{[r]})\bPhi\bPhi\trans\}$, 
     \begin{align*}
    &\E_{\subSsc}^{[r]} \left[\bar w_{j}^{[r]}\dot{b}(\bPhi\trans\bar\bgamma^{[r]}) \exp\{\bPhi\trans (\bar\balpha^{[r]} + \bar\bxi^{[r]}_{j} )\}\bPhi\bPhi\trans \right],\\
    \text{and} \quad &\E_{\subSsc}^{[r]} \left[ \bar w_{j}^{[r]}\ddot{b}\{\bPhi\trans(\bar\bgamma^{[r]} + \bar\bzeta_{j}^{[r]})\} \exp(\bPhi\trans \bar\balpha^{[r]} )\bPhi\bPhi\trans \right].
    \end{align*}
    \end{enumerate}
\end{assumption}

\begin{assumption}
\label{asu:2}
We assume that for $r \in \{0, 1\}$
\begin{enumerate} 
    \item [(i)]  $\log(d) \asymp \log(q)$, $n_{\scriptscriptstyle\iota,0} =O(n_{\scriptscriptstyle \iota,1})$, $\iota\in\{\Ssc, \Tsc\}$, and $n_{\subSsc, r} \asymp n_{\subTsc, r}$.
    \item [(ii)]
     ${K^4(s^{[r]}_{\rm nui}}+{s_\beta^{[r]} })\sqrt{\log(q) /n_r} = o(1)$, $\|\bbeta^{[r]}\|_1 = o\{\sqrt{n_r/\log(q)}\}$, and  $\max_{1\leq j \leq q}\|\bar\bOmega_j^{[r]}\|_1 = O(1)$, for $r = 0,1$.
     \item [(iii)] $\bar\bxi_j^{[r]}$ and 
   $\bar \bdelta_j^{[r]}$ are $l_v$ sparse, i.e., $\|\bar \bxi^j\|_v\leq R_{{\rm nui},v}^{[r]}$ and $\|\bar \bdelta^j\|_v\leq R_{{\rm nui},v}^{[r]}$,
   for some $v\in [0,1]$ satisfying $R_{{\rm nui},v}^{[r]}\{ \log(q)/n_r\}^{1/2 - v/2} = o(1)$.

    \item [(iv)] $\bar\bdelta$ is $l_v$ sparse for some $v \in [0,1]$ and $\|\bar\bdelta\|_v\leq R_{\delta, v}$, where 
    $0 < R_{\delta, v} = O[\{n_0/\log(q)\}^{1-v/2}]$.
\end{enumerate}    
\end{assumption}

\begin{assumption}
\label{asu:3}
We assume that  
\begin{enumerate}
    \item [(i)]
    Functions $b(\cdot)$ and $g(\cdot)$ are non-decreasing, 
     and for any $u$ and $v\in \mathbb{R}$, there exists a positive constant $L$ such that $|\dot{b}(u) - \dot{b}(v)| \leq L |u-v|$ and 
    $|\dot{g}(u) - \dot{g}(v)| \leq L |u-v|$.
       
    \item [(ii)] 
    $\max_{i \in \mathcal{I}^{[r]}\subSsc}\exp(|\bPhi_i\trans\bar \balpha^{[r]}|) = O_p(1)$ and 
    $\max_{i \in \mathcal{I}\subSsc^{[r]}, 1\leq j\leq q}\exp\{|\bPhi_i\trans (\bar \balpha^{[r]} + \bar \bxi^{[r]}_j)|\} = O_p(1)$, for $r\in \{0,1\}$.
\end{enumerate}
\end{assumption}

\begin{assumption}\label{asu:4}
We assume that for $r\in \{0,1\}$, the tuning parameters satisfy
\begin{enumerate}
    \item [(i)] $\lambda_{\theta_j}^{[r]} \asymp K\sqrt{\log(q)/n_{\subTsc, r}}$, for $j = 1, \ldots,q$, where $\lambda_{\theta_j}^{[r]}$ is the tuning parameter in the nodewise lasso defined in Supplementary Section  \ref{spp:precision};
    \item [(ii)] $\lambda_\alpha^{[r]}$, $\lambda_\gamma^{[r]}$, $\lambda^{[r]}$,  $\lambda^{[r]}_{\alpha_j}$, $\lambda^{[r]}_{\gamma_j}$  $\asymp \sqrt{\log(q)/n_{r}}$, for $j = 1, \ldots,q$.
\end{enumerate}
\label{asu:a4}
\end{assumption}

\begin{lemma}
\label{lemma:omega}
Under Assumptions \ref{asu:1}, \ref{asu:2} and \ref{asu:4}, we have 
$$
\|\Omegahat_{j}^{[r]} - \bar\bOmega_{j}^{[r]}\|_1 = O_p\left\{  K^2 (s_\beta^{[r]} + s_{\rm nui}^{[r]}) \sqrt{\log (q)/n_r} \right\},
$$
$$
\|\Omegahat_{j}^{[r]} - \bar\bOmega_{j}^{[r]}\|_2^2 = O_p \left\{
 K^4(s_\beta^{[r]} + s_{\rm nui}^{[r]}) \log (q)/n_r
\right\},
$$
and 
$$
\Big\{\hat{\varsigma}^{[r]}_{j} - \big(\bar\Omega_{j, j}^{[r]}\big)^{-1}\Big\}^2 = O_p \left\{K^4
 (s_\beta^{[r]} + s_{\rm nui}^{[r]}) \log (q)/n_r
\right\}, 
$$
where $\widehat\varsigma_j^{[r]}$ is the nodewise residual variance estimator defined in Supplementary Section \ref{spp:precision}. 
Moreover, 
$$
\Big|\Omegahat_{j}^{[r]}
\bar\bSigma^{[r]}
\big(\Omegahat_{j}^{[r]} \big)\trans
- \bar\Omega_{j, j}^{[r]}\Big| 
\leq \|\bar\bSigma^{[r]}\|_{\max}\|\Omegahat^{[r]}_{ j} - \bar\bOmega_{j}^{[r]}\|_1^2 \wedge \lambda_{\max}(\bar\bSigma^{[r]})^2\|\Omegahat^{[r]}_{ j} - \bar\bOmega_{j}^{[r]}\|_2^2 + 2|\hat{\varsigma}^{[r]}_j - \bar\Omega_{j, j}^{-1}|.
$$    
\end{lemma}
\begin{proof}
Lemma \ref{lemma:omega} can be proved using Lemma \ref{lemma:betatilde} and Theorem 3.2 in \cite{van2014asymptotically}.
\end{proof}

\begin{lemma}
\label{lemma:betatilde}
Under Assumptions \ref{asu:1} -\ref{asu:4}, 
we have 
$$
\|\widetilde \bbeta^{[r]}  - \bar\bbeta^{[r]}\|_2^2 = O_p\left\{(s_\beta^{[r]}+ s_{\rm nui}^{[r]}) \log (q)/n_{r} \right\},
$$
and
$$
\|\widetilde \bbeta^{[r]} - \bar\bbeta^{[r]}\|_1 = O_p\left\{ (s_{\beta}^{[r]} + s_{\rm nui}^{[r]}) \sqrt{\log(q)/n_{r} }\right\}.
$$
\end{lemma}

\begin{lemma}
\label{lemma:alphagamma}
Under Assumptions \ref{asu:1}- \ref{asu:4}, we have 
{\small
\begin{align*}
&\|\widehat\balpha^{[r]}_j - \bar\balpha^{[r]}_j\|_2^2+\|
\widehat\bgamma^{[r]}_j
- \bar\bgamma^{[r]}_j\|_2^2 = O_p\left[  \frac{K^4(s_{\rm nui}^{[r]} + s_\beta^{[r]}) \log (q)}{n_r}+  R_{{\rm nui},v}^{[r]}\left\{\frac{\log (q)}{n_r} \right\}^{1- \frac{v}{2}} \right],\\
& \|\widehat\balpha^{[r]}_j - \bar\balpha^{[r]}_j\|_1 +\|\widehat\bgamma^{[r]}_j
- \bar\bgamma^{[r]}_j\|_1  =O_p\left[ K^4(s_{\rm nui}^{[r]} + s_\beta^{[r]})\left\{\frac{ \log(q) }{n_r}\right\}^{\frac{1}{2}} +  R_{{\rm nui},v}^{[r]}\left\{\frac{\log(q)}{n_r} \right\}^{\frac{1-v}{2}}\right].
\end{align*}
}
\end{lemma}

\subsection{Proof of Proposition \ref{prop:1}}

We have 
    \begin{equation}
    \begin{aligned}
\partial_{\bbeta} \E[ \Lsc_r(\bbeta; \balpha^{[r]}, \bgamma^{[r]} ) ]  = & \E\subSsc [ \exp(\bPhi\trans \balpha^{[r]}) \X \{ b(\bPhi\trans\bgamma^{[r]}) -  Y \} \mid R = r] \\
  & + \E\subTsc[   \X\{ g(\X_i\trans \bbeta) - b(\bPhi\trans \bgamma^{[r]})  \} \mid R =r].       
    \end{aligned}
\label{aeq}    
\end{equation}
When $m_r^{\star}(\z)$ is correctly specified, i.e., $m_r^{\star}(\z) = b\big(\phi(\z)\trans\bgamma^{[r]} \big)$, we have that
$$\E\subSsc[ \exp(\bPhi\trans\balpha^{[r]}) \X \{ b(\bPhi\trans\bgamma^{[r]}) -  Y\} \mid R = r] = \bzero$$
 and the solution of $\E\subTsc[   \X\{ g(\X\trans \bbeta) - b(\bPhi\trans \bgamma^{[r]})  \} \mid R =r] = \bzero$
 is $\bar \bbeta^{[r]}$ according to its definition and the equality 
 $$
 \begin{aligned}
  \E\subTsc[   \X\{ g(\X_i\trans \bbeta) - b(\bPhi\trans \bgamma^{[r]})  \} \mid R =r] = &  \E\subTsc \big[ \E\subTsc[  \X\{ g(\X\trans \bbeta) - b(\bPhi\trans \bgamma^{[r]})  \} \mid \bPhi, R =r] \mid R =r \big] \\
  = & \E\subTsc \big[ \E\subTsc[  \X\{ g(\X\trans \bbeta) - Y  \} \mid \bPhi, R =r] \mid R =r \big] \\
  = & \E\subTsc[ \X\{ g(\X\trans \bbeta) - Y  \} \mid R =r]. 
 \end{aligned}
$$
 When $h_r^{\star}(\z)=\exp\big( \phi(\z)\trans\balpha^{[r]}\big)$, we have
\[
\E\subSsc[ \exp(\bPhi\trans\balpha^{[r]}) \X \{ b(\bPhi\trans\bgamma^{[r]}) - g(\X\trans \bbeta)\}\mid R = r ]+ \E \subTsc [  \X \{ g(\X\trans \bbeta) - b(\bPhi\trans \bgamma^{[r]}) \} \mid R =r ] = \bzero,
\]
and the remaining term $\E\subSsc[\exp(\bPhi\trans\balpha^{[r]}) \X \{ g(\X\trans \bbeta) -  Y \}\mid R = r] = \E\subTsc[ \X \{ g(\X\trans \bbeta) -  Y \}\mid R = r] =  \bzero$ leads to a valid solution for $\bar \bbeta^{[r]}$. As a result, the proposed loss function provides a doubly robust estimator of $\bar \bbeta^{[r]}$.

\subsection{Proof of Lemma \ref{lemma:betatilde}}
By the convexity of  $\Lsc_r(\bbeta; \widetilde\balpha^{[r]}, \widetilde\bgamma^{[r]})+\lambda^{[r]}\|\bbeta\|_1$, we have
$$
\partial_{\bbeta}\Lsc_r(\widetilde\bbeta^{[r]};  \widetilde\balpha^{[r]}, \widetilde\bgamma^{[r]} )\trans (\widetilde\bbeta^{[r]} - \bar\bbeta^{[r]}) + \lambda^{[r]}\|\widehat\bbeta^{[r]}\|_1 \leq\lambda^{[r]}\|\bar\bbeta^{[r]}\|_1, 
$$
where
$$
\partial_{\bbeta} \Lsc_r(\bbeta; \balpha, \bgamma )
=  \frac{{\rm d } \Lsc_r(\bbeta;\balpha, \bgamma ) }{{\rm d } \bbeta}   
= - \Eschat\subSsc^{ [r]}  \exp(\bPhi_i\trans\balpha) \X_i \{  Y_i - b(\bPhi_i\trans\bgamma)\} - \Eschat\subTsc^{ [r]}  \X_i\{ b(\bPhi_i\trans\bgamma) - g(\X_i\trans \bbeta) \}.
$$
We further have 
\begin{align}
\label{eq:l1_1}
&\langle\partial_{\bbeta}\Lsc_r(\widetilde\bbeta^{[r]};  \widetilde\balpha^{[r]}, \widetilde\bgamma^{[r]} )
- \partial_{\bbeta}\Lsc_r(\bar\bbeta^{[r]};  \widetilde\balpha^{[r]}, \widetilde\bgamma^{[r]} ) , (\widetilde\bbeta^{[r]}- \bar\bbeta^{[r]})  \rangle  + \lambda^{[r]}\|\widetilde\bbeta^{[r]}\|_1 \nonumber\\
\leq &
- \partial_{\bbeta}\Lsc_r(\bar\bbeta^{[r]};  \widetilde\balpha^{[r]}, \widetilde\bgamma^{[r]} ) \trans (\widetilde\bbeta^{[r]}- \bar\bbeta^{[r]})  +   \lambda^{[r]}\|\bar\bbeta^{[r]}\|_1,     
\end{align}
where
$$
\langle\partial_{\bbeta}\Lsc_r(\widetilde\bbeta^{[r]};  \widetilde\balpha^{[r]}, \widetilde\bgamma^{[r]} )
- \partial_{\bbeta}\Lsc_r(\bar\bbeta^{[r]};  \widetilde\balpha^{[r]}, \widetilde\bgamma^{[r]} ) , (\widetilde\bbeta^{[r]}- \bar\bbeta^{[r]})  \rangle 
=\Eschat\subTsc^{ [r]}  \{ g(\X_i\trans \widetilde\bbeta^{[r]})  - g(\X_i\trans \bar\bbeta^{[r]}) \} \X_i\trans ( \widetilde\bbeta^{[r]}- \bar\bbeta^{[r]}).
$$
We now deal with $\widetilde\balpha^{[r]}$ and $\widetilde\bgamma^{[r]}$ in $- \partial_{\bbeta}\Lsc_r(\bar\bbeta^{[r]};  \widetilde\balpha^{[r]}, \widetilde\bgamma^{[r]} ) \trans (\widetilde\bbeta^{[r]}- \bar\bbeta^{[r]}) $.
We have
$$
- \partial_{\bbeta} \Lsc_r(\bar\bbeta^{[r]}; \widetilde\balpha^{[r]}, \widetilde\bgamma^{[r]} )\trans (\widetilde\bbeta^{[r]} - \bar\bbeta^{[r]})
 = - \partial_{\bbeta} \Lsc_r(\bar\bbeta^{[r]}; \bar\balpha^{[r]}, \bar\bgamma^{[r]})\trans (\widetilde\bbeta^{[r]} - \bar\bbeta^{[r]})  + V_1 + V_2 + V_3
$$
where 
$$
V_1 = \Eschat\subTsc^{ [r]}  \{ b(\bPhi_i\trans\widetilde\bgamma^{[r]}) - b(\bPhi_i\trans\bar\bgamma^{[r]})\} \X_i \trans (\widetilde\bbeta^{[r]} - \bar\bbeta^{[r]})
= \Eschat\subTsc^{[r]} \dot{b}(\bPhi_i\trans \bgamma^*)\bPhi_i\trans(\widetilde\bgamma^{[r]} - \bar \bgamma^{[r]}) \X_i \trans (\widetilde\bbeta^{[r]} - \bar\bbeta^{[r]}), 
$$
\begin{align*}
V_2 &=  \Eschat\subSsc^{ [r]}  \{\exp(\bPhi_i\trans\widetilde\balpha^{[r]}) - \exp(\bPhi_i\trans\bar\balpha^{[r]})\} \{  Y_i - b(\bPhi_i\trans\bar\bgamma^{[r]})\} \X_i \trans(\widetilde\bbeta^{[r]} - \bar\bbeta^{[r]})\\
&= \Eschat\subSsc^{ [r]} \exp(\bPhi_i\trans\balpha^*)
 \{  Y_i - b(\bPhi_i\trans\bar\bgamma^{[r]})\} 
 \bPhi_i\trans (\widetilde\balpha^{[r]} - \bar\balpha^{[r]})
 \X_i \trans(\widetilde\bbeta^{[r]} - \bar\bbeta^{[r]}),     
\end{align*}
\begin{align*}
V_3 &=\Eschat\subSsc^{ [r]} \exp(\bPhi_i\trans\widetilde\balpha^{[r]})
\{b(\bPhi_i\trans\bar\bgamma^{[r]}) - b(\bPhi_i\trans\widetilde\bgamma^{[r]})\}\X_i \trans(\widetilde\bbeta^{[r]} - \bar\bbeta^{[r]}) \\
&=\Eschat\subSsc^{[r]} \exp(\bPhi_i\trans\widetilde\balpha^{[r]}) \dot{b}(\bPhi_i\trans \bgamma^*) \bPhi_i\trans (\bar\bgamma^{[r]} - \widetilde\bgamma^{[r]})
\X_i \trans(\widetilde\bbeta^{[r]} - \bar\bbeta^{[r]}), 
\end{align*}
and $\balpha^*$ and $\bgamma^*$ are on the line connecting $\widetilde\balpha^{[r]}$ and $\bar\balpha^{[r]}$, and 
$\widetilde\bgamma^{[r]}$ and $\bar\bgamma^{[r]}$, respectively.

We then upper bound $V_1$,  $V_2$, and  $V_3$ with high probability. 
For $V_1$, we have 
\begin{align*}
V_1  & \leq \left[\Eschat\subTsc^{[r]} \dot{b}(\bPhi_i\trans \bgamma^*)^2 \{\bPhi_i\trans(\widetilde\bgamma^{[r]} - \bar\bgamma^{[r]})\}^2\right]^{1/2}
\left[\Eschat\subTsc^{[r]} \{\X_i \trans (\widetilde\bbeta^{[r]} - \bar\bbeta^{[r]})\}^2\right]^{1/2} \\
&\leq_p C_{\gamma, 1}  \sqrt{s_{\rm nui}^{[r]}}\lambda_\gamma^{[r]}\left[\Eschat\subTsc^{[r]} \{\X_i \trans (\widetilde\bbeta^{[r]} - \bar\bbeta^{[r]})\}^2\right]^{1/2}, 
\end{align*}
where $C_{\gamma, 1}$ is a positive constant. 

Since $\max_{i \in \mathcal{I}\subSsc^{[r]}} \exp(|\bPhi_i\trans \bar\balpha^{[r]}|) = O_p(1)$ and $\max_{i \in \mathcal{I}\subSsc^{[r]}}  \|\bPhi_i\|_\infty$ is bounded, and $\|\widetilde \balpha^{[r]} - \bar\balpha^{[r]}\|_1 = O(s_{\rm nui}^{[r]}\lambda_{\alpha}^{[r]}) = o_p(1)$, then we have 
$\max_{i \in \mathcal{I}\subSsc^{[r]}} \exp(\bPhi_i\trans \balpha^*) = O_p(1)$. Hence, for $V_2$, 
we have 
\begin{align*}
V_2 &\leq 
\left[\Eschat\subSsc^{[r]} \exp(2\bPhi_i\trans\balpha^*)
 \{  Y_i - b(\bPhi_i\trans\bar\bgamma^{[r]})\} ^2
 \{\bPhi_i\trans (\widetilde\balpha^{[r]} - \bar\balpha^{[r]})\}^2\right]^{1/2}
\left[\Eschat\subSsc^{[r]} 
\{\X_i \trans(\widetilde\bbeta^{[r]} - \bar\bbeta^{[r]}) \}^2\right]^{1/2} \\
& \leq_p C_{\alpha, 1} \sqrt{s_{\rm nui}^{[r]}} \lambda_\alpha^{[r]}
\left[\Eschat\subSsc^{[r]} 
\{\X_i \trans(\widetilde\bbeta^{[r]} - \bar\bbeta^{[r]}) \}^2\right]^{1/2},
\end{align*}
where $C_{\alpha, 1}$  is a positive constant. 

For $V_3$, we further decompose it as 
$V_{31} + V_{32}$, where
\begin{align*}
V_{31}   & = \Eschat\subSsc^{[r]} \exp(\bPhi_i\trans\bar\balpha^{[r]}) \dot{b}(\bPhi_i\trans \bgamma^*) \bPhi_i\trans (\bar \bgamma^{[r]} - \widetilde\bgamma^{[r]})
\X_i \trans(\widetilde\bbeta^{[r]} - \bar\bbeta^{[r]}) \\
& \leq_p C_{\gamma, 2} \sqrt{s_{\rm nui}^{[r]}} \lambda_\gamma^{[r]}
\left[\Eschat\subSsc^{[r]} 
\{\X_i \trans(\widetilde\bbeta^{[r]} - \bar\bbeta^{[r]})\}^2 \right]^{1/2},
\end{align*}
\begin{align*}
V_{32}& = \Eschat\subSsc^{[r]} \{ \exp(\bPhi_i\trans\widetilde\balpha^{[r]})  - \exp(\bPhi_i\trans\bar\balpha^{[r]})\}  \dot{b}(\bPhi_i\trans \bgamma^*) \bPhi_i\trans (\bar \bgamma^{[r]} - \widetilde\bgamma^{[r]})
\X_i \trans(\widetilde\bbeta^{[r]} - \bar\bbeta^{[r]} )\\
&\leq \left[\Eschat\subSsc^{[r]} \exp(2\bPhi_i\trans \balpha^*) \dot{b}(\bPhi_i\trans \bgamma^*)^2 
\{\bPhi_i\trans (\bar\bgamma^{{r}} - \widetilde\bgamma^{[r]})\}^2 \{\bPhi_i\trans (\widetilde\balpha^{[r]} - \bar \balpha^{[r]})\}^2
\right]^{1/2} \left[\Eschat\subSsc^{[r]}  
\{\X_i \trans(\widetilde\bbeta^{[r]} - \bar\bbeta^{[r]})\}^2 \right]^{1/2} \\
&= o_p(V_{31}), 
\end{align*}
where $C_{\gamma, 2}$ is a positive constant. 
Hence, 
$$
V_3 \leq_p 2C_{\gamma, 2} \sqrt{s_{\rm nui}^{[r]}} \lambda_\gamma^{[r]}
\left[\Eschat\subSsc^{[r]} 
\X_i \trans(\widetilde\bbeta^{[r]} - \bar\bbeta^{[r]})^2 \right]^{1/2}.
$$

Coming back to (\ref{eq:l1_1}), we have 
\begin{align}
\label{eq:l1_2}
&\Eschat\subTsc^{[r]} \{ g(\X_i\trans \widetilde\bbeta^{[r]})  - g(\X_i\trans \bar\bbeta^{[r]}) \} \X_i\trans ( \widetilde\bbeta^{[r]} - \bar\bbeta^{[r]})  + \lambda^{[r]}\|\widetilde\bbeta^{[r]}\|_1 \nonumber\\
\leq_p &\|\partial_{\bbeta} \Lsc_r(\bar\bbeta^{[r]}; \bar\balpha^{[r]}, \bar\bgamma^{[r]})\|_\infty \|\widetilde\bbeta^{[r]} - \bar\bbeta^{[r]}\|_1 + \lambda^{[r]}\|\bar\bbeta^{[r]}\|_1
+ V_4, 
\end{align}
where 
\begin{align*}
V_4 =C_{\gamma, 1}  \sqrt{s_{\rm nui}^{[r]}}\lambda_\gamma^{[r]}\left[\Eschat\subTsc^{[r]} \{\X_i \trans (\widetilde\bbeta^{[r]} - \bar\bbeta^{[r]})\}^2\right]^{1/2}
+ \left\{ \sqrt{s_{\rm nui}^{[r]}} 
(C_{\alpha, 1}\lambda_\alpha^{[r]} + 2C_{\gamma, 2}  \lambda_\gamma^{[r]} ) \right\} \left[\Eschat\subSsc^{[r]} 
\{\X_i \trans(\widetilde\bbeta^{[r]} - \bar\bbeta^{[r]})\}^2 \right]^{1/2}.
\end{align*}
Furthermore, there exists a small enough positive constant $C_{\beta,1} <1$, such that 
$\|\partial_{\bbeta} \Lsc_r(\bar\bbeta^{[r]}; \bar\balpha^{[r]}, \bar\bgamma^{[r]} )\|_\infty  \leq_p C_{\beta,1}\lambda^{[r]}$ by Union Bound inequality and Bernstein inequality. 
Then by (\ref{eq:l1_2}), we have 
\begin{equation}
\label{eq:l1_3}
D + \lambda^{[r]}\|\widetilde\bbeta^{[r]}\|_1 
\leq_p 
C_{\beta,1}\lambda^{[r]} \|\widetilde\bbeta^{[r]} - \bar\bbeta^{[r]}\|_1 + \lambda^{[r]}\|\bar\bbeta^{[r]}\|_1
+ V_4,
\end{equation}
where
$$
D = \Eschat\subTsc^{[r]} \{ \dot{g}(\X_i\trans \bbeta^*) ( \widetilde\bbeta^{[r]} - \bar\bbeta^{[r]})\trans\X_i \X_i\trans ( \widetilde\bbeta^{[r]} - \bar\bbeta^{[r]})\},
$$
and $\bbeta^*$ is on the line connecting $\bar\bbeta^{[r]}$ and $\widetilde\bbeta^{[r]}$.

By triangle inequalities
\[
 \lambda^{[r]}\|(\bar\bbeta^{[r]})_{S_\beta^{[r]}} \|_1 \leq  \lambda^{[r]}\|(\widetilde\bbeta^{[r]})_{S_\beta^{[r]}}\|_1 +  \lambda^{[r]}\|(\widetilde\bbeta^{[r]} -   \bar\bbeta^{[r]})_{S_\beta^{[r]}}\|_1,
\]
and
\[
 \lambda^{[r]}\|(\widetilde\bbeta^{[r]} - \bar\bbeta^{[r]})_{(S_\beta^{[r]})^c}\|_1 \leq  \lambda^{[r]}\|(\widetilde\bbeta^{[r]})_{(S_\beta^{[r]})^c}\|_1 +  \lambda^{[r]} \|(\bar\bbeta^{[r]})_{(S_\beta^{[r]})^c} \|_1,
\]
we obtain
\begin{align*}
D + (1 - C_{\beta, 1})\lambda^{[r]} \|(\widetilde\bbeta^{[r]} - \bar\bbeta^{[r]})_{(S_\beta^{[r]})^c}\|_1
\leq_p & 2\|(\bar\bbeta^{[r]})_{S_\beta^c}\|_1
+ (1 +C_{\beta, 1})\lambda^{[r]} \|(\widetilde\bbeta^{[r]} - \bar\bbeta^{[r]})_{S_\beta^{[r]}}\|_1 + V_4 \\
= & (1 +C_{\beta, 1})\lambda^{[r]} \|(\widetilde\bbeta^{[r]} - \bar\bbeta^{[r]})_{S_\beta^{[r]}}\|_1 +  V_4.
\end{align*}
The last equality holds because we assume that $\bar\bbeta^{[r]}$ is $l_0$ sparse. 
Or equivalently, we obtain 
\begin{equation*}
D  + (1 - C_{\beta, 1})\lambda^{[r]} \|\widetilde\bbeta^{[r]} - \bar\bbeta^{[r]}\|_1 
\leq_p
2 \lambda^{[r]}\|(\widetilde\bbeta^{[r]} - \bar\bbeta^{[r]})_{S_\beta^{[r]}}\|_1 + V_4.
\end{equation*}
We now have two possible cases by comparing 
$\chi 2\lambda^{[r]} \|(\widetilde\bbeta^{[r]} - \bar\bbeta^{[r]})_{S_\beta^{[r]}}\|_1 $ and $(1 - \chi) V_4$, where constant $\chi \in (0,1)$.
We either have
\begin{equation}
\label{eq:l1_4}
(1 - \chi)\left\{D  + (1 - C_{\beta, 1})\lambda^{[r]} \|\widetilde\bbeta^{[r]} - \bar\bbeta^{[r]}\|_1  \right\}
\leq_p  2\lambda^{[r]} \|(\widetilde\bbeta^{[r]} - \bar\bbeta^{[r]})_{S_\beta^{[r]}}\|_1
\end{equation}
or
\begin{equation}
\label{eq:l1_5}
\chi\left\{ D  + (1 - C_{\beta, 1})\lambda^{[r]} \|\widetilde\bbeta^{[r]} - \bar\bbeta^{[r]}\|_1  \right\}
\leq_p  V_4.    
\end{equation}
When (\ref{eq:l1_4}) holds, since $D  >0$, we have
$$
(1 - C_{\beta, 1})\lambda^{[r]} \|\widetilde\bbeta^{[r]} - \bar\bbeta^{[r]}\|_1 
\leq_p (1-\chi)^{-1}  2\lambda^{[r]} \|(\widetilde\bbeta^{[r]} - \bar\bbeta^{[r]})_{S_\beta^{[r]}}\|_1,
$$
and consequently
$$
\|\widetilde\bbeta^{[r]} - \bar\bbeta^{[r]}\|_1^2 \leq_p 
C_{\beta,3}\|(\widetilde\bbeta^{[r]} - \bar\bbeta^{[r]})_{S_\beta^{[r]}}\|_1^2
\leq 
C_{\beta,3}s_\beta^{[r]} \|\widetilde\bbeta^{[r]} - \bar\bbeta^{[r]}\|_2^2,
$$
where $C_{\beta,3}$ is a positive constant. 
By the restricted eigenvalue condition on $\bar\bSigma^{[r]}$, we have 
$$
\|\widetilde\bbeta^{[r]} - \bar\bbeta^{[r]}\|_2^2 =O_p\left\{ s_\beta^{[r]} (\lambda^{[r]})^2\right\}, \quad \text{and} \quad \|\widetilde\bbeta^{[r]} - \bar\bbeta^{[r]}\|_1 =O_p \left(s_\beta^{[r]} \lambda^{[r]} \right).
$$

When (\ref{eq:l1_5}) holds, since $\lambda^{[r]} \asymp\lambda^{[r]}_\alpha \asymp\lambda^{[r]}_\gamma \asymp \sqrt{\log (q)/n_r}$,
there exists a positive constant $C_{\beta, 4}$
such that 
$$
1 - C_{\beta, 1}\leq_p  C_{\beta, 4}
\left\{\sqrt{s_\beta^{[r]}} 
\frac{\| \widetilde\bbeta^{[r]} - \bar\bbeta^{[r]} \|_2}{\| \widetilde\bbeta^{[r]} - \bar\bbeta^{[r]} \|_1 } + 
\sqrt{s_{\rm nui}^{[r]}} \left(  \| \Eschat\subSsc^{[r]} \X_i\X_i\trans  -\E\subSsc^{[r]} \X_i\X_i\trans \|_{\max}^{1/2} \vee \| \Eschat\subTsc^{[r]} \X_i\X_i\trans  -\E\subTsc^{[r]} \X_i\X_i\trans \|_{\max}^{1/2}  \right)\right\}.
$$
Since $s_{\rm nui}^{[r]} = o \{\sqrt{n_r/\log(q)}\}$, by Bernstein inequality and Union Bound inequality we can show that
there exists a small enough positive constant $$C_{\beta, 5} < (1-C_{\beta,1})/(2 C_{\beta, 4}) \wedge C_{\beta, 4}\sqrt{\lambda_{\min}(\bar\bSigma^{[r]} )/2} $$ such that 
with probability going to one that 
$$C_{\beta, 4} \sqrt{s_{\rm nui}^{[r]}}\left(  \| \Eschat\subSsc^{[r]} \X_i\X_i\trans  -\E\subSsc^{[r]} \X_i\X_i\trans \|_{\max}^{1/2} \vee \| \Eschat\subTsc^{[r]} \X_i\X_i\trans  -\E\subTsc^{[r]} \X_i\X_i\trans \|_{\max}^{1/2}  \right) \leq C_{\beta, 5}.$$ 
Hence, 
$$
\|\widetilde\bbeta^{[r]} - \bar\bbeta^{[r]} \|_1^2 = O_p(s_\beta^{[r]}\|\widetilde\bbeta^{[r]} - \bar\bbeta^{[r]} \|_2^2).
$$
Moreover, we have 
\begin{align*}
D &\geq \lambda_{\min}[ \E\subTsc^{[r]} \{ \dot{g}(\X_i\trans \bbeta^*) \X_i \X_i\trans\}]\|\widetilde\bbeta^{[r]} - \bar\bbeta^{[r]}\|_2^2\\ &- \| \Eschat\subTsc^{[r]} \{ \dot{g}(\X_i\trans \bbeta^*) \X_i \X_i\trans\} -  \E\subTsc^{[r]} \{ \dot{g}(\X_i\trans \bbeta^*) \X_i \X_i\trans\}\|_{\max}\|\widetilde\bbeta^{[r]} - \bar\bbeta^{[r]}\|_1^2\\   
& \geq_p C_{\beta, 6}\|\widetilde\bbeta^{[r]} - \bar\bbeta^{[r]}\|_2^2,
\end{align*}
where $C_{\beta, 6}$ is a positive constant. 
Since $(1 - C_{\beta, 1})\lambda^{[r]} \|\widetilde\bbeta^{[r]} - \bar\bbeta^{[r]}\|_1 \geq 0$, we also have 
$$
D \leq_p C( \sqrt{s_{\rm nui}^{[r]}} \lambda_\alpha^{[r]} + \sqrt{s_{\rm nui}^{[r]} } \lambda_\gamma^{[r]}) \|\widetilde\bbeta^{[r]} - \bar\bbeta^{[r]}\|_2, 
$$
where $C$ is a positive constant.
Consequently, we obtain 
$$
\|\widetilde\bbeta^{[r]} - \bar\bbeta^{[r]}\|_2  =O_p\left(\sqrt{s_{\rm nui}^{[r]}} \lambda_\alpha^{[r]} + \sqrt{s_{\rm nui}^{[r]} } \lambda_\gamma^{[r]}\right), 
$$
and 
$$
\|\widetilde\bbeta^{[r]} - \bar\bbeta^{[r]}\|_1  =O_p \left\{\left(  \sqrt{s_{\beta}^{[r]}} \sqrt{s_{\rm nui}^{[r]}} \lambda_\alpha^{[r]} + \sqrt{s_{\beta}^{[r]}} \sqrt{s_{\rm nui}^{[r]} } \lambda_\gamma^{[r]}\right) \wedge s_{\rm nui}^{[r]}  \frac{(\lambda_\alpha^{[r]})^2 + (\lambda_\gamma^{[r]})^2}{\lambda^{[r]}}\right\}.
$$

In conclusion, we have 
$$
\|\widetilde\bbeta^{[r]} - \bar\bbeta^{[r]}\|_2^2 = O_p \{s_\beta^{[r]} (\lambda^{[r]})^2 + s_{\rm nui}^{[r]}(\lambda_\alpha^{[r]})^2 + s_{\rm nui}^{[r]} (\lambda_\gamma^{[r]})^2\},
$$
and
$$
\|\widetilde\bbeta^{[r]} - \bar\bbeta^{[r]}\|_1 =O_p \left( s_\beta^{[r]} \lambda^{[r]} + s_{\rm nui}^{[r]}  \frac{(\lambda_\alpha^{[r]})^2 + (\lambda_\gamma^{[r]})^2}{\lambda^{[r]}}\right).
$$

\subsection{Proof of Lemma \ref{lemma:alphagamma}}
Recall $\widehat \balpha_j^{[r]} - \bar \balpha_j^{[r]} = \widetilde\balpha^{[r]} - \bar \balpha^{[r]} +\widehat\bxi_j^{[r]} - \bar\bxi_j^{[r]}$.
When the density ratio model is correctly specified, then 
$\bar\bxi_j^{[r]} = \boldsymbol{0}$, and $ \bar \balpha^{[r]}_j =  \bar \balpha^{[r]}$, for $j=1, \dots, q$.
 We allow $\bar\bxi_j^{[r]}$ to be soft sparse, and it can be approximated well by a sparse vector.
Let $\eta>0$ be a threshold and define the threshold subset 
\[
S_\eta = [k\in \{1, \dots, d\}: |(\bar\bxi_j^{[r]})_k| > \eta].
\] 
In practice, samples are divided into two sets with positive and negative only $\widehat w_{ij}^{[r]} = \widehat\bOmega_j^{[r]}\X_i$'s and solve for $\widehat \bxi_{j+}^{[r]}$ and $\widehat \bxi_{j-}^{[r]}$, respectively.
For simplicity of the proofs, 
we assume that $\widehat w_{ij}^{[r]}$'s are all positive, i.e., 
$\widehat w_{ij}^{[r]} >0$ and $\bar w_{ij}^{[r]} = \bar\bOmega^{[r]}_j\X_i > C_w >0$,  $\forall i, j$. 
The same error rates can be obtained when applying the stratification strategy described in Supplementary \ref{app:split}, using similar proof techniques as those for Lemma S.7 in \cite{zhou2022doubly}.

Since the loss function 
$$
\Eschat_{\Sscsub\Cupsub\Tscsub}^{[r]}\widehat w_{j}^{[r]}\dot{b}( \bPhi\trans\widetilde \bgamma^{[r]})\Fsc^{[r]}(\bxi;\widetilde \balpha^{[r]}) + \lambda_{\alpha_j}^{[r]} \|\bxi\|_1
$$
is convex in $\bxi$, then we have
$$
\Eschat_{\Sscsub\Cupsub\Tscsub}^{[r]}\widehat w_{j}^{[r]}\dot{b}( \bPhi\trans\widetilde \bgamma^{[r]})\left[ \rho_{\scriptscriptstyle \Tsc,r} S \exp\{\bPhi\trans (\widetilde\balpha^{[r]} + \widehat \bxi_j^{[r]})\} - \rho_{\scriptscriptstyle \Ssc,r} (1 - S)\right] \bPhi\trans (\widehat \bxi_j^{[r]} - \bar  \bxi_j^{[r]})+ \lambda_{\alpha_j}^{[r]} \|\widehat \bxi_j^{[r]}\|_1 \leq  \lambda_{\alpha_j}^{[r]} \| \bar\bxi_j^{[r]}\|_1.
$$
After a simple rearrangement, we have 
\begin{align}
\label{eq:basic1}
& D_{\alpha}^j(\widehat \bxi_j^{[r]},\bar \bxi_j^{[r]}; \widetilde\balpha^{[r]}, \widetilde\bgamma^{[r]}, \widetilde\bbeta^{[r]})  +\lambda_{\alpha_j}^{[r]} \|\widehat \bxi_j^{[r]}\|_1 \nonumber\\
\leq 
& - \Eschat_{\Sscsub\Cupsub\Tscsub}^{[r]}\widehat w_{j}^{[r]}\dot{b}( \bPhi\trans\widetilde \bgamma^{[r]})\left[ \rho_{\subTsc, r} S \exp\{\bPhi\trans (\widetilde\balpha^{[r]} + \bar \bxi_j^{[r]})\} - \rho_{\subSsc,r} (1 - S)\right] \bPhi\trans (\widehat \bxi_j^{[r]} - \bar  \bxi_j^{[r]}) +  \lambda_{\alpha_j}^{[r]} \| \bar\bxi_j^{[r]}\|_1, 
\end{align}
where 
\begin{align*}
&D_{\alpha}^j(\widehat \bxi_j^{[r]},\bar \bxi_j^{[r]}; \widetilde\balpha^{[r]}, \widetilde\bgamma^{[r]}, \widetilde\bbeta^{[r]})  \\
=& \Eschat\subSsc^{[r]}\widehat w_{j}^{[r]}\dot{b}( \bPhi\trans\widetilde\bgamma^{[r]} ) \left[ \rho_{\scriptscriptstyle \Tsc,r} S \exp\{\bPhi\trans (\widetilde\balpha^{[r]} + \widehat \bxi_j^{[r]})\} - \rho_{\scriptscriptstyle \Ssc,r} (1 - S)\right]  (\widehat \bxi_j^{[r]} - \bar  \bxi_j^{[r]})\trans\bPhi.
\end{align*}

We first deal with $\widetilde\balpha^{[r]}$, $\widetilde\bgamma^{[r]}$ and $\widetilde\bbeta^{[r]}$ in $D_{\alpha}^j(\widehat \bxi_j^{[r]},\bar \bxi_j^{[r]}; \widetilde\balpha^{[r]}, \widetilde\bgamma^{[r]}, \widetilde\bbeta^{[r]})$.
Recall that under  Assumption \ref{asu:2}, we have 
$\|\widetilde\balpha^{[r]} - \bar \balpha^{[r]}\|_1 = O_p(s_{\rm nui}^{[r]}\lambda_\alpha^{[r]}) = o_p(1)$ and $\|\widetilde\bgamma^{[r]} - \bar \bgamma^{[r]}\|_1 = O_p(s_{\rm nui}^{[r]} \lambda_\gamma^{[r]})= o_p(1)$.
Under (i) in Assumption \ref{asu:1} and (i) in Assumption \ref{asu:3}, 
there exists a positive constant $c_{\alpha,1}$ such that 
\begin{equation}\label{ineq:alphagamma}
\max_{i \in \mathcal{I}\subSsc^{[r]}}\exp\{\bPhi_i\trans(\widetilde\balpha^{[r]} - \bar \balpha^{[r]})\} \geq_p e^{-c_{\alpha,1}},  \ \ \text{and}  \ \
\dot{b}(\bPhi\trans \bar \bgamma^{[r]}) -\dot{b}(\bPhi\trans\widetilde\bgamma^{[r]})  = o_p(1).
\end{equation}
By Lemma \ref{lemma:omega}, under Assumption \ref{asu:2}, we have $\|\Omegahat_{j}^{[r]} - \bar\bOmega_{j}^{[r]}\|_1 =  O_p\left\{ K^2 (s_\beta^{[r]} + s_{\rm nui}^{[r]}) \sqrt{\log (q)/n_r} \right\} =  o_p(1)$.
By (i)  in Assumption \ref{asu:1} and (ii) in Assumption \ref{asu:2},
we have 
\begin{equation} \label{eq:weight}
\max_{i}|\widehat w_{ij}^{[r]} - \bar w_{ij}^{[r]}| =  O_p\left\{ K^3(s_\beta^{[r]} + s_{\rm nui}^{[r]}) \sqrt{\log (q)/n_r} \right\}  = o_p(1). 
\end{equation}
By (\ref{ineq:alphagamma}) and (\ref{eq:weight}), under 
(iv) in Assumption \ref{asu:2}, we have 
\begin{align}
\label{eq:basic2}
D_{\alpha}^j(\widehat \bxi_j^{[r]},\bar \bxi_j^{[r]}; \widetilde\balpha^{[r]}, \widetilde\bgamma^{[r]}, \widetilde\bbeta^{[r]}) 
\geq_p  c_{\xi, 1}
D_{\alpha}^j(\widehat \bxi_j^{[r]},\bar \bxi_j^{[r]}; \bar \balpha^{[r]}, \bar \bgamma^{[r]},  \bar\bbeta^{[r]}), 
\end{align}
where $c_{\xi, 1}$ is a positive constant, and 
\begin{align*}
D_{\alpha}^j(\widehat \bxi_j^{[r]},\bar \bxi_j^{[r]}; \bar \balpha^{[r]}, \bar \bgamma^{[r]}, \bar\bbeta^{[r]})
=  
\Eschat\subSsc^{[r]} \bar w_{j}^{[r]}
\dot{b}( \bPhi\trans\bar \bgamma^{[r]} )
 \left[\exp\{\bPhi\trans (\bar \balpha^{[r]} + \widehat \bxi_j^{[r]})\} -  \exp\{\bPhi\trans (\bar \balpha^{[r]} + \bar \bxi_j^{[r]})\}\right] \bPhi\trans (\widehat \bxi_j^{[r]} - \bar  \bxi_j^{[r]}).
\end{align*}

Second, we deal with $\widetilde\balpha^{[r]}$, $\widetilde\bgamma^{[r]}$ and $\widetilde\bbeta^{[r]}$ in 
$\Eschat_{\Sscsub\Cupsub\Tscsub}^{[r]}\widehat w_{j}^{[r]}\dot{b}( \bPhi\trans\widetilde \bgamma^{[r]})\left[ \rho_{\scriptscriptstyle \Tsc,r} S \exp\{\bPhi\trans (\widetilde\balpha^{[r]} + \bar \bxi_j^{[r]})\} - \rho_{\scriptscriptstyle \Ssc,r} (1 - S)\right]  \bPhi$.
We decompose 
\begin{align}\label{eq:basic3}
&- \Eschat_{\Sscsub\Cupsub\Tscsub}^{[r]}\widehat w_{j}^{[r]}\dot{b}( \bPhi\trans\widetilde \bgamma^{[r]})\left[ \rho_{\scriptscriptstyle \Tsc,r} S \exp\{\bPhi\trans (\widetilde\balpha^{[r]} + \bar \bxi_j^{[r]})\} - \rho_{\scriptscriptstyle \Ssc,r} (1 - S)\right] \bPhi\trans (\widehat \bxi_j^{[r]} - \bar  \bxi_j^{[r]}) \nonumber \\
=& \T_0\trans (\widehat \bxi_j^{[r]} - \bar  \bxi_j^{[r]}) + T_1 + T_2 + T_3 + T_4, 
\end{align}
where 
\[
\T_0 = \Eschat_{\Sscsub\Cupsub\Tscsub}^{[r]}\bar w_{j}^{[r]}\dot{b}( \bPhi\trans\bar \bgamma^{[r]})\left[ \rho_{\scriptscriptstyle\Ssc,r} (1 - S) -  \rho_{\scriptscriptstyle\Tsc, r} S \exp\{\bPhi\trans (\bar\balpha^{[r]} + \bar \bxi_j^{[r]})\} \right] \bPhi 
\]
\[
T_1 =
\Eschat\subSsc^{[r]}\bar w_{j}^{[r]} \dot{b}( \bPhi\trans\bar\bgamma^{[r]} )  \left[\exp\{\bPhi\trans (\bar \balpha^{[r]} + \bar \bxi_j^{[r]})\} - \exp\{\bPhi\trans (\widetilde\balpha^{[r]} + \bar \bxi_j^{[r]})\}\right](\widehat \bxi_j^{[r]} - \bar  \bxi_j^{[r]})\trans \bPhi, 
\]
\[
T_2 =
\Eschat_{\Sscsub\Cupsub\Tscsub}^{[r]} \bar w_{j}^{[r]}  \{ \dot{b}( \bPhi\trans\widetilde\bgamma^{[r]} ) - \dot{b}( \bPhi\trans\bar\bgamma^{[r]} )\} \left[\rho_{\scriptscriptstyle\Ssc,r} (1 - S) -  \rho_{\scriptscriptstyle\Tsc, r} S \exp\{\bPhi\trans (\widetilde\balpha^{[r]} + \bar \bxi_j^{[r]})\} \right] (\widehat \bxi_j^{[r]} - \bar  \bxi_j^{[r]})\trans\bPhi, 
\]
\[
T_3 =
\Eschat_{\Sscsub\Cupsub\Tscsub}^{[r]} {\e_j}\trans (\widehat\bOmega^{[r]} -\bar\bOmega^{[r]})\X \dot{b}( \bPhi\trans\bar\bgamma^{[r]} ) \left[\rho_{\scriptscriptstyle\Ssc,r} (1 - S) -  \rho_{\scriptscriptstyle\Tsc, r} S \exp\{\bPhi\trans (\widetilde\balpha^{[r]} + \bar \bxi_j^{[r]})\} \right] 
(\widehat \bxi_j^{[r]} - \bar  \bxi_j^{[r]})\trans \bPhi, 
\]
and 
\[
T_4 =
\Eschat_{\Sscsub\Cupsub\Tscsub}^{[r]} {\e_j}\trans (\widehat\bOmega^{[r]} -\bar\bOmega^{[r]})\X 
\{ \dot{b}( \bPhi\trans\widetilde\bgamma^{[r]} ) - \dot{b}( \bPhi\trans\bar\bgamma^{[r]})\}\left[\rho_{\scriptscriptstyle\Ssc,r} (1 - S) -  \rho_{\scriptscriptstyle\Tsc, r} S \exp\{\bPhi\trans (\widetilde\balpha^{[r]} + \bar \bxi_j^{[r]})\} \right]  (\widehat \bxi_j^{[r]} - \bar  \bxi_j^{[r]})\trans \bPhi.
\]

Note that by (\ref{ineq:alphagamma}) there exists $u\in (0,1)$ such that 
\begin{align}\label{ineq:alpha}
&\exp\{\bPhi_i\trans (\bar\balpha^{[r]} + \bar\bxi^{[r]}_j)\} - \exp\{\bPhi_i\trans (\widetilde \balpha^{[r]} + \bar\bxi^{[r]}_j)\} \nonumber
\\
=& \exp[\bPhi_i\trans \{u \bar\balpha^{[r]} + (1 - u)\widetilde \balpha^{[r]} + \bar\bxi^{[r]}_j \}](\bar\balpha^{[r]} - \widetilde \balpha^{[r]})\trans \bPhi_i \nonumber \\
=& \exp\{ \bPhi_i\trans (\bar\balpha^{[r]} + \bar\bxi^{[r]}_j)\} \exp\{(1-u)\bPhi_i\trans(\widetilde \balpha^{[r]} - \bar\balpha^{[r]}) \}
(\bar\balpha^{[r]} - \widetilde \balpha^{[r]})\trans \bPhi_i \nonumber \\
\leq_p & e^{c_{\alpha,1}} \exp\{ \bPhi_i\trans (\bar\balpha^{[r]} + \bar\bxi^{[r]}_j)\}(\bar\balpha^{[r]} - \widetilde \balpha^{[r]})\trans \bPhi_i.
\end{align}
By (\ref{ineq:alpha}) and Cauchy-Schwartz inequality, 
we have 
\begin{align}
\label{eq:T1}
T_1
& \leq e^{c_{\alpha,1}}
\left[\Eschat\subSsc^{[r]}(\bar w_{j}^{[r]})^2 \{\dot{b}( \bPhi\trans\bar\bgamma^{[r]})\}^2 \exp\{\bPhi\trans (\bar \balpha^{[r]} + \bar \bxi_j^{[r]})\}  \{(
\widetilde\balpha^{[r]} - \bar \balpha^{[r]})\trans \bPhi\}^2 \right]^{1/2} T_{1,1}^{1/2}\nonumber\\
& \leq_p c_{\xi, 2} \{K^2 s_{\rm nui}^{[r]}(\lambda_{\alpha}^{[r]})^2\}^{1/2}T_{1,1}^{1/2},
\end{align}
where
$$
T_{1,1} = \Eschat\subSsc^{[r]} \exp\{\bPhi\trans (\bar \balpha^{[r]} + \bar \bxi_j^{[r]})\}  
\{(\widehat \bxi_j^{[r]} - \bar  \bxi_j^{[r]})\trans \bPhi\}^2,
$$
and
$c_{\xi, 2}$ is a positive constant. The last inequality holds  
since $\max_{i,j}|\bar w_{ij}^{[r]}| = O_p(K)$, $\max_{i}\dot{b}( \bPhi_i\trans\bar\bgamma^{[r]}) = O_p(1)$, $\max_{i,j}\exp\{\bPhi_i\trans (\bar \balpha^{[r]} + \bar \bxi_j^{[r]})\} = O_p(1)$, which are implied by Assumptions \ref{asu:1} and \ref{asu:3}. 

For $T_2$, We decompose it as
$\T_{2,1}\trans (\widehat \bxi_j^{[r]} - \bar  \bxi_j^{[r]}) + T_{2,2}$, 
where 
 \[
\T_{2,1} = \Eschat_{\Tscsub}^{[r]}\bar w_{j}^{[r]} \{\dot{b}( \bPhi\trans\widetilde\bgamma^{[r]}) - \dot{b}( \bPhi\trans\bar\bgamma^{[r]})\}\bPhi 
- \Eschat_{\Sscsub}^{[r]} \bar w_j^{[r]} \{\dot{b}( \bPhi\trans\widetilde\bgamma^{[r]}) - \dot{b}( \bPhi\trans\bar \bgamma^{[r]})\} \exp\{\bPhi\trans (\bar\balpha^{[r]} + \bar \bxi^{[r]}_j)\}\bPhi, 
\]
and 
\[
T_{2,2} = \Eschat\subSsc^{[r]} \bar w^{[r]}_{j}\{ \dot{b}( \bPhi\trans\widetilde\bgamma^{[r]}) -  \dot{b}( \bPhi\trans\bar\bgamma^{[r]} )\} \left[ \exp\{\bPhi\trans (\bar\balpha^{[r]}  + \bar\bxi^{[r]} _{j}) -   \exp\{\bPhi\trans (\widetilde \balpha^{[r]} + \bar\bxi^{[r]}_{j})\} \right] (\widehat \bxi_j^{[r]} - \bar  \bxi_j^{[r]}) \trans \bPhi.
\]
Under (i)  in Assumption \ref{asu:3},
by  Cauchy-Schwartz inequality, 
we have 
\begin{align}
\label{eq:T2}
&\T_{2,1}\trans (\widehat \bxi_j^{[r]} - \bar  \bxi_j^{[r]})\nonumber\\
\leq& T_{2,3}^{1/2}
\left[ \Eschat_{\Tscsub}^{[r]}(\bar w^{[r]}_j)^2  L^2
\{(\widetilde \bgamma^{[r]} - \bar\bgamma^{[r]})\trans \bPhi\}^2 
\right]^{1/2} \nonumber\\
+& T_{1,1}^{1/2} \left[\Eschat_{\Sscsub}^{[r]}(\bar w^{[r]}_{j})^2\exp\{\bPhi\trans (\bar \balpha^{[r]} + \bar \bxi^{[r]}_{j})\} L^2
\{(\widetilde \bgamma^{[r]} - \bar\bgamma^{[r]})\trans \bPhi\}^2  \right]^{1/2}
 \nonumber\\
\leq_p& \{c_{\xi,3} K^2 s_{\rm nui}^{[r]} (\lambda_\gamma^{[r]})^2\}^{1/2}(T_{2,3}^{1/2} + T_{1,1}^{1/2}),
\end{align}
where
$T_{2,3}=\Eschat_{\Tscsub}^{[r]}   \{(\widehat \bxi_{j}^{[r]} - \bar\bxi^{[r]}_{j})\trans \bPhi\}^2$,
and $c_{\xi,3}$ is a positive constant. 
By  (\ref{ineq:alpha}) and Cauchy-Schwartz inequality, under (ii) in Assumption \ref{asu:2}, we further have
\begin{align*}
T_{2,2} =o_p \left[\{K^2 s_{\rm nui}^{[r]}(\lambda_{\alpha}^{[r]})^2\}^{1/2} T_{1,1}^{1/2}
\right].
\end{align*}

We then decompose
$T_3$ as $\T_{3,1}\trans(\widehat \bxi_j^{[r]} - \bar  \bxi_j^{[r]}) + T_{3,2}$,
where \[
\T_{3,1}
= \Eschat_{\Tscsub}^{[r]} \{\widehat w_{j}^{[r]} - \bar w^{[r]}_{j}\} \dot{b}( \bPhi\trans\bar\bgamma^{[r]})
\bPhi  - 
\Eschat_{\Sscsub}^{[r]} \{\widehat w_{j}^{[r]} - \bar w^{[r]}_{j}\} \dot{b}( \bPhi\trans\bar\bgamma^{[r]})
 \exp\{\bPhi\trans (\bar\balpha^{[r]} + \bar\bxi^{[r]}_{j})\}  \bPhi 
\]
and 
\[
T_{3,2} = \Eschat\subSsc^{[r]} \{\widehat w_{j}^{[r]} -  \bar w^{[r]}_{j}\}
\dot{b}(  \bPhi\trans\bar\bgamma^{[r]}  ) \left[ \exp\{\bPhi\trans (\bar\balpha^{[r]} + \bar\bxi^{[r]}_{j}) -   \exp\{\bPhi\trans (\widetilde\balpha^{[r]} + \bar\bxi^{[r]}_{j})\} \right] (\widehat \bxi_j^{[r]} - \bar  \bxi_j^{[r]})\trans \bPhi.
\]
By Lemma \ref{lemma:omega}, under (ii) in Assumption \ref{asu:1}, we have 
\begin{align}
\label{eq:T3}
\T_{3,1}\trans(\widehat \bxi_j^{[r]} - \bar  \bxi_j^{[r]})  
\leq_p \left\{c_{\xi, 4}  K^4(s_\beta^{[r]} + s_{\rm nui}^{[r]}) \log(q)/n_r \right\}^{1/2}(T_{2,3}^{1/2} + T_{1,1}^{1/2}), 
\end{align}
where $c_{\xi, 4}$ is a positive constant, and 
$$
T_{3,2} = o_p\left\{ K^4 (s_\beta^{[r]} + s_{\rm nui}^{[r]}) \log (q)/n_r T_{1,1} \right\}^{1/2}.
$$
Further, by (\ref{eq:weight}), we have 
\begin{equation}\label{eq:T4}
  T_4 = o_p(T_2).  
\end{equation}

Hence, by (\ref{eq:basic1}),(\ref{eq:basic2}),  (\ref{eq:basic3}), (\ref{eq:T1}), (\ref{eq:T2}), (\ref{eq:T3}) and (\ref{eq:T4}), there exists a positive constant $c_{\xi, 5}$ such that
\begin{align}\label{eq:basic4}
c_{\xi,1}D_{\alpha}^j(\widehat \bxi_j^{[r]},\bar \bxi_j^{[r]}; \bar\balpha^{[r]}, \bar\bgamma^{[r]}, \bar\bbeta^{[r]})  +\lambda_{\alpha_j}^{[r]} \|\widehat \bxi_j^{[r]}\|_1 
& \leq_p \|\T_0\|_\infty\|\widehat \bxi_j^{[r]} - \bar  \bxi_j^{[r]}\|_1
+  \lambda_{\alpha_j}^{[r]} \| \bar\bxi_j^{[r]}\|_1 \nonumber\\
& + c_{\xi, 5} \left\{ K^4(s_\beta^{[r]} + s_{\rm nui}^{[r]}) \log (q)/n_r\right\}^{1/2}  (T_{2,3}^{1/2} + T_{1,1}^{1/2}).
\end{align}
We then upper bound $\|\T_0\|_\infty$ with high probability. Since 
$\bar w_{j}^{[r]} \dot{b}( \bPhi\trans\bar\bgamma^{[r]} )\bPhi $ from the target population and $\bar w_{j}^{[r]} \dot{b}( \bPhi\trans\bar\bgamma^{[r]} ) \exp\{\bPhi\trans (\bar\balpha^{[r]} + \bar\bxi^{[r]}_j)\} \bPhi $ from the source population are  sub-Gaussian, by the definition of $\bar\bxi^{[r]}_j$,  union-bound inequality and Bernstein inequality, 
there exists a small enough positive constant $c_{\xi, 6}$, such that 
$\|\T_{0}\|_\infty \leq_p c_{\xi, 6}\lambda_{\alpha_j}^{[r]}$. 
Hence, by (\ref{eq:basic4}), we obtain
\begin{align}
c_{\xi,1}D_{\alpha}^j(\widehat \bxi_j^{[r]},\bar \bxi_j^{[r]}; \bar\balpha^{[r]}, \bar\bgamma^{[r]}, \bar\bbeta^{[r]})  +\lambda_{\alpha_j}^{[r]} \|\widehat \bxi_j^{[r]}\|_1 
& \leq_p c_{\xi, 6} \lambda_{\alpha_j}^{[r]}  \|\widehat \bxi_j^{[r]} - \bar  \bxi_j^{[r]}\|_1
+  \lambda_{\alpha_j}^{[r]} \| \bar\bxi_j^{[r]}\|_1 \nonumber\\
& + c_{\xi, 5} \left\{ K^4(s_\beta^{[r]} + s_{\rm nui}^{[r]}) \log (q)/n_r\right\}^{1/2}  (T_{2,3}^{1/2} + T_{1,1}^{1/2}).    
\end{align}

By triangle inequalities
\[
 \lambda_{\alpha_j}^{[r]}\|(\bar \bxi_j^{[r]})_{S_\eta} \|_1 \leq  \lambda_{\alpha_j}^{[r]}\|(\widehat \bxi_j^{[r]})_{S_\eta}\|_1 +  \lambda_{\alpha_j}^{[r]}\|(\widehat \bxi_j^{[r]}- \bar  \bxi_j^{[r]})_{S_\eta}\|_1,
\]
and
\[
 \lambda_{\alpha_j}^{[r]}\|(\widehat \bxi_j^{[r]} - \bar  \bxi_j^{[r]})_{S_\eta^c}\|_1 \leq  \lambda_{\alpha_j}^{[r]}\|(\widehat \bxi_j^{[r]})_{S_\eta^c} \|_1 +  \lambda_{\alpha_j}^{[r]} \|(\bar \bxi_j^{[r]})_{S_\eta^c} \|_1,
\]
we obtain
\begin{align*}
& c_{\xi, 1}D_{\alpha}^j(\widehat \bxi_j^{[r]},\bar \bxi_j^{[r]}; \bar \balpha^{[r]}, \bar \bgamma^{[r]}, \bar\bbeta^{[r]}) + (1 - c_{\xi, 6})\lambda_{\alpha_j}^{[r]}\|(\widehat \bxi_j^{[r]} - \bar \bxi_j^{[r]})_{S_\eta^c}\|_1\\
\leq_p&  
2\lambda_{\alpha_j}^{[r]} \| (\bar\bxi_j^{[r]})_{S_\eta^c}\|_1 
+ (1 + c_{\xi, 6})\lambda_{\alpha_j}^{[r]}\|(\widehat \bxi_j^{[r]} - \bar \bxi_j^{[r]} )_{S_\eta}\|_1 
+c_{\xi, 5} \left\{ K^4(s_\beta^{[r]} + s_{\rm nui}^{[r]}) \log (q)/n_r\right\}^{1/2}  (T_{2,3}^{1/2} + T_{1,1}^{1/2}),  
\end{align*}
or equivalently, 
\begin{align}
\label{eq:l2_1}
& c_{\xi, 1}D_{\alpha}^j(\widehat \bxi_j^{[r]},\bar \bxi_j^{[r]}; \bar \balpha^{[r]}, \bar \bgamma^{[r]}, \bar\bbeta^{[r]}) + (1 - c_{\xi, 6})\lambda_{\alpha_j}^{[r]}\|\widehat \bxi_j^{[r]} - \bar \bxi_j^{[r]}\|_1\\ \nonumber
\leq_p&  
2\lambda_{\alpha_j}^{[r]} \| (\bar\bxi_j^{[r]})_{S_\eta^c}\|_1 
+ 2\lambda_{\alpha_j}^{[r]}\|(\widehat \bxi_j^{[r]} - \bar \bxi_j^{[r]})_{S_\eta}\|_1 
+c_{\xi, 5} \left\{ K^4(s_\beta^{[r]} + s_{\rm nui}^{[r]}) \log (q)/n_r\right\}^{1/2}  (T_{2,3}^{1/2} + T_{1,1}^{1/2}).  
\end{align}
Note that since $\bar\bxi_j^{[r]}$ is allowed to be soft sparse, $\| (\bar\bxi_j^{[r]})_{S_\eta^c}\|_1 $ can be nonzero. 

We have two possible cases by comparing $\chi\{2\lambda_{\alpha_j}^{[r]} \| (\bar\bxi_j^{[r]})_{S_\eta^c}\|_1 
+ 2\lambda_{\alpha_j}^{[r]}\|(\widehat \bxi_j^{[r]} - \bar \bxi_j^{[r]})_{S_\eta}\|_1 \}$ and $(1 - \chi)c_{\xi, 5} \left\{ K^4(s_\beta^{[r]} + s_{\rm nui}^{[r]}) \log (q)/n_r\right\}^{1/2}  (T_{2,3}^{1/2} + T_{1,1}^{1/2})$, for a $\chi \in (0, 1)$.
We have either
\begin{equation}
\label{eq:l2_3}
\chi \widetilde D_{\alpha}^j \leq_p 
c_{\xi, 5} \left\{ K^4(s_\beta^{[r]} + s_{\rm nui}^{[r]}) \log (q)/n_r\right\}^{1/2}  (T_{2,3}^{1/2} + T_{1,1}^{1/2}),
\end{equation}
or
\begin{equation}
\label{eq:l2_4}
(1 - \chi)\widetilde D_{\alpha}^j\leq_p 
2\lambda_{\alpha_j}^{[r]} \| (\bar\bxi_j^{[r]})_{S_\eta^c}\|_1 
+ 2\lambda_{\alpha_j}^{[r]}\|(\widehat \bxi_j^{[r]} - \bar \bxi_j^{[r]})_{S_\eta}\|_1,   
\end{equation}
where 
$$
\widetilde D_{\alpha}^j = c_{\xi, 1}D_{\alpha}^j(\widehat \bxi_j^{[r]},\bar \bxi_j^{[r]}; \bar \balpha^{[r]}, \bar \bgamma^{[r]}, \bar\bbeta^{[r]}) + (1 - c_{\xi, 6})\lambda_{\alpha_j}^{[r]}\|\widehat \bxi_j^{[r]} - \bar \bxi_j^{[r]}\|_1
$$
When equation (\ref{eq:l2_4}) holds, since 
$ c_{\xi, 1}D_{\alpha}^j(\widehat \bxi_j^{[r]},\bar \bxi_j^{[r]}; \bar \balpha^{[r]}, \bar \bgamma^{[r]}, \bar\bbeta^{[r]}) \geq  0$, we have 
\[
(1 - c_{\xi, 6})\lambda_{\alpha_j}^{[r]}\|(\widehat \bxi_j^{[r]} - \bar \bxi_j^{[r]})_{S_\eta^c}\|_1
\leq
2(1 - \chi)^{-1}\lambda_{\alpha_j}^{[r]} \| (\bar\bxi_j^{[r]})_{S_\eta^c}\|_1  + 
(1 -\chi)^{-1}(1 + c_{\xi, 6})\lambda_{\alpha_j}^{[r]}\|(\widehat \bxi_j^{[r]} - \bar \bxi_j^{[r]})_{S_\eta}\|_1.
\]
By equation (43) in \cite{negahban2012unified}, we have  
\[
D_{\alpha}^j(\widehat \bxi_j^{[r]},\bar \bxi_j^{[r]}; \bar\balpha^{[r]}, \bar \bgamma^{[r]}, \bar\bbeta^{[r]}) \geq C_1\|\widehat \bxi_j^{[r]}
- \bar \bxi_j^{[r]}\|^2_2  - C_2 \frac{\log(p+q)}{n}\|(\bar \bxi_j^{[r]})_{S_\eta^c}\|_1^2
\]
for some positive constant $C_1$ and $C_2$.
Then by Theorem 1 in \cite{negahban2012unified}, we  have 
\[
\|\widehat \bxi_j^{[r]}
- \bar \bxi_j^{[r]}\|^2_2
\leq_p C_4 (\lambda_{\alpha_j}^{[r]})^2 |S_\eta| +  C_3 \lambda_{\alpha_j}^{[r]} \left\{ \frac{\log(d)}{n} \|(\bar \bxi_j^{[r]})_{S_\eta^c}\|_1^2  + \|(\bar \bxi_j^{[r]})_{S_\eta^c}\|_1\right\}
\]
for some positive constant $C_3$ and $C_4$.
By the proof of Corollary 3 in \cite{negahban2012unified}, 
under (iii) in Assumption \ref{asu:2},
we have 
\[
|S_\eta| \leq R_{{\rm nui},v}^{[r]} \eta^{-v}, 
\]
and 
\[
\|(\bar\bxi_j^{[r]})_{S_\eta^c}\|_1\leq R_{{\rm nui},v}^{[r]} \eta^{1-v}.
\]
Setting $\eta\asymp\sqrt{n_r^{-1}\log(d)} \asymp \sqrt{n_r^{-1}\log(q)} $, we have 
\begin{equation}
\label{eq:l2_5}
\|\widehat \bxi_j^{[r]}
- \bar \bxi_j^{[r]}\|^2_2
= O_p \left[ R_{{\rm nui},v}^{[r]} \left\{\frac{\log (q)}{n_r} \right\}^{1- v/2} \right].
\end{equation}

When (\ref{eq:l2_3}) holds, since $(1 - c_{\xi, 6})\lambda_{\alpha_j}^{[r]}\|\widehat \bxi_j^{[r]} - \bar \bxi_j^{[r]}\|_1 >0$, we have 
$$
c_{\xi, 1}D_{\alpha}^j(\widehat \bxi_j^{[r]},\bar \bxi_j^{[r]}; \bar \balpha^{[r]}, \bar \bgamma^{[r]}, \bar\bbeta^{[r]}) \leq \chi^{-1}c_{\xi, 5} \left\{ K^4(s_\beta^{[r]} + s_{\rm nui}^{[r]}) \log (q)/n_r\right\}^{1/2}  (T_{2,3}^{1/2} + T_{1,1}^{1/2}).
$$
Under Assumptions \ref{asu:1} and \ref{asu:3},
we have 
$$
\|\widehat \bxi_j^{[r]}
- \bar \bxi_j^{[r]}\|^2_2
=  O_p \left\{ K^4(s_\beta^{[r]} + s_{\rm nui}^{[r]}) \log (q)/n_r\right\}, 
$$
and 
$$
\|\widehat \bxi_j^{[r]}
- \bar \bxi_j^{[r]}\|_1
=  O_p \left\{ K^4(s_\beta^{[r]} + s_{\rm nui}^{[r]}) \sqrt{\log (q)/n_r}\right\}.
$$

In conclusion, we have 
$$
\|\widehat \balpha_j^{[r]}
- \bar \balpha_j^{[r]}\|^2_2
=  O_p \left\{ K^4(s_\beta^{[r]} + s_{\rm nui}^{[r]}) \log (q)/n_r + R_{{\rm nui},v}^{[r]} \left\{\frac{\log (q)}{n_r} \right\}^{1- v/2} \right\}, 
$$
and 
$$
\|\widehat \balpha_j^{[r]}
- \bar \balpha_j^{[r]}\|_1
=  O_p \left\{ K^4(s_\beta^{[r]} + s_{\rm nui}^{[r]}) \sqrt{\log (q)/n_r}+ R_{{\rm nui},v}^{[r]} \left\{\frac{\log (q)}{n_r} \right\}^{1/2- v/2} \right\}.
$$

Following similar derivations, under Assumptions \ref{asu:1}-\ref{asu:4}, we have 
$$
\|\widehat \bgamma_j^{[r]}
- \bar \bgamma_j^{[r]}\|^2_2
=  O_p \left\{ K^4(s_\beta^{[r]} + s_{\rm nui}^{[r]}) \log (q)/n_r + R_{{\rm nui},v}^{[r]} \left\{\frac{\log (q)}{n_r} \right\}^{1- v/2} \right\}, 
$$
and 
$$
\|\widehat \bgamma_j^{[r]}
- \bar \bgamma_j^{[r]}\|_1
=  O_p \left\{ K^4(s_\beta^{[r]} + s_{\rm nui}^{[r]}) \sqrt{\log (q)/n_r}+ R_{{\rm nui},v}^{[r]} \left\{\frac{\log (q)}{n_r} \right\}^{1/2- v/2} \right\}.
$$

\subsection{Proof of Lemma \ref{equ:def:delta}}

Recall that 
\begin{equation*}
 \partial_{\bbeta}\Lsc_r(\bbeta;\balpha, \bgamma )  = -\Eschat\subSsc^{[r]} \exp(\bPhi\trans\balpha) \X \{  Y - b(\bPhi\trans\bgamma)\} - \Eschat\subTsc^{[r]} \X\{ b(\bPhi\trans\bgamma) - g(\X\trans \bbeta) \}, 
\end{equation*}
and 
\begin{equation*}
  \partial_{\bbeta}^2\Lsc_r(\bbeta; \balpha, \bgamma ) = \Eschat\subTsc^{[r]} \dot{g}(\X\trans \bbeta)\X\X\trans. 
\end{equation*}
We decompose 
\begin{align*}
\widehat{\beta}^{[r]}_{{\rm Deb},j} - \bar\beta_{j}^{[r]} &= \widetilde \beta_j^{[r]} - \bar\beta_{j}^{[r]} +  \e_j\trans\Omegahat^{[r]}
\left[\Eschat\subSsc^{[r]} \exp(\bPhi\trans\widehat\balpha_j^{[r]}) \X \{ Y - b(\bPhi\trans\widehat\bgamma_j^{[r]})\} + \Eschat\subTsc^{[r]} \X\{b(\bPhi\trans\widehat\bgamma_j^{[r]}) - g(\X\trans \bbetatilde^{[r]}) \}\right] \\
& = \widetilde \beta_j^{[r]} - \bar\beta_{j}^{[r]}
 - \e_j\trans\bar\bOmega^{[r]} \partial_{\bbeta}\Lsc_r(\bar\bbeta^{[r]}; \bar\balpha_j^{[r]}, \bar\bgamma_j^{[r]} ) + T_1 + T_2 + T_3, 
\end{align*}
where 
$$
T_1 = -\e_j\trans \left( \Omegahat^{[r]} -  \bar\bOmega^{[r]} \right) \partial_{\bbeta}\Lsc_r(\bar\bbeta^{[r]}; \bar\balpha_j^{[r]}, \bar\bgamma_j^{[r]} ) , 
$$
$$
T_2 = - \e_j\trans  \bar\Omega^{[r]} \left\{\partial_{\bbeta}\Lsc_r(\widetilde\bbeta^{[r]}; \widehat\balpha_j^{[r]}, \widehat\bgamma_j^{[r]})   - \partial_{\bbeta}\Lsc_r(\bar\bbeta^{[r]}; \bar\balpha_j^{[r]}, \bar\bgamma_j^{[r]} )\right\}, 
$$
$$
T_3 = -\e_j\trans  \left( \Omegahat^{[r]} -  \bar\bOmega^{[r]} \right) 
\left\{\partial_{\bbeta}\Lsc_r(\widetilde\bbeta^{[r]}; \widehat\balpha_j^{[r]}, \widehat\bgamma_j^{[r]})   - \partial_{\bbeta}\Lsc_r(\bar\bbeta^{[r]}; \bar\balpha_j^{[r]}, \bar\bgamma_j^{[r]} )\right\}.
$$
By Lemma \ref{lemma:omega} and Proposition \ref{prop:1}, there exists a constant  $c>0$, such that 
\begin{align*}
|T_1| & \leq \|\e_j\trans \left( \Omegahat^{[r]} -  \bar\bOmega^{[r]} \right) \|_1 \|\partial_{\bbeta}\Lsc(\bar\bbeta^{[r]}; \bar\balpha^j, \bar\bgamma^j)\|_\infty   \\
& \leq_p  c K^2 (s_\beta^{[r]} + s_{\rm nui}^{[r]}) \frac{\log (q)}{n_r}.
\end{align*}
We decompose $T_2$ as
$T_{2,1} + T_{2,2}$, where 
\begin{align*}
T_{2,1} =& -\e_j\trans  \bar\Omega^{[r]}\left\{\partial_{\bbeta}\Lsc_r(\widetilde\bbeta^{[r]}; \widehat\balpha^{[r]}_j, \widehat\bgamma^{[r]}_j)   - \partial_{\bbeta}\Lsc_r(\bar\bbeta^{[r]}; \widehat\balpha^{[r]}_j, \widehat\bgamma^{[r]}_j) \right\}\\
 =& -\e_j\trans   \bar\Omega^{[r]}
\partial_{\bbeta}^2\Lsc_r(\bar\bbeta^{[r]}; \widehat\balpha^{[r]}_j, \widehat\bgamma^{[r]}_j) (\widetilde\bbeta^{[r]} - \bar\bbeta^{[r]}) 
-\e_j\trans  \bar\Omega^{[r]}\{\partial_{\bbeta}^2\Lsc_r(\bbeta^*; \widehat\balpha^j, \widehat\bgamma^j) - \partial_{\bbeta}^2\Lsc_r(\bar\bbeta^{[r]}; \widehat\balpha^j, \widehat\bgamma^j)\}(\widetilde\bbeta^{[r]} - \bar\bbeta^{[r]}) \\
 = & -\widetilde \beta_j^{[r]} + \bar\beta_{j}^{[r]}
- \e_j\trans  \bar\Omega^{[r]} (\widehat\bSigma^{[r]} - \bar \bSigma^{[r]})(\widetilde\bbeta^{[r]} - \bar\bbeta^{[r]})\\
&-\e_j\trans   \bar\Omega^{[r]} \{\partial_{\bbeta}^2\Lsc_r(\bbeta^*; \widehat\balpha^j, \widehat\bgamma^j) - \partial_{\bbeta}^2\Lsc_r(\bar\bbeta^{[r]}; \widehat\balpha^j, \widehat\bgamma^j)\}(\widetilde\bbeta^{[r]} - \bar\bbeta^{[r]}),
\end{align*} 
$T_{2,2} = -\e_j\trans  \bar\Omega^{[r]} \left\{\partial_{\bbeta}\Lsc_r(\bar\bbeta^{[r]}; \widehat\balpha^{[r]}_j, \widehat\bgamma^{[r]}_j)
 - \partial_{\bbeta}\Lsc_r(\bar\bbeta^{[r]}; \bar\balpha_j^{[r]}, \bar\bgamma_j^{[r]})\right\}$, 
and $\bbeta^*$ is on the line connecting $\bar\bbeta^{[r]}$ and $\widetilde \bbeta^{[r]}$.
We further have 
\begin{align*}
T_{2,1} - \bar\beta_{j}^{[r]} + \widetilde \beta_j^{[r]}
& \leq 
\|\e_j\trans \bar\bOmega^{[r]} \|_1\|\widehat\bSigma^{[r]} - \bar\bSigma^{[r]}\|_{\max}\|\widetilde\bbeta^{[r]} - \bar\bbeta^{[r]}\|_1
+ \sup_i|\e_j\trans  \bar\bOmega^{[r]}\X_i| L
|\Eschat\subTsc^{[r]} \{\X\trans (\widetilde\bbeta^{[r]} - \bar\bbeta^{[r]})\}^2|\\
& =
O_p\left\{
(s_{\beta}^{[r]} + s^{[r]}_{\rm nui})\log (q)/n_r
\right\} + O_p\left\{K (s_{\beta}^{[r]}  + s_{\rm nui}^{[r]})\log (q)/n_r\right \}.
\end{align*}
The last inequality holds under (i) in Assumption \ref{asu:1}, (ii) in Assumption \ref{asu:2}, (i) in Assumption \ref{asu:3},  Lemma \ref{lemma:betatilde}, and 
$\|\widehat\bSigma^{[r]} - \bar\bSigma^{[r]}\|_{\max} = O_p\{\sqrt{\log(q)/n_r}\}$
given the proof of Theorem 3.3 in \cite{van2014asymptotically}.

Using the moment condition (14), under Assumption 1, (ii) and (iii) in Assumption \ref{asu:2}, (i) in Assumption \ref{asu:3}, by Lemma \ref{lemma:alphagamma},
we have  
\begin{align*}
T_{2,2} 
\leq & 
\big\|\Eschat\subSsc^{[r]} \e_j\trans \bar\bOmega^{[r]}\X\exp(\bPhi\trans\bar\balpha_j^{[r]})\dot{b}( \bPhi\trans\bar\bgamma_j^{[r]})\bPhi-\Eschat\subTsc^{[r]} \e_j\trans \bar\bOmega^{[r]}\X \dot{b}( \bPhi\trans\bar\bgamma_j^{[r]})\bPhi\big\|_{\infty} \|\widehat \bgamma_j^{[r]} - \bar\bgamma_j^{[r]}\|_1 \\
+& \|\Eschat\subSsc^{[r]}  \e_j\trans \bar\bOmega^{[r]}\X \exp(\bPhi \trans\bar\balpha_j^{[r]})\{  Y_i - b(\bPhi\trans\bar\bgamma_j^{[r]})\} \bPhi \|_\infty\|\widehat\balpha_j^{[r]} - \bar\balpha_j^{[r]}\|_1 + T_{2,3} + T_{2,4} + T_{2,5}\\
= & O_p\left[ K^4(s_{\beta}^{[r]} + s_{\rm nui}^{[r]})\frac{ \log(q) }{n_r}+  R_{{\rm nui},v}^{[r]}\left\{\frac{\log(q)}{n_r} \right\}^{1-\frac{v}{2}}\right]+ T_{2,3} + T_{2,4} + T_{2,5},
\end{align*}
where
$$
T_{2,3} \leq L \Eschat\subTsc^{[r]} \e_j\trans \bar\bOmega^{[r]}\X \{\bPhi\trans(\widehat \bgamma_j^{[r]} - \bar\bgamma_j^{[r]}) \}^2 = O_p(K \|\widehat \bgamma_j^{[r]} - \bar\bgamma_j^{[r]}\|_2^2), 
$$
\begin{align*}
T_{2, 4} &\leq
\Eschat\subSsc^{[r]} \e_j\trans \bar\bOmega^{[r]}\X
\exp\{|\bPhi\trans (\widehat\balpha_j^{[r]} - \bar\balpha_j^{[r]})|\} \dot{b}(\bPhi\trans\widehat\bgamma_j^{[r]})
\{\bPhi\trans(\widehat \bgamma_j^{[r]} - \bar\bgamma_j^{[r]}) \}\{\bPhi\trans(\widehat \balpha_j^{[r]} - \bar\balpha_j^{[r]}) \} \\
&+ 
L\Eschat\subSsc^{[r]} \e_j\trans \bar\bOmega^{[r]}\X\exp(\bPhi\trans\bar\balpha_j^{[r]})\{\bPhi\trans(\widehat \bgamma_j^{[r]} - \bar\bgamma_j^{[r]}) \}^2 \\
& = O_p(K \|\widehat \bgamma_j^{[r]} - \bar\bgamma_j^{[r]}\|_2^2 + K \|\widehat \bgamma_j^{[r]} - \bar\bgamma_j^{[r]}\|_2 \|\widehat \balpha_j^{[r]} - \bar\balpha_j^{[r]}\|_2),    
\end{align*}
and 
\begin{align*}
T_{2,5} &\leq  \Eschat\subSsc^{[r]}  \e_j\trans \bar\bOmega^{[r]}\X \exp\{|\bPhi\trans (\widehat\balpha_j^{[r]} - \bar\balpha_j^{[r]})|\}\{  Y_i - b(\bPhi\trans\widehat\bgamma_j^{[r]})\} \{\bPhi\trans(\widehat\balpha_j^{[r]} - \bar\balpha_j^{[r]}) \}^2\\
&+ L\Eschat\subSsc^{[r]}  \e_j\trans \bar\bOmega^{[r]}\X 
\exp(\bPhi\trans\bar\balpha_j^{[r]})
\{\bPhi\trans(\widehat \bgamma_j^{[r]} - \bar\bgamma_j^{[r]}) \}\{\bPhi\trans(\widehat \balpha_j^{[r]} - \bar\balpha_j^{[r]}) \}
\\
&= O_p(K \|\widehat \balpha_j^{[r]} - \bar\balpha_j^{[r]}\|_2^2 + K \|\widehat \bgamma_j^{[r]} - \bar\bgamma_j^{[r]}\|_2 \|\widehat \balpha_j^{[r]} - \bar\balpha_j^{[r]}\|_2).   
\end{align*}

Since $T_3 = o_p(T_k)$, $k=1, 2$, then 
by Lemma \ref{lemma:alphagamma}, we have \begin{align}
\begin{split}
\label{eq:asyexp_full}
\widehat{\beta}_{{\rm Deb},j}^{[r]} - \bar\beta_{j}^{[r]}
& = -\e_j\trans\bar\bOmega^{[r]} \partial_{\bbeta}\Lsc_r(\bar\bbeta^{[r]}; \bar\balpha_j^{[r]}, \bar\bgamma_j^{[r]} )\\ 
&+ O_p\left[  \frac{K^5(s_{\beta}^{[r]} + s_{\rm nui}^{[r]})^2} \log (q){n_r}+  KR_{{\rm nui},v}^{[r]}\left\{\frac{\log (q)}{n_r} \right\}^{1- \frac{v}{2}} \right].
\end{split}
\end{align}

\subsection{Proof of Theorem \ref{thm:beta_thresh}}
Since $-\e_j\trans\bar\bOmega^{[r]} \partial_{\bbeta}\Lsc_r(\bar\bbeta^{[r]}; \bar\balpha_j^{[r]}, \bar\bgamma_j^{[r]} )$ is mean-zero and sub-exponential with sub-exponential norms uniformly bounded for all $j$ under Assumption 1, (i) in Assumption \ref{asu:1},  (ii) in \ref{asu:2} and (ii) in \ref{asu:3}, by Bernstein inequality and Union Bound inequality, we have 
$$
\|-\e_j\trans\bar\bOmega^{[r]} \partial_{\bbeta}\Lsc_r(\bar\bbeta^{[r]}; \bar\balpha_j^{[r]}, \bar\bgamma_j^{[r]} )\|_\infty = O_p\left\{\sqrt{\frac{\log(q)}{n_r}} \right\}.
$$
By (\ref{eq:asyexp_full}), we further have 
\begin{align*}
\|\widehat \bbeta_{\rm Deb}^{[r]} - \bar\bbeta^{[r]}\|_\infty
= O_p \left(\sqrt{\log (q)}\left[n_r^{-1/2}+     
 \frac{K^5(s_{\beta}^{[r]} + s_{\rm nui}^{[r]}) \log (q)}{n_r}+  KR_{{\rm nui},v}^{[r]}\left\{\frac{\log (q)}{n_r} \right\}^{1- \frac{v}{2}} 
\right]\right).
\end{align*}
Then by the proof of Theorem 4.3 in \cite{battey2018distributed},  we obtain the results in Theorem 1.

\subsection{Proof of Lemma \ref{lemma:KTr}}
Here we consider a general scenario where $\bar \bbeta^{[0]}$, $\bar \bbeta^{[1]}$, and $\bar\bdelta$ are all $l_v$ sparse, for a fixed $v \in [0,1]$.
Note that there exist positive constants $C$ and $c$ such that 
$$
\|\widehat{\bbeta}^{[0]}_{\rm Deb}-\widehat\bbeta_{\rm Thr}^{[1]} - \bar\bdelta\|_\infty \leq  \|\widehat{\bbeta}^{[0]}_{\rm Deb}- \bar\bbeta^{[0]} \|_\infty + \|\widehat\bbeta_{\rm Thr}^{[1]}   - \bar\bbeta^{[1]} \|_\infty \leq   C \left\{\sqrt{\log(q)} (n_0^{-1/2} + \Delta)\right\}, 
$$
where
$$
\Delta = \max_{r=0, 1}\left\{  K^5 (s_{\beta}^{[r]} + s_{\rm nui}^{[r]}) \frac{\log(q)}{n_r} \right\},
$$ with probability at least $1- c/q$.
Define $S_{\delta, v} = \{j \in \{1, \dots, q\}:  |\bar\delta_j| \geq \sqrt{ \log(q)/n_0} \}$.
Let $\tau_{\scriptscriptstyle{\rm KTr}} = C_1\sqrt{ \log(q)} (n_0^{-1/2} + \Delta)\geq C \{\sqrt{\log(q)} (n_0^{-1/2} + \Delta)\}+ \sqrt{ \log(q)/n_0}$.
Following the proof of Theorem 4.3 in \cite{battey2018distributed}, we can first show that
$\|(\widehat{\bbeta}^{[0]}_{\rm Deb})_{S_{\delta, v}^c}-(\widehat\bbeta_{\rm Thr}^{[1]})_{S_{\delta, v}^c}\|_\infty \leq \tau_{\scriptscriptstyle{\rm KTr}}$, $\widehat\bdelta_{S_{\delta, v}^c} = \boldsymbol{0}$ and consequently $\|\widehat\bdelta_{S_{\delta, v}^c} - \bar\bdelta_{S_{\delta, v}^c}\|_2 =  \|\bar\bdelta_{S_{\delta, v}^c}\|_2$.
Since $\|\bar\bdelta\|_v \leq R_{\delta,v}$ and $ R_{\delta,v} =O[\{n_0/\log(q)\}^{1-v/2}]$, we have 
$$
\|\bar\bdelta_{S_{\delta, v}^c}\|_2^2 = \sum_{j \in S_{\delta, v}^c} | \bar\delta_j |^{v} | \bar\delta_j|^{2-v} = O\left[\left\{\frac{\log(q)}{n_0}\right\}^{1-v/2} R_{\delta, v}\right],
$$
$$
\|\bar\bdelta_{S_{\delta, v}^c}\|_1 = \sum_{j \in S_{\delta, v}^c} | \bar\delta_j |^{v} | \bar\delta_j|^{1-v} = O\left[\left\{\frac{\log(q)}{n_0}\right\}^{(1-v)/2} R_{\delta, v}\right],
$$
and 
$$
\|\widehat\bdelta_{S_{\delta, v}^c} - \bar\bdelta_{S_{\delta, v}^c}\|_\infty =  \|\bar\bdelta_{S_{\delta, v}^c}\|_\infty -O \left\{\sqrt{ \log(q)/n_0}\right\}.
$$

For $j\in S_{\delta, v}$, if $|\bar\delta_j| \geq 2\tau_{\scriptscriptstyle{\rm KTr}}$, we have $|\widehat \beta_{{\rm Deb},j}^{[0]} - \widehat \beta_{{\rm Thr}, j}^{[1]}| \geq |\bar\delta_j| - C \{\sqrt{\log(q)} (n_0^{-1/2} + \Delta)\} \geq \tau_{\scriptscriptstyle{\rm KTr}}$. Hence, 
$|\widehat\delta_j - \bar\delta_j| = |\widehat \beta_{{\rm Deb},j}^{[0]} - \widehat \beta_{{\rm Thr}, j}^{[1]} - \bar\delta_j| \leq \tau_{\scriptscriptstyle{\rm KTr}}$.
While if $|\bar\delta_j| < 2\tau_{\scriptscriptstyle{\rm KTr}}$, we have $|\widehat\delta_j - \bar\delta_j| \leq \max\{|\bar\delta_j|, |\widehat \beta_{{\rm Deb}, j}^{[0]} - \widehat\beta_{{\rm Thr},j}^{[1]} - \bar\delta_j|\} \leq 2\tau_{\scriptscriptstyle{\rm KTr}}$.
Hence, we have 
$$
\|\widehat\bdelta_{S_{\delta, v}} -  \bar\bdelta_{S_{\delta, v}}\|_2 \leq 2\sqrt{|S_{\delta,v}|}\tau_{\scriptscriptstyle{\rm KTr}},
$$
and
$$
\|\widehat\bdelta_{S_{\delta, v}} -  \bar\bdelta_{S_{\delta, v}}\|_1 \leq 2|S_{\delta,v}|\tau_{\scriptscriptstyle{\rm KTr}}.
$$
Follow the proof of Corollary 3 in \cite{negahban2012unified}, we have 
$$
|S_{\delta, v}| \leq \left\{\sqrt{\frac{\log(q)}{n_0}} \right\}^{-v}R_{\delta, v}.
$$
Hence, we obtain 
$$
\|\widehat\bdelta -  \bar\bdelta\|_2 =O_p\left[\left\{\frac{\log(q)}{n_0}\right\}^{-v/4} R_{\delta, v}^{1/2} \left\{ \sqrt{\frac{\log(q)}{n_0}} + \sqrt{\log(q)}\Delta \right\}\right],
$$
$$
\|\widehat\bdelta -  \bar\bdelta\|_\infty =O_p\left\{\sqrt{\frac{\log(q)}{n_0}} + \sqrt{\log(q)}\Delta\right\},
$$
and 
$$
\|\widehat\bdelta -  \bar\bdelta\|_1 = O_p\left[\left\{\frac{\log(q)}{n_0}\right\}^{-v/2} R_{\delta, v} \left\{ \sqrt{\frac{\log(q)}{n_0}} + \sqrt{\log(q)}\Delta \right\}\right].
$$

\subsection{Proof of Theorem \ref{thm:makeup}}

First of all, we have 
\begin{align*}
&\Big\|\widehat{\bbeta}^{[0]}_{{\rm Deb},(2)}-\widehat\bbeta^{[0]}_{{\rm Thr} ,(1)}\Big\|_2^2-\Big\|\widehat{\bbeta}^{[0]}_{{\rm Deb},(2)}-\widehat\bbeta^{[0]}_{{\rm KTr} ,(1)}\Big\|_2^2 \\
=& 2\left(\widehat\bbeta^{[0]}_{{\rm KTr} ,(1)}-\widehat\bbeta^{[0]}_{{\rm Thr} ,(1)}\right)\trans\left(\widehat \bbeta^{[0]}_{{\rm Deb},(2)} - \bar\bbeta^{[0]}\right)+\left\|\bar\bbeta^{[0]}-\widehat\bbeta^{[0]}_{{\rm Thr} ,(1)}\right\|_2^2-\left\|\bar\bbeta^{[0]}-\widehat\bbeta^{[0]}_{{\rm KTr} ,(1)}\right\|_2^2.
\end{align*}
With binary selection , i.e., $a=\infty$ in (13), when $\Big\|\widehat{\bbeta}^{[0]}_{{\rm Deb},(2)}-\widehat\bbeta^{[0]}_{{\rm Thr} ,(1)}\Big\|_2^2-\Big\|\widehat{\bbeta}^{[0]}_{{\rm Deb},(2)} -\widehat\bbeta^{[0]}_{{\rm KTr} ,(1)}\Big\|_2^2 \geq 0$, we have 
\begin{align}\label{eq:Ktr}
   \left\|\bar\bbeta^{[0]}-\widehat\bbeta^{[0]}_{{\rm KTr} ,(1)}\right\|_2^2 \leq 2\left(\widehat\bbeta^{[0]}_{{\rm KTr} ,(1)}-\widehat\bbeta^{[0]}_{{\rm Thr} ,(1)}\right)\trans\left(\widehat \bbeta^{[0]}_{{\rm Deb},(2)} - \bar\bbeta^{[0]}\right)+\left\|\bar\bbeta^{[0]}-\widehat\bbeta^{[0]}_{{\rm Thr} ,(1)}\right\|_2^2.
\end{align}
The price of transferability detection is $\sqrt{ 2\left(\widehat\bbeta^{[0]}_{{\rm KTr} ,(1)}-\widehat\bbeta^{[0]}_{{\rm Thr} ,(1)}\right)\trans\left(\widehat \bbeta^{[0]}_{(2)} - \bar\bbeta^{[0]}\right)}$ in terms of $l_2$ error. 

By the proof of Theorem \ref{thm:beta_thresh}, we have
\begin{align*}
(\widehat \beta^{[0]}_{{\rm Deb},(2)})_j - \bar\beta^{[0]}_j = V_{0,j}+ V_{1, j}, 
\end{align*}
where 
\begin{align*}
V_{0,j}=&  - \e_j\trans\bar\bOmega^{[0]} \partial_{\bbeta}\Lsc_{0, (2)}(\bar\bbeta^{[0]}; \bar\balpha_j^{[0]}, \bar\bgamma_j^{[0]} ), \\
V_{1,j} = & O_p (\Delta_0).
\end{align*}
Let $\V_k = (V_{k,1}, \dots, V_{k, q})\trans$, for $k=0, 1$.
Due to data splitting, 
we know $(\widehat\bbeta^{[0]}_{{\rm KTr} ,(1)}-\widehat\bbeta^{[0]}_{{\rm Thr} ,(1)})\trans\V_0$ has 
mean zero. Moreover, we have
$$
\| \widehat\bbeta^{[0]}_{{\rm KTr} ,(1)}-\widehat\bbeta^{[0]}_{{\rm Thr} ,(1)}\|_2^2 \| \mathbb{E}(\V_0\V_0\trans) \|_2 = O_p( \| \widehat\bbeta^{[0]}_{{\rm KTr} ,(1)}-\widehat\bbeta^{[0]}_{{\rm Thr} ,(1)}\|_2^2 )
$$
under the assumptions that 
$\|\bar\bOmega^{[0]}\|_2$, and $\|\E^{[0]}_{\iota}(\X_i\X_i\trans)\|_2$ for $\iota = \mathcal{S}, \mathcal{T}$ are bounded and $\exp(\bPhi_i\trans\bar\balpha^{[0]}_j) $ is uniformly bounded for all $i$'s and $j$'s with probability going to one. Hence, by the Central Limit Theorem, we have 
$(\widehat\bbeta^{[0]}_{{\rm KTr} ,(1)}-\widehat\bbeta^{[0]}_{{\rm Thr} ,(1)})\trans\V_0 = O_p( n_0^{-1/2}\| \widehat\bbeta^{[0]}_{{\rm KTr} ,(1)}-\widehat\bbeta^{[0]}_{{\rm Thr} ,(1)}\|_2)$.
Regarding $\V_1 = \X O_p(\Delta_0 )$, by Lemma \ref{lemma:omega} and Theorem 1,  we have 
\begin{align*}
|(\widehat\bbeta^{[0]}_{{\rm KTr} ,(1)}-\widehat\bbeta^{[0]}_{{\rm Thr} ,(1)} )\trans \V_1| 
= O_p\{ \Delta_0   \|\widehat\bbeta^{[0]}_{{\rm KTr} ,(1)}-\widehat\bbeta^{[0]}_{{\rm Thr} ,(1)}  \|_2\}.
\end{align*}
Hence, we have 
\begin{equation}\label{eq:detectErr}
2\left(\widehat\bbeta^{[0]}_{{\rm KTr} ,(1)}-\widehat\bbeta^{[0]}_{{\rm Thr} ,(1)}\right)\trans\left(\widehat \bbeta^{[0]}_{{\rm Deb},(2)} - \bar\bbeta^{[0]}\right) = O_p \{(n_0^{-1/2} + \Delta_0)\|\widehat\bbeta^{[0]}_{{\rm KTr} ,(1)}-\widehat\bbeta^{[0]}_{{\rm Thr} ,(1)}  \|_2\}.    
\end{equation}
Moreover, by Theorem \ref{thm:beta_thresh} and Lemma \ref{lemma:KTr}, we have
\begin{align*}
&\|\widehat\bbeta^{[0]}_{{\rm KTr} ,(1)}-\widehat\bbeta^{[0]}_{{\rm Thr} ,(1)} \|_2 \\ 
=& O_p \left(
\max_{r =0, 1}\left[ \sqrt{s^{[r]}_{\beta}\log(q)} \left\{n_r^{-1/2} + \Delta_r \right\} \right]
+ \left\{\frac{\log(q)}{n_0}\right\}^{-v/4} R_{\delta, v}^{1/2} \sqrt{\log(q)} \left\{ n_0^{-1/2} + \Delta \right\}\right).    
\end{align*}

When $\Big\|\widehat{\bbeta}^{[0]}_{{\rm Deb},(2)}-\widehat\bbeta^{[0]}_{{\rm Thr} ,(1)}\Big\|_2^2-\Big\|\widehat{\bbeta}^{[0]}_{{\rm Deb},(2)} -\widehat\bbeta^{[0]}_{{\rm KTr} ,(1)}\Big\|_2^2 < 0$,
we have 
\begin{align}\label{eq:Thr}
   \left\|\bar\bbeta^{[0]}-\widehat\bbeta^{[0]}_{{\rm KTr} ,(1)}\right\|_2^2 > 2\left(\widehat\bbeta^{[0]}_{{\rm KTr} ,(1)}-\widehat\bbeta^{[0]}_{{\rm Thr} ,(1)}\right)\trans\left(\widehat \bbeta^{[0]}_{{\rm Deb},(2)} - \bar\bbeta^{[0]}\right)+\left\|\bar\bbeta^{[0]}-\widehat\bbeta^{[0]}_{{\rm Thr} ,(1)}\right\|_2^2.
\end{align}
By (\ref{eq:Ktr}), (\ref{eq:Thr}), and (\ref{eq:detectErr}), we prove Theorem \ref{thm:makeup}.

\section{Additional Methodological Details}
\label{sec:app:method-details}

This section collects the methodological details of AIMS that are not expanded in the main text. We first give the nodewise precision-matrix estimator used in the debiasing step, then describe the weight-stratified calibration used when the calibration weights change sign, and finally summarize the tuning rules used by the algorithms.

\subsection{Nodewise Precision-Matrix Estimator}
\label{spp:precision}
The nodewise precision-matrix row $\Omegahat^{[r]}_{j}$ is given by  $$\Omegahat^{[r]}_{j}=\big(-\widehat \btheta_{j,1}^{[r]},\ldots, - \widehat \btheta_{j,j-1}^{[r]}, 1, - \widehat \btheta_{j,j+1}^{[r]},\ldots, - \widehat \btheta_{j,q}^{[r]} \big)/\hat{\varsigma}^{[r]}_j,$$ where
\begin{align*}
\widehat{\btheta}^{[r]}_j=&\underset{\btheta \in \mathbb{R}^{q-1}}{\arg \min }~\Eschat_{\scriptscriptstyle \Tsc}^{[r]} \dot{g}\big(\X\trans \widetilde\bbeta^{[r]}\big) (X_{j} - \X_{\textit{-}j} \trans \btheta)^2 + 2 \lambda^{[r]}_{\theta_j} \|\btheta\|_1 ,\\
\hat{\varsigma}^{[r]}_j =& \Eschat_{\scriptscriptstyle \Tsc}^{[r]} \dot{g}\big(\X\trans \widetilde\bbeta^{[r]}\big) \big(X_{j} - \X_{\textit{-}j} \trans\widehat{\btheta}^{[r]}_j\big)^2 + 2 \lambda^{[r]}_{\theta_j} \big\|\widehat{\btheta}^{[r]}_j\big\|_1,
\end{align*}
and $\X_{\textit{-}j} = (X_1,\ldots,X_{j-1},X_{j+1},\ldots,X_q)\trans$. The tuning parameter  $\lambda^{[r]}_{\theta_j}$ is chosen via five-fold cross-validation from the range $\big[0.1 \sqrt{\frac{\log q}{n_{\scriptscriptstyle \Tsc,r}}}
,2\sqrt{\frac{\log q}{n_{\scriptscriptstyle \Tsc,r}}} \big]$.

\subsection{Weight-Stratified Calibration}
\label{app:split}
As mentioned in Remark \ref{rmk:split},  the weights $\widehat w_{j}^{[r]}$'s may not always be nonnegative.  For instance, if $\X$ has a mean-zero symmetric distribution (except for the intercept), the weights $\widehat w_{j}^{[r]}$'s can have approximately equal probabilities of being positive or negative. As a result, the calibration loss may become irregular and ill-posed. To address this issue, we use a weight-stratified calibration strategy.

Instead of directly fitting (11), we define the following estimators: 
\begin{equation}
\begin{split}
\widehat\bxi^{[r]}_{j+}&=\argmin{\bdelta\in\mathbb{R}^{q+p}}\Eschat_{\Sscsub\Cupsub\Tscsub}^{[r]} \I(\widehat w_{j}^{[r]} >0)|\widehat w_{j}^{[r]}| \dot{b}( \bPhi\trans\widetilde \bgamma^{[r]})\Fsc^{[r]}(\bxi;\widetilde \balpha^{[r]}) + \lambda_{\alpha_j+}^{[r]} \|\bxi\|_1   \, ;\\
\widehat\bxi^{[r]}_{j-}&=\argmin{\bdelta\in\mathbb{R}^{q+p}}\Eschat_{\Sscsub\Cupsub\Tscsub}^{[r]} \I(\widehat w_{j}^{[r]} \leq 0)|\widehat w_{j}^{[r]}| \dot{b}( \bPhi\trans\widetilde \bgamma^{[r]})\Fsc^{[r]}(\bxi;\widetilde \balpha^{[r]}) + \lambda_{\alpha_j-}^{[r]} \|\bxi\|_1 \, ;\\
\widehat\bzeta^{[r]}_{j+}&=\argmin{\bzeta\in\mathbb{R}^{q+p}}\Eschat_{\Sscsub}^{[r]} \I(\widehat w_{j}^{[r]} >0)| \widehat w_{j}^{[r]}| \exp(\bPhi \trans\widetilde\balpha^{[r]})\Gsc(\bzeta;\widetilde \bgamma^{[r]})+ \lambda^{[r]}_{\gamma_j+} \|\bzeta\|_1  \, ;\\
\widehat\bzeta^{[r]}_{j-}&=\argmin{\bzeta\in\mathbb{R}^{q+p}}\Eschat_{\Sscsub}^{[r]} \I(\widehat w_{j}^{[r]} \leq 0)| \widehat w_{j}^{[r]}| \exp(\bPhi \trans\widetilde\balpha^{[r]})\Gsc(\bzeta;\widetilde \bgamma^{[r]})+ \lambda^{[r]}_{\gamma_j-} \|\bzeta\|_1 .
\end{split}   
\label{equ:cal:sign}
\end{equation}
We now compute  $\widehat\balpha^{[r]}_{j+}=\widetilde\balpha^{[r]}+\widehat\bxi^{[r]}_{j+}$, $\widehat\balpha^{[r]}_{j-}=\widetilde\balpha^{[r]}+\widehat\bxi^{[r]}_{j-}$,  $\widehat\bgamma^{[r]}_{j+} = \widetilde \bgamma^{[r]} + \widehat\bzeta^{[r]}_{j+}$, and $\widehat\bgamma^{[r]}_{j-} = \widetilde \bgamma^{[r]} + \widehat\bzeta^{[r]}_{j-}$. Using these, we derive the one-step debiased estimator $\widehat{\beta}_{{\rm Deb},j}^{[r]}$ as: 
\[
\widehat{\beta}_{{\rm Deb},j}^{[r]} =\e_j\trans\widetilde{\bbeta}^{[r]} +\e_j\trans\Omegahat^{[r]}\left[\Eschat\subSsc^{[r]} \widehat h^j(\bPhi) \X \{  Y - \widehat r^j(\bPhi)\} + \Eschat_{\Tsc}^{[r]} \X\{ \widehat r^j(\bPhi) - g(\X\trans \bbetatilde^{[r]}) \}\right].
\]
where $\widehat h^j(\bPhi)=\I(\widehat w_{j}^{[r]} > 0)\exp \{\bPhi\trans \widehat\balpha^{[r]}_{j+} \} + \I(\widehat w_{j}^{[r]} \leq0)\exp \{\bPhi \trans \widehat\balpha^{[r]}_{j-} \}$ and $\widehat r^j(\bPhi)=\I(\widehat w_{j}^{[r]} > 0)b(\bPhi \trans \widehat\bgamma^{[r]}_{j+}) +  \I(\widehat w_{j}^{[r]} \leq 0)b(\bPhi\trans \widehat\bgamma^{[r]}_{j-} )$.

\subsection{Tuning Rules for AIMS}
\label{sec:method:tune}
\label{sec:supp:tuning}

We summarize the tuning parameters used in AIMS and describe their practical selection rules. The tuning parameters fall into three groups: regularization parameters for sparse nuisance and target estimation, bootstrap-calibrated thresholds for the bias-correction and transfer steps, and the aggregation temperature used for negative transfer protection. All tuning procedures are label-free with respect to the target population.

\begin{table}[htb!]
\centering
\footnotesize
\begin{tabular}{p{2.0cm} p{3.2cm} p{3.0cm} p{4.1cm}}
\hline\hline
Parameter & Role & Theoretical order & Practical choice \\
\hline
$\lambda^{[r]}$ 
& $\ell_1$ penalty for the preliminary target estimator $\widetilde\bbeta^{[r]}$ 
& $\sqrt{\log(q)/n_r}$ 
& Selected by five-fold cross-validation on a grid centered at the theoretical rate. \\

$\lambda_{\alpha}^{[r]}$ 
& $\ell_1$ penalty for the preliminary density ratio model 
& $\sqrt{\log(d)/n_r}$ 
& Selected by five-fold cross-validation for the density ratio loss in \eqref{eq:initial alpha}. \\

$\lambda_{\gamma}^{[r]}$ 
& $\ell_1$ penalty for the preliminary conditional mean imputation model 
& $\sqrt{\log(d)/n_{\scriptscriptstyle\Ssc,r}}$ 
& Selected by five-fold cross-validation using the labeled source samples in subgroup $r$. \\

$\lambda_{\theta_j}^{[r]}$
& $\ell_1$ penalty for the nodewise precision-matrix estimator
& $\sqrt{\log(q)/n_{\scriptscriptstyle\Tsc,r}}$
& Selected by five-fold cross-validation on the target covariates. \\

$\lambda_{\alpha_j}^{[r]}$ 
& $\ell_1$ penalty for the calibrated density ratio nuisance correction for coordinate $j$ 
& Same order as $\lambda_{\alpha}^{[r]}$ 
& Chosen by Gaussian multiplier bootstrap using quantile level $q_\tau$. \\

$\lambda_{\gamma_j}^{[r]}$ 
& $\ell_1$ penalty for the calibrated imputation nuisance correction for coordinate $j$ 
& Same order as $\lambda_{\gamma}^{[r]}$ 
& Chosen by Gaussian multiplier bootstrap using quantile level $q_\tau$. \\

$\tau^{[r]}$ 
& Threshold for sparsifying the dense debiased estimator $\widehat\bbeta_{\rm Deb}^{[r]}$ 
& $\sqrt{\log(q)}\{n_r^{-1/2}+\Delta_r\}$ 
& Implemented as $\tau^{[r]}=C_{\tau}\{\log(q)/n_{\scriptscriptstyle\Ssc,r}\}^{1/2}$. \\

$\tau_{\rm KTr}$ 
& Threshold for estimating the sparse majority-minority difference in Algorithm~\ref{alg:2} 
& $\sqrt{\log(q)}\{n_0^{-1/2}+\Delta\}$ 
& Chosen by Gaussian multiplier bootstrap using quantile level $q_\tau$. \\

$a$ 
& Temperature parameter in the exponential aggregation weight in Algorithm~\ref{alg:3} 
& Fixed positive constant 
& Fixed at a default value in the main comparison; $a=\infty$ corresponds to hard selection between the minority-only and knowledge transfer estimators. \\
\hline\hline
\end{tabular}
\caption{Summary of tuning parameters in AIMS. Here $n_r=n_{\scriptscriptstyle\Ssc,r}\wedge n_{\scriptscriptstyle\Tsc,r}$, $d$ is the dimension of the nuisance feature map $\phi(\Z)$, and $\Delta_r$ is the higher-order nuisance-estimation error term defined in Section~\ref{sec:theory}.}
\label{tab:tuning}
\end{table}

For the bootstrap-calibrated quantities, we generate independent standard normal multipliers $\epsilon_i$'s and compute
\begin{equation*}
\begin{aligned}
\Msc_{\alpha_j}^{[r]}(\bepsilon) = & \left\|\Eschat\subSsc^{[r]}\epsilon\widehat w_{j}^{[r]} \exp(\bPhi\trans\widetilde\balpha^{[r]})\dot{b}( \bPhi\trans\widetilde \bgamma^{[r]})\bPhi -\Eschat\subTsc^{[r]} \epsilon\widehat w_{j}^{[r]}\dot{b}( \bPhi\trans\widetilde \bgamma^{[r]} )\bPhi \right\|_{\infty};\\
\Msc_{\gamma_j}^{[r]}(\bepsilon) =& \left\|\Eschat\subSsc^{[r]}\epsilon\widehat w_{j}^{[r]}\exp(\bPhi \trans\widetilde\balpha^{[r]})\bPhi\{  Y - b(\bPhi\trans\widetilde \bgamma^{[r]})\}\right\|_{\infty}; \\
\Msc_{\kappa}(\epsilon) = & \max_{j=1,\ldots,q} \Bigg|  \e_j\trans \Omegahat^{[0]} \left[\Eschat\subSsc^{[0]} \epsilon \widehat h^j(\bPhi) \X \{  Y - \widehat r^j(\bPhi)\} + \Eschat\subTsc^{[0]} \epsilon \X\{ \widehat r^j(\bPhi) - g(\X\trans \bbetatilde^{[0]}) \}\right]  \\
& - \e_j\trans \Omegahat^{[1]} \left[\Eschat\subSsc^{[1]} \epsilon\widehat h^j(\bPhi) \X \{  Y - \widehat r^j(\bPhi)\} + \Eschat\subTsc^{[1]} \epsilon \X \{ \widehat r^j(\bPhi) - g(\X\trans \bbetatilde^{[1]}) \}\right] \Bigg|. 
\end{aligned}
\end{equation*}
With $B=500$ multiplier draws, we set $\lambda_{\alpha_j}^{[r]}$, $\lambda_{\gamma_j}^{[r]}$, and $\tau_{\rm KTr}$ to the empirical $q_\tau$ quantiles of $\Msc_{\alpha_j}^{[r]}$, $\Msc_{\gamma_j}^{[r]}$, and $\Msc_{\kappa}$, respectively. In the main simulation comparison, we use $C_\tau=2$, $q_\tau=0.8$, and $a=5$. Supplementary S.3.3 reports sensitivity to $q_\tau$ and $C_\tau$, as well as aggregation-temperature variants with $a\in\{1,2,5,10,\infty\}$.

\section{Benchmark Implementations and Additional Results}
\label{sec:app:bench-results}

This section contains the benchmark algorithms used in the numerical comparisons and the supplementary result tables that are omitted from the main text for readability. We first describe the benchmark implementations, then give the additional simulation tables, and finally report the complete T2D benchmark table and supplementary \(\Delta\Delta G\) results.

\subsection{Benchmark Methods}
\label{app:benchmark}

In this section, we give the details of the benchmark methods. The importance weighting methods with $\ell_1$ penalty (IW) and with weighted $\ell_1$ penalty (IW$_{\rm aLasso}$) are presented in Algorithm \ref{alg:iw}.  The imputation-only methods with $\ell_1$ penalty (IM) and with weighted $\ell_1$ penalty (IM$_{\rm aLasso}$) are given in Algorithm \ref{alg:im}. The TransGLM method with two variants according to our setup is presented in Algorithm \ref{alg:transGLM}. The tuning parameters $\lambda$'s for these benchmark methods are selected by five-fold cross-validation. 

For CORAL, we use only the labeled minority-source sample and the unlabeled minority-target covariates. Let $\widehat\Sigma_{\scriptscriptstyle \Ssc,0}$ and $\widehat\Sigma_{\scriptscriptstyle \Tsc,0}$ be the empirical covariance matrices of the non-intercept target-model covariates in the minority source and target samples, respectively. After centering the minority-source covariates, CORAL applies the transformation
\[
(\widehat\Sigma_{\scriptscriptstyle \Ssc,0}+\lambda_{\rm coral} I)^{-1/2}
(\widehat\Sigma_{\scriptscriptstyle \Tsc,0}+\lambda_{\rm coral} I)^{1/2},
\]
with $\lambda_{\rm coral}=1$, and then fits an $\ell_1$-penalized logistic target model using the aligned labeled minority-source sample. The intercept is expressed on the target-centered covariate scale.

For TransFusion, we use a source-labeled adaptation because labeled target outcomes are unavailable. The labeled minority-source sample is used as the pseudo-target task, and the labeled majority-source sample is used as the auxiliary task. We fit the TransFusion block-design penalized logistic regression of \citet{he2024transfusion}, which parameterizes each auxiliary-task coefficient as a shared pseudo-target coefficient plus a sparse task contrast. The reported TransFusion estimator is the resulting task-averaged coefficient with its intercept calibrated on the labeled minority-source sample.

\begin{algorithm}[H]
    \caption{Importance Weighting with $\ell_1$ penalty (IW) and with weighted $\ell_1$ penalty (IW$_{\rm aLasso}$).}  
    \textbf{Input:} $\Dscr^{[0]} = \{\D_i =(S_i Y_i, \X_i\trans, \W_i\trans, S_i, R_i)\trans:  R_i = 0,~i \in [n] \}$ and the preliminary estimator $\widetilde\balpha^{[0]}$. 
    
    \textbf{IW:} Obtain $\widehat \bbeta_{\rm iw}$ as
\[
\widehat \bbeta_{\rm iw} = \argmin{\bbeta \in \R^q } \Eschat\subSsc^{[0]} \exp(\bPhi\trans\widetilde\balpha^{[0]})  \{  Y \X\trans \bbeta - G(\X\trans \bbeta)\} + \lambda\|\bbeta\|_1 \, .
\]
    \textbf{IW$_{\rm aLasso}$:} Obtain $\widehat \bbeta_{\rm adaiw}$ as
\[
\widehat \bbeta_{\rm adaiw} = \argmin{\bbeta \in \R^q } \Eschat\subSsc^{[0]} \exp(\bPhi\trans\widetilde\balpha^{[0]})  \{  Y \X\trans \bbeta - G(\X\trans \bbeta)\} + \lambda  \sum_{j=1}^q b_j |\beta_j| \, ,
\]
where 
$b_j = 1/|\hat \beta_{{\rm r},j}|$ with
\[
(\hat \beta_{{\rm r},1},\ldots,\hat \beta_{{\rm r},q})\trans = \argmin{\bbeta \in \R^q } \Eschat\subSsc^{[0]} \exp(\bPhi\trans\widetilde\balpha^{[0]})  \{  Y \X\trans \bbeta - G(\X\trans \bbeta)\} + n^{-2/3} \|\bbeta\|_2^2 \, .
\]

\textbf{Output:} $\widehat \bbeta_{\rm iw}$ and $\widehat \bbeta_{\rm adaiw}$.
\label{alg:iw}
\end{algorithm}

\begin{algorithm}[H]
    \caption{Imputation-only method with $\ell_1$ penalty (IM) and with weighted $\ell_1$ penalty (IM$_{\rm aLasso}$)}  
    \textbf{Input:}$\Dscr^{[0]} = \{\D_i =(S_i Y_i, \X_i\trans, \W_i\trans, S_i, R_i)\trans:  R_i = 0,~i \in [n] \}$  and the preliminary estimator  $\widetilde\bgamma^{[0]}$. 
    
    \textbf{IM:} Obtain $\widehat \bbeta_{\rm im}$ as
\[
\widehat \bbeta_{\rm im} = \argmin{\bbeta \in \R^q }  \Eschat\subTsc^{[0]}  \{ b(\bPhi\trans\widetilde\bgamma^{[0]}) \X\trans \bbeta - G(\X\trans \bbeta) \}  + \lambda\|\bbeta\|_1 \, .
\]   

    \textbf{IM$_{\rm aLasso}$:} Obtain $\widehat \bbeta_{\rm adaim}$ as
\[
\widehat \bbeta_{\rm adaim} = \argmin{\bbeta \in \R^q }  \Eschat\subTsc^{[0]}  \{ b(\bPhi\trans\widetilde\bgamma^{[0]}) \X\trans \bbeta - G(\X\trans \bbeta) \}  + \lambda  \sum_{j=1}^q b_j |\beta_j| \, .
\] 
where 
$b_j = 1/|\hat \beta_{{\rm r},j}|$ with
\[
 (\hat \beta_{{\rm r},1},\ldots,\hat \beta_{{\rm r},q})\trans =  \argmin{\bbeta \in \R^q }  \Eschat\subTsc^{[0]}  \{ b(\bPhi\trans\widetilde\bgamma^{[0]}) \X\trans \bbeta - G(\X\trans \bbeta) \} + n^{-2/3} \|\bbeta\|_2^2 \, .
\]
\textbf{Output:} $\widehat \bbeta_{\rm im}$ and $\widehat \bbeta_{\rm adaim}$.
\label{alg:im}
\end{algorithm}

\begin{algorithm}[H]
    \caption{TransGLM and TransGLM$_{\rm iw}$}  
    \textbf{Input:} $\Dscr^{[1,1]} = \{\D_i =(S_i Y_i, \X_i\trans, \W_i\trans, S_i, R_i)\trans:  S_i = 1, R_i = 1,~i \in [n] \}$, $\Dscr^{[1,0]} = \{\D_i =(S_i Y_i, \X_i\trans, \W_i\trans, S_i, R_i)\trans:  S_i = 1, R_i = 0,~i \in [n] \}$ 
    and the preliminary estimators $\widetilde\balpha^{[r]}, r =0,1$. 

    \textbf{TransGLM:}   Obtain $\widehat{\bbeta}_{\rm trans}$ by applying the TransGLM algorithm \citep{tian2022transfer} using $\Dscr^{[1,0]}$ as the target sample and $\Dscr^{[1,1]}$ as the source sample.
    
        \textbf{TransGLM$_{\rm iw}$:} Obtain $\widehat{\bbeta}_{\rm trans,iw}$ by applying the TransGLM algorithm \citep{tian2022transfer} with the following modifications:
    \begin{itemize}
        \item Use $\Dscr^{[1,0]}$ as the target sample, with each observation weighted by $\exp(\bPhi\trans\widetilde{\balpha}^{[0]})$.
        \item Use $\Dscr^{[1,1]}$ as the source sample, with each observation weighted by $\exp(\bPhi\trans\widetilde{\balpha}^{[1]})$.
    \end{itemize}
    
\textbf{Output:}  $\widehat \bbeta_{\rm trans}$ and $\widehat \bbeta_{\rm trans,iw}$.
\label{alg:transGLM}
\end{algorithm}

\subsection{Exact Data-Generating Mechanisms for the Simulation Studies}
\label{sec:sim:dgp}

The exact data-generating mechanisms used in Section 5 are as follows. Let $\Z=(Z_1,\ldots,Z_{p+q})\trans$, where $Z_1=1$ and $Z_j$, $j\geq 2$, are independently generated from a standard normal distribution truncated to $(-1.5,1.5)$. We set $\X=(Z_1,\ldots,Z_q)\trans$, $\W=(Z_{q+1},\ldots,Z_{q+p})\trans$, $\phi(\Z)=\Z$, and use the logistic link $g(a)=b(a)={\rm expit}(a)$. For the minority group, define
\[
\bgamma^{[0]}=
(0,\ 0.8,\ -0.6,\ 0.5,\ 0.4,\ \bzero_{q-5}\trans,\ 0.70,\ -0.70,\ 0.50,\ \bzero_{p-3}\trans)\trans,
\]
and
\[
\balpha^{[0]}=
(0,\ 0.55,\ -0.55,\ 0.4125,\bzero_{q-4}\trans,
1.03125,\ -0.825,\ 0.61875,\bzero_{p-3}\trans)\trans .
\]
The majority outcome coefficient is fixed as a local perturbation of the minority coefficient:
\[
\bgamma^{[1]}=
\bgamma^{[0]}+
(0,\ 0.10\mathbf{1}_4\trans,\ \bzero_{p+q-5}\trans)\trans.
\]
The majority source--target sampling coefficient is also fixed:
\[
\balpha^{[1]}=
\balpha^{[0]}+
(0,\ 0.01,\ -0.01,\ 0.0075,\ \bzero_{q-4}\trans,\ 0.045,\ -0.0375,\ 0.03,\ \bzero_{p-3}\trans)\trans.
\]
To introduce controlled nonlinear misspecification, let
\[
u_Y(\Z)=0.50 Z_2Z_{q+1}+0.25(Z_3^2-\mu_2),
\]
and
\[
u_S(\Z)=0.35 Z_2Z_{q+1}-0.175 Z_3Z_{q+2}+0.175(Z_4^2-\mu_2),
\]
where $\mu_2$ is the second moment of the truncated normal covariate, approximated by the empirical second moment in each generated sample. For $r\in\{0,1\}$, we generate the outcome $Y$ in the following three settings:
\[
\begin{array}{ll}
\text{Setting I:} &
\begin{aligned}[t]
\P(Y=1\mid \Z,R=r)&={\rm expit}\{\Z\trans\bgamma^{[r]}\},\\
\P(S=1\mid \Z,R=r)&={\rm expit}\{\Z\trans\balpha^{[r]}\};
\end{aligned}\\[0.6em]
\text{Setting II:} &
\begin{aligned}[t]
\P(Y=1\mid \Z,R=r)&={\rm expit}\{\Z\trans\bgamma^{[r]}+u_Y(\Z)\},\\
\P(S=1\mid \Z,R=r)&={\rm expit}\{\Z\trans\balpha^{[r]}\};
\end{aligned}\\[0.6em]
\text{Setting III:} &
\begin{aligned}[t]
\P(Y=1\mid \Z,R=r)&={\rm expit}\{\Z\trans\bgamma^{[r]}\},\\
\P(S=1\mid \Z,R=r)&={\rm expit}\{\Z\trans\balpha^{[r]}+u_S(\Z)\}.
\end{aligned}
\end{array}
\]

\subsection{Complete Binary-Outcome Simulation Results}\label{sec:app:sim}

This subsection provides additional binary-outcome simulation results that complement the main figures in Section~\ref{sec:sim}. The first two tables give the complete benchmark comparisons for the sample-size and dimensionality grids, including methods omitted from the main figures for readability. The remaining tables vary the amount of minority target-unlabeled covariate data, examine AIMS through ablation and tuning-sensitivity analyses, and report the majority-side estimator used as auxiliary information. Unless otherwise stated, the tables use the simulation design in Section~\ref{sec:sim}, with $q=100$ and default AIMS tuning constants $C_\tau=2$, $q_\tau=0.8$, and $a=5$, except in the tables where $n_{\scriptscriptstyle \Tsc,0}$, $q_\tau$, or $C_\tau$ is explicitly varied.

Table~\ref{tab:sim:k10:sample:full} expands Figure~\ref{fig:sim:k10:sample} to the full benchmark set along the sample-size axis. The ranking is consistent with the figure: AIMS has the smallest error in every reported cell, and TransGLM$_{\rm iw}$ is typically the strongest external competitor. The importance-weighting-only and imputation-only estimators improve as $n_{\scriptscriptstyle \Ssc,0}$ increases, but they remain less accurate than AIMS in most cells, underscoring the value of combining weighting, imputation, and majority-guided transfer information.

\begin{table}[htb!]
\small
\centering
\setlength{\tabcolsep}{4pt}
\renewcommand{\arraystretch}{0.95}
\begin{tabular}{lcccc}
\hline\hline
\multicolumn{5}{c}{Setting I}
\\
Method & \multicolumn{4}{c}{$n_{\scriptscriptstyle \Ssc,0}$}
\\
 & $300$ & $400$ & $500$ & $600$
\\
\hline
AIMS & \textbf{0.315 (0.177)} & \textbf{0.259 (0.173)} & \textbf{0.218 (0.152)} & \textbf{0.255 (0.157)}
\\
IW & 0.785 (0.235) & 0.665 (0.180) & 0.618 (0.196) & 0.574 (0.205)
\\
IW$_{\rm aLasso}$ & 1.740 (1.428) & 1.426 (1.205) & 1.101 (0.830) & 0.815 (0.411)
\\
IM & 0.977 (0.224) & 0.819 (0.155) & 0.741 (0.188) & 0.697 (0.180)
\\
IM$_{\rm aLasso}$ & 0.726 (0.249) & 0.521 (0.157) & 0.421 (0.164) & 0.382 (0.148)
\\
IM$_{\rm RF}$ & 1.577 (0.165) & 1.561 (0.143) & 1.553 (0.137) & 1.539 (0.119)
\\
IM$_{\rm XGB}$ & 1.802 (0.510) & 1.449 (0.332) & 1.295 (0.359) & 1.183 (0.315)
\\
CORAL & 1.678 (0.271) & 1.622 (0.269) & 1.572 (0.221) & 1.540 (0.210)
\\
TransFusion & 0.867 (0.210) & 0.849 (0.188) & 0.835 (0.174) & 0.826 (0.158)
\\
TransGLM & 0.986 (0.219) & 0.956 (0.197) & 0.938 (0.177) & 0.922 (0.165)
\\
TransGLM$_{\rm iw}$ & 0.428 (0.235) & 0.366 (0.173) & 0.359 (0.159) & 0.334 (0.145)
\\
\hline
\multicolumn{5}{c}{Setting II}
\\
Method & \multicolumn{4}{c}{$n_{\scriptscriptstyle \Ssc,0}$}
\\
 & $300$ & $400$ & $500$ & $600$
\\
\hline
AIMS & \textbf{0.303 (0.177)} & \textbf{0.267 (0.166)} & \textbf{0.228 (0.154)} & \textbf{0.228 (0.137)}
\\
IW & 0.731 (0.236) & 0.653 (0.212) & 0.583 (0.177) & 0.533 (0.177)
\\
IW$_{\rm aLasso}$ & 1.557 (1.264) & 1.215 (0.746) & 1.010 (0.582) & 0.775 (0.515)
\\
IM & 0.832 (0.221) & 0.680 (0.164) & 0.599 (0.142) & 0.565 (0.162)
\\
IM$_{\rm aLasso}$ & 0.644 (0.268) & 0.489 (0.174) & 0.356 (0.144) & 0.321 (0.124)
\\
IM$_{\rm RF}$ & 1.406 (0.176) & 1.396 (0.148) & 1.382 (0.147) & 1.366 (0.126)
\\
IM$_{\rm XGB}$ & 1.477 (0.498) & 1.191 (0.343) & 1.047 (0.293) & 0.933 (0.262)
\\
CORAL & 1.585 (0.320) & 1.561 (0.287) & 1.505 (0.279) & 1.479 (0.235)
\\
TransFusion & 0.831 (0.222) & 0.817 (0.189) & 0.801 (0.191) & 0.789 (0.158)
\\
TransGLM & 0.931 (0.242) & 0.906 (0.214) & 0.869 (0.196) & 0.863 (0.171)
\\
TransGLM$_{\rm iw}$ & 0.412 (0.225) & 0.376 (0.185) & 0.337 (0.165) & 0.314 (0.154)
\\
\hline
\multicolumn{5}{c}{Setting III}
\\
Method & \multicolumn{4}{c}{$n_{\scriptscriptstyle \Ssc,0}$}
\\
 & $300$ & $400$ & $500$ & $600$
\\
\hline
AIMS & \textbf{0.314 (0.190)} & \textbf{0.288 (0.158)} & \textbf{0.269 (0.158)} & \textbf{0.258 (0.143)}
\\
IW & 0.706 (0.266) & 0.607 (0.207) & 0.581 (0.213) & 0.522 (0.186)
\\
IW$_{\rm aLasso}$ & 1.663 (1.176) & 1.204 (0.800) & 0.948 (0.565) & 0.839 (0.490)
\\
IM & 0.923 (0.248) & 0.793 (0.193) & 0.714 (0.177) & 0.603 (0.173)
\\
IM$_{\rm aLasso}$ & 0.711 (0.295) & 0.484 (0.194) & 0.403 (0.136) & 0.323 (0.140)
\\
IM$_{\rm RF}$ & 1.522 (0.172) & 1.498 (0.163) & 1.510 (0.140) & 1.472 (0.112)
\\
IM$_{\rm XGB}$ & 1.674 (0.565) & 1.353 (0.418) & 1.234 (0.324) & 1.033 (0.264)
\\
CORAL & 1.643 (0.305) & 1.546 (0.291) & 1.528 (0.249) & 1.463 (0.221)
\\
TransFusion & 0.847 (0.217) & 0.825 (0.208) & 0.835 (0.180) & 0.797 (0.147)
\\
TransGLM & 0.977 (0.258) & 0.928 (0.210) & 0.930 (0.190) & 0.884 (0.159)
\\
TransGLM$_{\rm iw}$ & 0.414 (0.228) & 0.382 (0.241) & 0.330 (0.123) & 0.347 (0.131)
\\
\hline\hline
\end{tabular}
\caption{Binary-outcome sample-size grid. The columns vary the labeled minority-source sample size $n_{\scriptscriptstyle \Ssc,0}\in\{300,400,500,600\}$, while $p=400$, $q=100$, $n_{\scriptscriptstyle \Ssc,1}=2000$, $n_{\scriptscriptstyle \Tsc,0}=2000$, and $n_{\scriptscriptstyle \Tsc,1}=3000$ are fixed. Each cell reports the Monte Carlo mean (empirical standard deviation) of $\|\widehat\bbeta-\bar\bbeta^{[0]}\|_2^2$ over $100$ replications; boldface indicates the smallest Monte Carlo mean within each setting and column.}
\label{tab:sim:k10:sample:full}
\end{table}

Table~\ref{tab:sim:k10:dimension:full} gives the corresponding full benchmark comparison along the dimensionality axis in Figure~\ref{fig:sim:k10:dimension}. Several benchmarks are less competitive at larger $p$: the random-forest and boosting imputation variants have noticeably larger errors, and TransGLM without importance weighting is dominated by its weighted version. AIMS remains stable across dimensions and gives the smallest error throughout the grid.

\begin{table}[htb!]
\small
\centering
\setlength{\tabcolsep}{4pt}
\renewcommand{\arraystretch}{0.95}
\begin{tabular}{lccccc}
\hline\hline
\multicolumn{6}{c}{Setting I}
\\
Method & \multicolumn{5}{c}{$p$}
\\
 & $50$ & $100$ & $200$ & $400$ & $800$
\\
\hline
AIMS & \textbf{0.286 (0.161)} & \textbf{0.283 (0.185)} & \textbf{0.285 (0.160)} & \textbf{0.259 (0.173)} & \textbf{0.257 (0.186)}
\\
IW & 0.697 (0.270) & 0.672 (0.240) & 0.629 (0.217) & 0.665 (0.180) & 0.691 (0.243)
\\
IW$_{\rm aLasso}$ & 1.457 (1.142) & 1.188 (0.715) & 1.138 (0.862) & 1.426 (1.205) & 1.241 (0.983)
\\
IM & 0.724 (0.202) & 0.737 (0.211) & 0.752 (0.183) & 0.819 (0.155) & 0.923 (0.192)
\\
IM$_{\rm aLasso}$ & 0.486 (0.187) & 0.471 (0.204) & 0.473 (0.152) & 0.521 (0.157) & 0.639 (0.209)
\\
IM$_{\rm RF}$ & 1.510 (0.148) & 1.472 (0.148) & 1.490 (0.126) & 1.561 (0.143) & 1.598 (0.146)
\\
IM$_{\rm XGB}$ & 1.094 (0.324) & 1.154 (0.368) & 1.243 (0.287) & 1.449 (0.332) & 1.742 (0.417)
\\
CORAL & 1.646 (0.280) & 1.574 (0.271) & 1.569 (0.257) & 1.622 (0.269) & 1.595 (0.256)
\\
TransFusion & 0.860 (0.185) & 0.826 (0.189) & 0.813 (0.165) & 0.849 (0.188) & 0.816 (0.192)
\\
TransGLM & 0.987 (0.212) & 0.941 (0.213) & 0.922 (0.185) & 0.956 (0.197) & 0.934 (0.204)
\\
TransGLM$_{\rm iw}$ & 0.387 (0.147) & 0.357 (0.184) & 0.355 (0.161) & 0.366 (0.173) & 0.376 (0.181)
\\
\hline
\multicolumn{6}{c}{Setting II}
\\
Method & \multicolumn{5}{c}{$p$}
\\
 & $50$ & $100$ & $200$ & $400$ & $800$
\\
\hline
AIMS & \textbf{0.303 (0.144)} & \textbf{0.299 (0.166)} & \textbf{0.263 (0.145)} & \textbf{0.267 (0.166)} & \textbf{0.228 (0.153)}
\\
IW & 0.563 (0.172) & 0.645 (0.470) & 0.569 (0.188) & 0.653 (0.212) & 0.602 (0.211)
\\
IW$_{\rm aLasso}$ & 1.050 (0.693) & 1.177 (0.726) & 1.100 (0.849) & 1.215 (0.746) & 1.172 (0.711)
\\
IM & 0.559 (0.152) & 0.609 (0.174) & 0.599 (0.161) & 0.680 (0.164) & 0.728 (0.174)
\\
IM$_{\rm aLasso}$ & 0.395 (0.145) & 0.427 (0.167) & 0.410 (0.155) & 0.489 (0.174) & 0.556 (0.203)
\\
IM$_{\rm RF}$ & 1.261 (0.128) & 1.310 (0.145) & 1.317 (0.132) & 1.396 (0.148) & 1.412 (0.167)
\\
IM$_{\rm XGB}$ & 0.826 (0.235) & 0.941 (0.288) & 1.007 (0.302) & 1.191 (0.343) & 1.435 (0.430)
\\
CORAL & 1.531 (0.262) & 1.531 (0.260) & 1.475 (0.256) & 1.561 (0.287) & 1.540 (0.311)
\\
TransFusion & 0.800 (0.160) & 0.832 (0.197) & 0.775 (0.166) & 0.817 (0.189) & 0.815 (0.221)
\\
TransGLM & 0.871 (0.166) & 0.902 (0.210) & 0.844 (0.198) & 0.906 (0.214) & 0.887 (0.228)
\\
TransGLM$_{\rm iw}$ & 0.353 (0.186) & 0.369 (0.176) & 0.331 (0.151) & 0.376 (0.185) & 0.347 (0.173)
\\
\hline
\multicolumn{6}{c}{Setting III}
\\
Method & \multicolumn{5}{c}{$p$}
\\
 & $50$ & $100$ & $200$ & $400$ & $800$
\\
\hline
AIMS & \textbf{0.249 (0.173)} & \textbf{0.265 (0.158)} & \textbf{0.243 (0.151)} & \textbf{0.288 (0.158)} & \textbf{0.254 (0.159)}
\\
IW & 0.627 (0.227) & 0.607 (0.216) & 0.651 (0.218) & 0.607 (0.207) & 0.665 (0.258)
\\
IW$_{\rm aLasso}$ & 1.287 (0.797) & 1.340 (0.926) & 1.266 (0.936) & 1.204 (0.800) & 1.312 (1.119)
\\
IM & 0.625 (0.202) & 0.681 (0.181) & 0.731 (0.197) & 0.793 (0.193) & 0.863 (0.198)
\\
IM$_{\rm aLasso}$ & 0.440 (0.182) & 0.424 (0.180) & 0.471 (0.166) & 0.484 (0.194) & 0.585 (0.226)
\\
IM$_{\rm RF}$ & 1.395 (0.158) & 1.411 (0.146) & 1.471 (0.127) & 1.498 (0.163) & 1.543 (0.144)
\\
IM$_{\rm XGB}$ & 0.970 (0.306) & 1.064 (0.309) & 1.241 (0.342) & 1.353 (0.418) & 1.642 (0.459)
\\
CORAL & 1.590 (0.307) & 1.527 (0.279) & 1.580 (0.246) & 1.546 (0.291) & 1.568 (0.246)
\\
TransFusion & 0.850 (0.212) & 0.798 (0.191) & 0.851 (0.183) & 0.825 (0.208) & 0.828 (0.182)
\\
TransGLM & 0.939 (0.231) & 0.893 (0.203) & 0.954 (0.210) & 0.928 (0.210) & 0.932 (0.203)
\\
TransGLM$_{\rm iw}$ & 0.403 (0.228) & 0.394 (0.211) & 0.375 (0.168) & 0.382 (0.241) & 0.397 (0.254)
\\
\hline\hline
\end{tabular}
\caption{Binary-outcome dimensionality grid. The columns vary the dimension $p\in\{50,100,200,400,800\}$, while $n_{\scriptscriptstyle \Ssc,0}=400$, $q=100$, $n_{\scriptscriptstyle \Ssc,1}=2000$, $n_{\scriptscriptstyle \Tsc,0}=2000$, and $n_{\scriptscriptstyle \Tsc,1}=3000$ are fixed. Each cell reports the Monte Carlo mean (empirical standard deviation) of $\|\widehat\bbeta-\bar\bbeta^{[0]}\|_2^2$ over $100$ replications; boldface indicates the smallest Monte Carlo mean within each setting and column.}
\label{tab:sim:k10:dimension:full}
\end{table}

Table~\ref{tab:sim:k10:nt0:sensitivity} varies the minority target-unlabeled sample size while holding $n_{\scriptscriptstyle \Ssc,0}=400$ and $p=400$ fixed. This comparison separates the role of unlabeled target covariates from the labeled-minority sample-size and dimensionality effects considered above. The ordering is stable over $n_{\scriptscriptstyle \Tsc,0}\in\{1000,2000,3000\}$: AIMS is the leading method in all three settings, TransGLM$_{\rm iw}$ is the closest external benchmark, and TransFusion is less accurate under this design.

\begin{table}[htb!]
\small
\centering
\setlength{\tabcolsep}{4pt}
\renewcommand{\arraystretch}{0.95}
\begin{tabular}{lccc}
\hline\hline
\multicolumn{4}{c}{Setting I}
\\
Method & \multicolumn{3}{c}{$n_{\scriptscriptstyle \Tsc,0}$}
\\
 & $1000$ & $2000$ & $3000$
\\
\hline
AIMS & \textbf{0.273 (0.175)} & \textbf{0.259 (0.173)} & \textbf{0.259 (0.154)}
\\
IM$_{\rm aLasso}$ & 0.528 (0.228) & 0.521 (0.157) & 0.519 (0.191)
\\
TransGLM$_{\rm iw}$ & 0.411 (0.243) & 0.366 (0.173) & 0.377 (0.171)
\\
TransFusion & 0.833 (0.208) & 0.849 (0.188) & 0.810 (0.180)
\\
\hline
\multicolumn{4}{c}{Setting II}
\\
Method & \multicolumn{3}{c}{$n_{\scriptscriptstyle \Tsc,0}$}
\\
 & $1000$ & $2000$ & $3000$
\\
\hline
AIMS & \textbf{0.280 (0.165)} & \textbf{0.267 (0.166)} & \textbf{0.260 (0.168)}
\\
IM$_{\rm aLasso}$ & 0.474 (0.220) & 0.489 (0.174) & 0.468 (0.179)
\\
TransGLM$_{\rm iw}$ & 0.390 (0.211) & 0.376 (0.185) & 0.386 (0.207)
\\
TransFusion & 0.810 (0.196) & 0.817 (0.189) & 0.799 (0.177)
\\
\hline
\multicolumn{4}{c}{Setting III}
\\
Method & \multicolumn{3}{c}{$n_{\scriptscriptstyle \Tsc,0}$}
\\
 & $1000$ & $2000$ & $3000$
\\
\hline
AIMS & \textbf{0.294 (0.179)} & \textbf{0.288 (0.158)} & \textbf{0.276 (0.161)}
\\
IM$_{\rm aLasso}$ & 0.509 (0.206) & 0.484 (0.194) & 0.502 (0.185)
\\
TransGLM$_{\rm iw}$ & 0.395 (0.174) & 0.382 (0.241) & 0.378 (0.159)
\\
TransFusion & 0.847 (0.191) & 0.825 (0.208) & 0.848 (0.191)
\\
\hline\hline
\end{tabular}
\caption{Sensitivity to the minority target-unlabeled sample size. The columns vary $n_{\scriptscriptstyle \Tsc,0}\in\{1000,2000,3000\}$, while $n_{\scriptscriptstyle \Ssc,0}=400$, $p=400$, $q=100$, $n_{\scriptscriptstyle \Ssc,1}=2000$, and $n_{\scriptscriptstyle \Tsc,1}=3000$ are fixed. Each cell reports the Monte Carlo mean (empirical standard deviation) of $\|\widehat\bbeta-\bar\bbeta^{[0]}\|_2^2$ over $100$ replications; boldface indicates the smallest Monte Carlo mean within each setting and column.}
\label{tab:sim:k10:nt0:sensitivity}
\end{table}

Table~\ref{tab:sim:k10:aims:ablation} compares the intermediate AIMS estimators and aggregation variants at the reference design point. The rows $\widetilde\bbeta^{[0]}$ and $\widehat\bbeta_{\rm Deb}^{[0]}$ are the preliminary minority estimator and the dense debiased estimator from the covariate shift correction step. AIMS$_{\rm min\textit{-}o}$ is the minority-only thresholded estimator $\widehat\bbeta_{\rm Thr}^{[0]}$ from Algorithm~\ref{alg:math}, AIMS$_{{\rm maj}\textit{-}{\rm g}}$ is the majority-guided transfer estimator $\widehat\bbeta_{\rm KTr}^{[0]}$ from Algorithm~\ref{alg:2}, and the AIMS rows are the final aggregated estimators $\widehat\bbeta_{\rm AIMS}^{[0]}$ from Algorithm~\ref{alg:3}, with the unlabeled AIMS row using the default $a=5$. The first three rows have larger errors than the majority-guided and aggregated estimators, showing the benefit of incorporating majority-group information in this local-transfer design. Across the finite temperatures $a\in\{1,2,5,10\}$, aggregation is stable; the hard-selection limit $a=\infty$ is slightly less stable.

\begin{table}[htb!]
\small
\centering
\setlength{\tabcolsep}{5pt}
\renewcommand{\arraystretch}{0.98}
\begin{tabular}{lccc}
\hline\hline
Method & Setting I & Setting II & Setting III
\\
\hline
$\widetilde\bbeta^{[0]}$ & 0.554 (0.193) & 0.522 (0.149) & 0.553 (0.172)
\\
$\widehat\bbeta_{\rm Deb}^{[0]}$ & 1.879 (0.411) & 1.923 (0.344) & 2.014 (0.686)
\\
AIMS$_{\rm min\textit{-}o}$ & 0.708 (0.214) & 0.632 (0.114) & 0.711 (0.168)
\\
AIMS$_{{\rm maj}\textit{-}{\rm g}}$ & \textbf{0.242 (0.150)} & 0.274 (0.161) & 0.263 (0.173)
\\
AIMS & 0.247 (0.147) & \textbf{0.260 (0.150)} & \textbf{0.262 (0.157)}
\\
AIMS ($a=1$) & 0.278 (0.123) & 0.275 (0.128) & 0.292 (0.135)
\\
AIMS ($a=2$) & 0.256 (0.134) & 0.263 (0.138) & 0.270 (0.147)
\\
AIMS ($a=10$) & 0.255 (0.158) & 0.263 (0.157) & 0.266 (0.166)
\\
AIMS ($a=\infty$) & 0.278 (0.186) & 0.282 (0.175) & 0.281 (0.183)
\\
\hline\hline
\end{tabular}
\caption{AIMS ablation at the reference binary-outcome design ($n_{\scriptscriptstyle \Ssc,0}=400$, $p=400$, $q=100$, $n_{\scriptscriptstyle \Ssc,1}=2000$, $n_{\scriptscriptstyle \Tsc,0}=2000$, $n_{\scriptscriptstyle \Tsc,1}=3000$, $C_\tau=2$, and $q_\tau=0.8$). Each cell reports the Monte Carlo mean (empirical standard deviation) of $\|\widehat\bbeta-\bar\bbeta^{[0]}\|_2^2$ over $100$ replications; boldface indicates the smallest Monte Carlo mean within each column.}
\label{tab:sim:k10:aims:ablation}
\end{table}

Table~\ref{tab:sim:k10:majority:reference} reports the corresponding majority-side estimators along the dimensionality axis, with errors measured relative to $\bar\bbeta^{[1]}$. The rows $\widetilde\bbeta^{[1]}$, $\widehat\bbeta_{\rm Deb}^{[1]}$, and $\widehat\bbeta_{\rm Thr}^{[1]}$ are the preliminary, dense debiased, and thresholded majority estimators from the covariate shift correction step. The dense debiased vector has larger squared $\ell_2$ error because it is not sparsified, while the preliminary and thresholded sparse estimators are of the same order. The transfer step uses $\widehat\bbeta_{\rm Thr}^{[1]}$ as the offset in Algorithm~\ref{alg:2}; its error remains stable as $p$ varies.

\begin{table}[htb!]
\small
\centering
\setlength{\tabcolsep}{4pt}
\renewcommand{\arraystretch}{0.95}
\begin{tabular}{lccccc}
\hline\hline
\multicolumn{6}{c}{Setting I}
\\
Method & \multicolumn{5}{c}{$p$}
\\
 & $50$ & $100$ & $200$ & $400$ & $800$
\\
\hline
$\widetilde\bbeta^{[1]}$ & 0.202 (0.101) & 0.187 (0.078) & 0.194 (0.115) & 0.199 (0.095) & 0.189 (0.086)
\\
$\widehat\bbeta_{\rm Deb}^{[1]}$ & 0.680 (0.179) & 0.688 (0.208) & 0.673 (0.200) & 0.671 (0.184) & 0.650 (0.177)
\\
$\widehat\bbeta_{\rm Thr}^{[1]}$ & 0.250 (0.181) & 0.237 (0.180) & 0.240 (0.174) & 0.224 (0.165) & 0.206 (0.168)
\\
\hline
\multicolumn{6}{c}{Setting II}
\\
Method & \multicolumn{5}{c}{$p$}
\\
 & $50$ & $100$ & $200$ & $400$ & $800$
\\
\hline
$\widetilde\bbeta^{[1]}$ & 0.216 (0.111) & 0.202 (0.112) & 0.193 (0.099) & 0.184 (0.092) & 0.162 (0.089)
\\
$\widehat\bbeta_{\rm Deb}^{[1]}$ & 0.735 (0.226) & 0.734 (0.275) & 0.697 (0.196) & 0.684 (0.177) & 0.642 (0.150)
\\
$\widehat\bbeta_{\rm Thr}^{[1]}$ & 0.301 (0.199) & 0.277 (0.207) & 0.251 (0.187) & 0.232 (0.183) & 0.198 (0.167)
\\
\hline
\multicolumn{6}{c}{Setting III}
\\
Method & \multicolumn{5}{c}{$p$}
\\
 & $50$ & $100$ & $200$ & $400$ & $800$
\\
\hline
$\widetilde\bbeta^{[1]}$ & 0.209 (0.114) & 0.193 (0.094) & 0.177 (0.086) & 0.192 (0.078) & 0.180 (0.076)
\\
$\widehat\bbeta_{\rm Deb}^{[1]}$ & 0.712 (0.327) & 0.678 (0.183) & 0.656 (0.179) & 0.648 (0.179) & 0.623 (0.171)
\\
$\widehat\bbeta_{\rm Thr}^{[1]}$ & 0.270 (0.215) & 0.247 (0.164) & 0.213 (0.171) & 0.257 (0.187) & 0.210 (0.158)
\\
\hline\hline
\end{tabular}
\caption{Majority-side diagnostic performance along the dimensionality axis ($q=100$, $n_{\scriptscriptstyle \Ssc,0}=400$, $n_{\scriptscriptstyle \Tsc,0}=2000$, $n_{\scriptscriptstyle \Ssc,1}=2000$, and $n_{\scriptscriptstyle \Tsc,1}=3000$). Each cell reports the Monte Carlo mean (empirical standard deviation) of $\|\widehat\bbeta-\bar\bbeta^{[1]}\|_2^2$ over $100$ replications.}
\label{tab:sim:k10:majority:reference}
\end{table}

Tables~\ref{tab:sim:k10:qboot:sensitivity} and \ref{tab:sim:k10:ctau:sensitivity} summarize tuning sensitivity. Varying the bootstrap quantile $q_\tau$ has only a modest effect over the range $0.6$--$0.9$, indicating that the calibration is not driven by a single quantile choice. The threshold constant $C_\tau$ has a more visible bias-variance tradeoff: very large thresholds are conservative and inflate the error, while moderate thresholds perform best in this reference design.

\begin{table}[htb!]
\small
\centering
\setlength{\tabcolsep}{5pt}
\renewcommand{\arraystretch}{0.98}
\begin{tabular}{lcccc}
\hline\hline
Setting & \multicolumn{4}{c}{$q_\tau$}
\\
 & $0.6$ & $0.7$ & $0.8$ & $0.9$
\\
\hline
Setting I & 0.273 (0.159) & 0.267 (0.160) & 0.247 (0.147) & \textbf{0.243 (0.143)}
\\
Setting II & 0.276 (0.147) & 0.262 (0.150) & 0.260 (0.150) & \textbf{0.255 (0.147)}
\\
Setting III & 0.279 (0.175) & 0.270 (0.170) & \textbf{0.262 (0.157)} & 0.264 (0.149)
\\
\hline\hline
\end{tabular}
\caption{Sensitivity to the bootstrap quantile $q_\tau$ at the reference binary-outcome design with $n_{\scriptscriptstyle \Ssc,0}=400$, $p=400$, $q=100$, $n_{\scriptscriptstyle \Tsc,0}=2000$, and $C_\tau=2$. Each cell reports the Monte Carlo mean (empirical standard deviation) of $\|\widehat\bbeta-\bar\bbeta^{[0]}\|_2^2$ over $100$ replications; boldface indicates the smallest Monte Carlo mean within each row.}
\label{tab:sim:k10:qboot:sensitivity}
\end{table}

\begin{table}[htb!]
\small
\centering
\setlength{\tabcolsep}{5pt}
\renewcommand{\arraystretch}{0.98}
\begin{tabular}{lccccc}
\hline\hline
Setting & \multicolumn{5}{c}{$C_\tau$}
\\
 & $1.0$ & $1.5$ & $2.0$ & $2.5$ & $3.0$
\\
\hline
Setting I & 0.212 (0.123) & \textbf{0.163 (0.134)} & 0.247 (0.147) & 0.349 (0.157) & 0.464 (0.163)
\\
Setting II & 0.211 (0.107) & \textbf{0.170 (0.130)} & 0.257 (0.153) & 0.393 (0.162) & 0.466 (0.147)
\\
Setting III & 0.229 (0.119) & \textbf{0.186 (0.141)} & 0.264 (0.159) & 0.393 (0.177) & 0.476 (0.161)
\\
\hline\hline
\end{tabular}
\caption{Sensitivity to the threshold constant $C_\tau$ at the reference binary-outcome design with $n_{\scriptscriptstyle \Ssc,0}=400$, $p=400$, $q=100$, $n_{\scriptscriptstyle \Tsc,0}=2000$, and $q_\tau=0.8$. Each cell reports the Monte Carlo mean (empirical standard deviation) of $\|\widehat\bbeta-\bar\bbeta^{[0]}\|_2^2$ over $100$ replications; boldface indicates the smallest Monte Carlo mean within each row.}
\label{tab:sim:k10:ctau:sensitivity}
\end{table}

\subsection{Continuous-Outcome Simulation Results}\label{sec:app:sim:continuous}

To assess the robustness of the empirical findings to the outcome type, we repeat the sample-size and dimensionality grids in Section~\ref{sec:sim} with a continuous outcome; the results are reported in Tables~\ref{tab:sim:continuous:sample} and \ref{tab:sim:continuous:dimension}, respectively. All covariate distributions, coefficient vectors \(\balpha^{[r]}\) and \(\bgamma^{[r]}\), nonlinear functions \(u_Y(\Z)\) and \(u_S(\Z)\), and source-sampling mechanisms are kept the same as in the binary-outcome simulation. The reported continuous-outcome tables use \(q=100\), \(n_{\scriptscriptstyle \Ssc,1}=2000\), \(n_{\scriptscriptstyle \Tsc,1}=3000\), and \(n_{\scriptscriptstyle \Tsc,0}=2000\). Along the sample-size axis, \(p=400\) and \(n_{\scriptscriptstyle \Ssc,0}\in\{300,400,500,600\}\); along the dimensionality axis, \(n_{\scriptscriptstyle \Ssc,0}=400\) and \(p\in\{50,100,200,400,800\}\). We use the same implementation and tuning constants as in the main simulation, namely \(C_\tau=2\), \(q_\tau=0.8\), \(B=500\), and aggregation temperature \(a=5\), and report Monte Carlo averages over 100 replications.

Specifically, conditional on \(\Z\) and \(R=r\), we generate \(Y=\Z\trans\bgamma^{[r]}+\epsilon\) in Settings I and III and \(Y=\Z\trans\bgamma^{[r]}+u_Y(\Z)+\epsilon\) in Setting II, where \(\epsilon\sim N(0,1)\). Thus, Setting II keeps the density ratio model correctly specified while misspecifying the imputation model, and Setting III keeps the imputation model correctly specified while misspecifying the density ratio model. Tables~\ref{tab:sim:continuous:sample} and \ref{tab:sim:continuous:dimension} show that the continuous-outcome results are consistent with the binary-outcome findings: AIMS has the smallest error across the reported sample-size and dimensionality grids, while TransGLM$_{\rm iw}$ is the closest external competitor.

\begin{table}[htb!]
\small
\centering
\setlength{\tabcolsep}{4pt}
\renewcommand{\arraystretch}{0.95}
\begin{tabular}{lcccc}
\hline\hline
\multicolumn{5}{c}{Setting I}
\\
Method & \multicolumn{4}{c}{$n_{\scriptscriptstyle \Ssc,0}$}
\\
 & $300$ & $400$ & $500$ & $600$
\\
\hline
AIMS & \textbf{0.093 (0.053)} & \textbf{0.067 (0.046)} & \textbf{0.062 (0.037)} & \textbf{0.064 (0.045)}
\\
IW & 0.554 (0.243) & 0.436 (0.159) & 0.385 (0.159) & 0.321 (0.137)
\\
IW$_{\rm aLasso}$ & 0.801 (0.374) & 0.625 (0.266) & 0.578 (0.233) & 0.471 (0.195)
\\
IM & 0.323 (0.200) & 0.232 (0.132) & 0.193 (0.118) & 0.171 (0.087)
\\
IM$_{\rm aLasso}$ & 0.197 (0.079) & 0.158 (0.054) & 0.124 (0.042) & 0.107 (0.037)
\\
IM$_{\rm RF}$ & 1.900 (0.214) & 1.748 (0.181) & 1.643 (0.159) & 1.599 (0.138)
\\
IM$_{\rm XGB}$ & 1.172 (0.223) & 0.941 (0.170) & 0.793 (0.160) & 0.707 (0.151)
\\
CORAL & 2.052 (0.262) & 2.016 (0.243) & 2.026 (0.229) & 2.045 (0.208)
\\
TransFusion & 0.969 (0.152) & 0.957 (0.123) & 0.949 (0.108) & 0.949 (0.105)
\\
TransGLM & 1.079 (0.181) & 1.045 (0.140) & 1.035 (0.116) & 1.028 (0.125)
\\
TransGLM$_{\rm iw}$ & 0.295 (0.148) & 0.238 (0.133) & 0.234 (0.135) & 0.201 (0.108)
\\
\hline
\multicolumn{5}{c}{Setting II}
\\
Method & \multicolumn{4}{c}{$n_{\scriptscriptstyle \Ssc,0}$}
\\
 & $300$ & $400$ & $500$ & $600$
\\
\hline
AIMS & \textbf{0.118 (0.066)} & \textbf{0.098 (0.045)} & \textbf{0.095 (0.043)} & \textbf{0.094 (0.053)}
\\
IW & 0.506 (0.211) & 0.425 (0.195) & 0.350 (0.148) & 0.284 (0.117)
\\
IW$_{\rm aLasso}$ & 0.851 (0.411) & 0.665 (0.281) & 0.565 (0.246) & 0.447 (0.204)
\\
IM & 0.249 (0.161) & 0.197 (0.135) & 0.145 (0.069) & 0.128 (0.061)
\\
IM$_{\rm aLasso}$ & 0.254 (0.089) & 0.226 (0.066) & 0.199 (0.058) & 0.184 (0.048)
\\
IM$_{\rm RF}$ & 1.664 (0.190) & 1.511 (0.163) & 1.400 (0.145) & 1.353 (0.114)
\\
IM$_{\rm XGB}$ & 0.920 (0.169) & 0.739 (0.152) & 0.595 (0.123) & 0.527 (0.118)
\\
CORAL & 2.193 (0.315) & 2.172 (0.298) & 2.192 (0.274) & 2.214 (0.225)
\\
TransFusion & 1.146 (0.165) & 1.134 (0.140) & 1.118 (0.123) & 1.118 (0.109)
\\
TransGLM & 1.144 (0.184) & 1.111 (0.143) & 1.096 (0.129) & 1.088 (0.124)
\\
TransGLM$_{\rm iw}$ & 0.272 (0.130) & 0.230 (0.150) & 0.209 (0.112) & 0.173 (0.093)
\\
\hline
\multicolumn{5}{c}{Setting III}
\\
Method & \multicolumn{4}{c}{$n_{\scriptscriptstyle \Ssc,0}$}
\\
 & $300$ & $400$ & $500$ & $600$
\\
\hline
AIMS & \textbf{0.085 (0.054)} & \textbf{0.072 (0.040)} & \textbf{0.070 (0.042)} & \textbf{0.062 (0.038)}
\\
IW & 0.519 (0.162) & 0.464 (0.159) & 0.348 (0.121) & 0.325 (0.116)
\\
IW$_{\rm aLasso}$ & 0.783 (0.288) & 0.703 (0.297) & 0.546 (0.234) & 0.473 (0.187)
\\
IM & 0.327 (0.179) & 0.258 (0.147) & 0.189 (0.112) & 0.173 (0.118)
\\
IM$_{\rm aLasso}$ & 0.191 (0.069) & 0.155 (0.055) & 0.121 (0.040) & 0.106 (0.032)
\\
IM$_{\rm RF}$ & 1.811 (0.209) & 1.725 (0.158) & 1.609 (0.163) & 1.531 (0.156)
\\
IM$_{\rm XGB}$ & 1.123 (0.213) & 0.958 (0.171) & 0.773 (0.167) & 0.684 (0.161)
\\
CORAL & 2.024 (0.257) & 2.040 (0.221) & 2.027 (0.234) & 1.980 (0.206)
\\
TransFusion & 0.961 (0.154) & 0.997 (0.136) & 0.976 (0.117) & 0.957 (0.108)
\\
TransGLM & 1.040 (0.176) & 1.074 (0.161) & 1.038 (0.139) & 1.012 (0.118)
\\
TransGLM$_{\rm iw}$ & 0.311 (0.136) & 0.293 (0.124) & 0.242 (0.106) & 0.231 (0.115)
\\
\hline
\hline\hline
\end{tabular}
\caption{Continuous-outcome sample-size grid. The columns vary the labeled minority-source sample size $n_{\scriptscriptstyle \Ssc,0}\in\{300,400,500,600\}$, while $p=400$, $q=100$, $n_{\scriptscriptstyle \Ssc,1}=2000$, $n_{\scriptscriptstyle \Tsc,0}=2000$, and $n_{\scriptscriptstyle \Tsc,1}=3000$ are fixed. Each cell reports the Monte Carlo mean (empirical standard deviation) of $\|\widehat\bbeta-\bar\bbeta^{[0]}\|_2^2$ over 100 replications. Boldface indicates the smallest Monte Carlo mean within each setting and column.}
\label{tab:sim:continuous:sample}
\end{table}

\begin{table}[htb!]
\small
\centering
\setlength{\tabcolsep}{4pt}
\renewcommand{\arraystretch}{0.95}
\begin{tabular}{lccccc}
\hline\hline
\multicolumn{6}{c}{Setting I}
\\
Method & \multicolumn{5}{c}{$p$}
\\
 & $50$ & $100$ & $200$ & $400$ & $800$
\\
\hline
AIMS & \textbf{0.093 (0.048)} & \textbf{0.100 (0.103)} & \textbf{0.086 (0.047)} & \textbf{0.067 (0.046)} & \textbf{0.073 (0.049)}
\\
IW & 0.446 (0.160) & 0.445 (0.166) & 0.444 (0.192) & 0.436 (0.159) & 0.436 (0.160)
\\
IW$_{\rm aLasso}$ & 0.640 (0.269) & 0.663 (0.309) & 0.643 (0.262) & 0.625 (0.266) & 0.613 (0.263)
\\
IM & 0.196 (0.128) & 0.219 (0.137) & 0.217 (0.152) & 0.232 (0.132) & 0.313 (0.176)
\\
IM$_{\rm aLasso}$ & 0.161 (0.051) & 0.158 (0.048) & 0.158 (0.047) & 0.158 (0.054) & 0.157 (0.061)
\\
IM$_{\rm RF}$ & 1.591 (0.180) & 1.630 (0.169) & 1.684 (0.177) & 1.748 (0.181) & 1.859 (0.189)
\\
IM$_{\rm XGB}$ & 0.743 (0.185) & 0.828 (0.186) & 0.870 (0.190) & 0.941 (0.170) & 1.095 (0.204)
\\
CORAL & 2.029 (0.220) & 2.021 (0.225) & 2.047 (0.251) & 2.016 (0.243) & 2.047 (0.218)
\\
TransFusion & 0.958 (0.121) & 0.965 (0.140) & 0.954 (0.138) & 0.957 (0.123) & 0.957 (0.125)
\\
TransGLM & 1.052 (0.142) & 1.066 (0.148) & 1.044 (0.149) & 1.045 (0.140) & 1.053 (0.136)
\\
TransGLM$_{\rm iw}$ & 0.243 (0.118) & 0.238 (0.114) & 0.250 (0.134) & 0.238 (0.133) & 0.245 (0.127)
\\
\hline
\multicolumn{6}{c}{Setting II}
\\
Method & \multicolumn{5}{c}{$p$}
\\
 & $50$ & $100$ & $200$ & $400$ & $800$
\\
\hline
AIMS & \textbf{0.119 (0.060)} & \textbf{0.117 (0.088)} & \textbf{0.116 (0.075)} & \textbf{0.098 (0.045)} & \textbf{0.105 (0.047)}
\\
IW & 0.408 (0.163) & 0.409 (0.165) & 0.405 (0.153) & 0.425 (0.195) & 0.399 (0.149)
\\
IW$_{\rm aLasso}$ & 0.637 (0.299) & 0.729 (0.329) & 0.649 (0.269) & 0.665 (0.281) & 0.658 (0.283)
\\
IM & 0.163 (0.115) & 0.171 (0.131) & 0.171 (0.104) & 0.197 (0.135) & 0.223 (0.112)
\\
IM$_{\rm aLasso}$ & 0.258 (0.066) & 0.258 (0.067) & 0.244 (0.069) & 0.226 (0.066) & 0.188 (0.056)
\\
IM$_{\rm RF}$ & 1.353 (0.163) & 1.388 (0.159) & 1.451 (0.161) & 1.511 (0.163) & 1.607 (0.153)
\\
IM$_{\rm XGB}$ & 0.580 (0.158) & 0.616 (0.155) & 0.668 (0.157) & 0.739 (0.152) & 0.848 (0.148)
\\
CORAL & 2.156 (0.256) & 2.187 (0.289) & 2.207 (0.294) & 2.172 (0.298) & 2.209 (0.270)
\\
TransFusion & 1.126 (0.129) & 1.141 (0.160) & 1.126 (0.149) & 1.134 (0.140) & 1.128 (0.142)
\\
TransGLM & 1.109 (0.143) & 1.128 (0.167) & 1.106 (0.158) & 1.111 (0.143) & 1.116 (0.148)
\\
TransGLM$_{\rm iw}$ & 0.237 (0.127) & 0.208 (0.103) & 0.226 (0.116) & 0.230 (0.150) & 0.225 (0.107)
\\
\hline
\multicolumn{6}{c}{Setting III}
\\
Method & \multicolumn{5}{c}{$p$}
\\
 & $50$ & $100$ & $200$ & $400$ & $800$
\\
\hline
AIMS & \textbf{0.088 (0.043)} & \textbf{0.082 (0.044)} & \textbf{0.079 (0.044)} & \textbf{0.072 (0.040)} & \textbf{0.070 (0.049)}
\\
IW & 0.430 (0.181) & 0.431 (0.172) & 0.438 (0.187) & 0.464 (0.159) & 0.430 (0.138)
\\
IW$_{\rm aLasso}$ & 0.631 (0.285) & 0.660 (0.310) & 0.661 (0.308) & 0.703 (0.297) & 0.656 (0.285)
\\
IM & 0.201 (0.125) & 0.226 (0.158) & 0.216 (0.137) & 0.258 (0.147) & 0.292 (0.157)
\\
IM$_{\rm aLasso}$ & 0.156 (0.050) & 0.162 (0.056) & 0.157 (0.047) & 0.155 (0.055) & 0.151 (0.055)
\\
IM$_{\rm RF}$ & 1.527 (0.152) & 1.565 (0.167) & 1.651 (0.179) & 1.725 (0.158) & 1.762 (0.189)
\\
IM$_{\rm XGB}$ & 0.727 (0.163) & 0.779 (0.175) & 0.861 (0.192) & 0.958 (0.171) & 1.044 (0.193)
\\
CORAL & 2.020 (0.237) & 2.013 (0.266) & 2.002 (0.244) & 2.040 (0.221) & 1.988 (0.237)
\\
TransFusion & 0.997 (0.129) & 0.975 (0.136) & 0.974 (0.132) & 0.997 (0.136) & 0.963 (0.124)
\\
TransGLM & 1.069 (0.148) & 1.053 (0.154) & 1.048 (0.148) & 1.074 (0.161) & 1.043 (0.149)
\\
TransGLM$_{\rm iw}$ & 0.259 (0.122) & 0.265 (0.122) & 0.276 (0.155) & 0.293 (0.124) & 0.280 (0.109)
\\
\hline
\hline\hline
\end{tabular}
\caption{Continuous-outcome dimensionality grid. The columns vary the dimension $p\in\{50,100,200,400,800\}$, while $n_{\scriptscriptstyle \Ssc,0}=400$, $q=100$, $n_{\scriptscriptstyle \Ssc,1}=2000$, $n_{\scriptscriptstyle \Tsc,0}=2000$, and $n_{\scriptscriptstyle \Tsc,1}=3000$ are fixed. Each cell reports the Monte Carlo mean (empirical standard deviation) of $\|\widehat\bbeta-\bar\bbeta^{[0]}\|_2^2$ over 100 replications. Boldface indicates the smallest Monte Carlo mean within each setting and column.}
\label{tab:sim:continuous:dimension}
\end{table}

\subsection{Additional T2D Benchmark Results}
\label{app:t2d:full}

Table \ref{tab:eval:beta:full} reports the complete T2D real-data comparison, including the adaptive-Lasso variants, RF-based imputation, TransGLM variants, and the MAP-based phenotyping baseline that are omitted from the main-text table for readability.

\begin{table}[htb!]
\small
\centering
\resizebox{\textwidth}{!}{
\begin{tabular}{lccccccccccccc}
\hline \hline\\[-2.5ex]
   & AIMS & IW & IW$_{\rm aLasso}$ & IM & IM$_{\rm aLasso}$ & IM$_{\rm RF}$ & IM$_{\rm XGB}$ & CORAL & TransFusion & TransGLM & TransGLM$_{\rm iw}$ & IM$_{\rm PheNorm}$ & IM$_{\rm MAP}$ \\
\hline\\[-2.5ex]
BSS  & \textbf{0.36} & -0.11 & -0.22 & 0.23 & 0.24 & -1.00 & -1.07 & -0.11 & 0.01 & 0.00 & 0.00 & -0.20 & -0.29\\
GOF  & \textbf{0.16} & -0.85 & -1.87 & -0.02 & -0.02 & -1.36 & -1.46 & -0.88 & -0.25 & -0.78 & -0.82 & -0.45 & -0.54\\
AUC & \textbf{0.89} & 0.64 & 0.62 & 0.81 & 0.83 & 0.78 & 0.81 & 0.62 & 0.81 & 0.71 & 0.70 & 0.82 & 0.82\\
\hline\hline\\[-2ex]
\end{tabular}}

\caption{\label{tab:eval:beta:full}Full predictive performance comparison for the T2D risk models evaluated on the validation data. This table includes additional benchmark variants omitted from Table \ref{tab:eval:beta} for readability.}
\end{table}

\subsection{Additional \texorpdfstring{\(\Delta\Delta G\)}{Delta Delta G} Results on the Target Majority Subgroup}
\label{app:ddg:full}

Table~\ref{tab:eval:ddg2} compares the preliminary estimator $\widetilde \bbeta^{[r]}$ (AIMS$_{\rm Init}$), the dense debiased estimator $\widehat{\bbeta}^{[r]}_{\rm Dense\_Deb}$ (AIMS$_{\rm Dense\_Deb}$), and the thresholded estimator $\widehat\bbeta_{\rm Thr}^{[r]}$ (AIMS$_{\rm Thr}$) for the majority subgroup. The thresholded estimator achieved the strongest Pearson and Spearman correlations with the held-out target-majority $\Delta\Delta G$ labels, while the preliminary estimator yielded slightly smaller RMSE and MAE. The dense debiased estimator performed worst under the prediction-error metrics RMSE and MAE, which is consistent with its role as an intermediate dense debiased estimator rather than a final prediction rule.

\begin{table}[htb!]
\footnotesize
\centering
\begin{tabular}{lrrr}
\hline \hline\\[-2.5ex]
   & AIMS$_{\rm Init}$ & AIMS$_{\rm Dense\_Deb}$ & AIMS$_{\rm Thr}$ \\
\hline\\[-2.5ex] 
RMSE  & 2.24  & 3.03 & 2.31\\
MAE  & 1.54  & 2.20 & 1.57\\
Pearson & 0.06 & 0.14 & 0.20 \\
Spearman & 0.09 & 0.20  & 0.31 \\
\hline\hline\\[-2ex] 
\end{tabular}
\caption{\label{tab:eval:ddg2} Predictive performance of the preliminary, dense debiased, and thresholded AIMS estimators on the target majority subgroup, evaluated using held-out ground-truth $\Delta\Delta G$ labels.}
\end{table}

\clearpage
\bibliography{library}

\end{document}